\documentclass[apsrev4-2, twocolumn,footinbib,superscriptaddress,floatfix,bibliography]{revtex4-2}
\usepackage{times}
\usepackage{float}
\usepackage{multirow}
\usepackage[dvipsnames]{xcolor}
\usepackage{amsmath}
\usepackage{amsthm}
\usepackage{amssymb}
\usepackage{amsbsy}
\usepackage{enumitem}
\usepackage{eufrak}
\usepackage{wasysym}
\usepackage[english]{babel}
\usepackage[T1]{fontenc}
\usepackage[utf8]{inputenc} 
\usepackage{graphicx}
\usepackage[colorlinks,bookmarks=false,citecolor=blue,linkcolor=red,urlcolor=blue]{hyperref}
\usepackage{ulem}
\usepackage{pstricks}
\usepackage{rotating}			       % creates rotated page (landscape) 
\usepackage[caption=false]{subfig}	
\usepackage{mathtools}
\usepackage{tabularx,hhline,makecell}
\usepackage{tabularx,enumitem}

\newcolumntype{P}[1]{>{\centering\arraybackslash}p{#1}}

\newcommand{\ns}{N_{\rm S}}

\begin{document}
%%%%%%%%%%%%%%%%%%%%%%%%%%%%%%%%%%%%%

%%%%%%%%%%%%%%%%%%%%%%%%%%%%%%%%%%%%%

\title{
Eight-color chiral spin liquid in the $S=1$ bilinear-biquadratic model with 
Kitaev interactions 
}

%%%%%%%%%%%%%%%%%%%%%%%%%%%%%%%%%%%%%

%%%%%%%%%%%%%%%%%%%%%%%%%%%%%%%%%%%%%
\author{Rico Pohle}
%%%%%%%%%%%%%%%%%%%%%%%%%%%%%%%%%%%%%
\affiliation{Graduate School of Science and Technology, Keio University, 
Yokohama 223-8522, Japan}
\affiliation{Department of Applied Physics, University of Tokyo, Hongo,  Bunkyo-ku,
Tokyo, 113-8656, Japan}
%%%%%%%%%%%%%%%%%%%%%%%%%%%%%%%%%%%%%
\author{Nic Shannon}
%%%%%%%%%%%%%%%%%%%%%%%%%%%%%%%%%%%%%
\affiliation{Theory of Quantum Matter Unit, Okinawa Institute of Science and 
Technology Graduate University, Onna-son, Okinawa 904-0412, Japan} 
%%%%%%%%%%%%%%%%%%%%%%%%%%%%%%%%%%%%%
\author{Yukitoshi Motome}
%%%%%%%%%%%%%%%%%%%%%%%%%%%%%%%%%%%%%
\affiliation{Department of Applied Physics, University of Tokyo, Hongo,  Bunkyo-ku,
Tokyo, 113-8656, Japan}

%%%%%%%%%%%%%%%%%%%%%%%%%%%%%%%%%%%%%
\date{\today}
%%%%%%%%%%%%%%%%%%%%%%%%%%%%%%%%%%%%%

%%%%%%%%%%%%%%%%%%%%%%%%%%%%%%%%%%%%%
\begin{abstract}
%%%%%%%%%%%%%%%%%%%%%%%%%%%%%%%%%%%%%

Multipolar spin systems provide a rich ground for the emergence of unexpected 
states of matter due to their enlarged spin degree of freedom. 
In this study, with a specific emphasis on $S=1$ magnets, we explore the 
interplay between spin nematic states and spin liquids.
Based on  the foundations laid in the prior work 
\mbox{[R. Pohle {\it et al.}, Phys. Rev. B {\bf 107}, L140403 (2023)]}, we investigate 
the $S=1$ Kitaev model with bilinear-biquadratic interactions, which stabilizes, 
next to Kitaev spin liquid, spin nematic and triple-$q$ phases, also an 
exotic chiral spin liquid. 
Through a systematic reduction of the spin degree of freedom -- from 
$\mathbb{CP}^{2}$ to $\mathbb{CP}^{1}$ and ultimately to a discrete 
eight-color model -- we provide an intuitive 
understanding of the nature and origin of this chiral spin liquid.  
We find that the chiral spin liquid is characterized by an extensive 
ground-state degeneracy, bound by a residual entropy, extremely short-ranged 
correlations, a nonzero scalar spin chirality marked by $\mathbb{Z}_{2}$ flux order, 
and a gapped continuum of excitations. 
Our work contributes not only to the specific exploration of $S=1$ Kitaev magnets 
but also to the broader understanding of the importance of multipolar spin
degree of freedom on the ground state and excitation properties in quantum 
magnets.

%%%%%%%%%%%%%%%%%%%%%%%%%%%%%%%%%%%%%
\end{abstract}
%%%%%%%%%%%%%%%%%%%%%%%%%%%%%%%%%%%%%

%%%%%%%%%%%%%%%%%%%%%%%%%%%%%%%%%%%%%%%%%%
\maketitle
%%%%%%%%%%%%%%%%%%%%%%%%%%%%%%%%%%%%%%%%%%

%%%%%%%%%%%%%%%%%%%%%%%%%%%%%%%%%%%%%%%%%%
%
% 					INTRODUCTION
%
%%%%%%%%%%%%%%%%%%%%%%%%%%%%%%%%%%%%%%%%%%
\section{Introduction}									\label{sec:Intro}
%%%%%%%%%%%%%%%%%%%%%%%%%%%%%%%%%%%%%%%%%%
%
%

%% What are Multipolar Mott insulators and why are they interesting? 
%
Multipolar systems exhibit higher-order moments, such as 
quadrupoles and octupoles, even in absence of monopoles and dipoles. 
The inherent fluctuations and allowed higher-order interactions in these systems provide 
a rich landscape of unconventional phases, evident from multipolar order observed in 
quantum magnets~\cite{Chen2010, Fu2015, Hirai2020, Maharaj2020, Paramekanti2020}, 
unconventional superconductors~\cite{Thalmeier2008, Matsubayashi2012, Freyer2018, Lee2018, Sim2020, Patri2022}, 
and topological insulators~\cite{Benalcazar2017, Romhanyi2019}.
Particular exotic examples, including
multipolar quantum spin ice~\cite{Sibille2015, Gaudet2019, Sibille2020}, 
$\mathbb{CP}^{2}$ skyrmion crystals~\cite{Akagi2021, Amari2022, Zhang2023}, and 
unconventional orders in the higher-spin Kondo lattice model~\cite{Lai2018, Masui2022},
demonstrate the diverse range of intriguing quantum phenomena present in 
multipolar condensed matter.

%% spin nematics and why they are interesting
%
Multipolar magnets made of $S=1$ moments are particularly intriguing, as they 
strike a delicate balance with a spin length that is ``small enough'' to exhibit strong quantum 
effects, while simultaneously being ``large enough'' to give rise to onsite quadrupole moments
\cite{Papanicolaou1988, Tsunetsugu2006, Lauchli2006, Penc2011}.
Quadrupoles break spin-rotation symmetry by selecting an axis without specifying 
a particular direction, closely resembling properties observed in 
classical liquid crystals \cite{Kelker1973, Kelker1988, deGennes1993}.
This unique characteristic makes them an ideal playground for exploring 
topological defects \cite{Mermin1966, Alexander2012},
investigating out-of-equilibrium effects \cite{Chuang1991, Bowick1994},
and drawing potential analogies to gravity~\cite{Volovik1990, Volovik2009, Chojnacki2023}.

Despite being theoretically proposed almost 60 years ago \cite{Blume1969, Matveev1973}, 
the experimental exploration of $S=1$ magnets has faced substantial challenges, 
primarily since their quadrupolar order remains hidden from conventional magnetic probes like 
neutron scattering~\cite{Andreev1984, Smerald2013}.
Fortunately, recent advancements in experimental techniques such as 
Raman scattering~\cite{Michaud2011, Valentine2020},  
resonant inelastic X-ray scattering (RIXS)~\cite{Savary2015, Takahashi2021, Nag2022}, and
the exploration of physical properties like the magnetocaloric effect~\cite{Kohama2019}, 
combined with the development of powerful
numerical techniques~\cite{HaoZhang2021, Remund2022, Dahlbom2022a, Dahlbom2022b, Iwazaki2023, SUNNY}, 
have expanded our capabilities to interpret thermodynamic and dynamic properties
of multipolar magnets in real materials.
The successful application of these diverse methods has significantly advanced 
our understanding of various $S=1$ compounds, including 
NiGa$_2$S$_4$~\cite{Nakatsuji2005, Stoudenmire2009, Valentine2020}, 
FeI$_2$~\cite{Bai2021, Dahlbom2024}, and 
Ba$_2$FeSi$_2$O$_7$~\cite{Do2023}.

%
%%%%%%%%%%%%%%%%%%%%%%%%%%%%%%%%%%%%%
%. Fig. -- Ground state phase diagrams 
%%%%%%%%%%%%%%%%%%%%%%%%%%%%%%%%%%%%%
\begin{figure*}[t]
	\centering
	\includegraphics[width=0.99\textwidth]{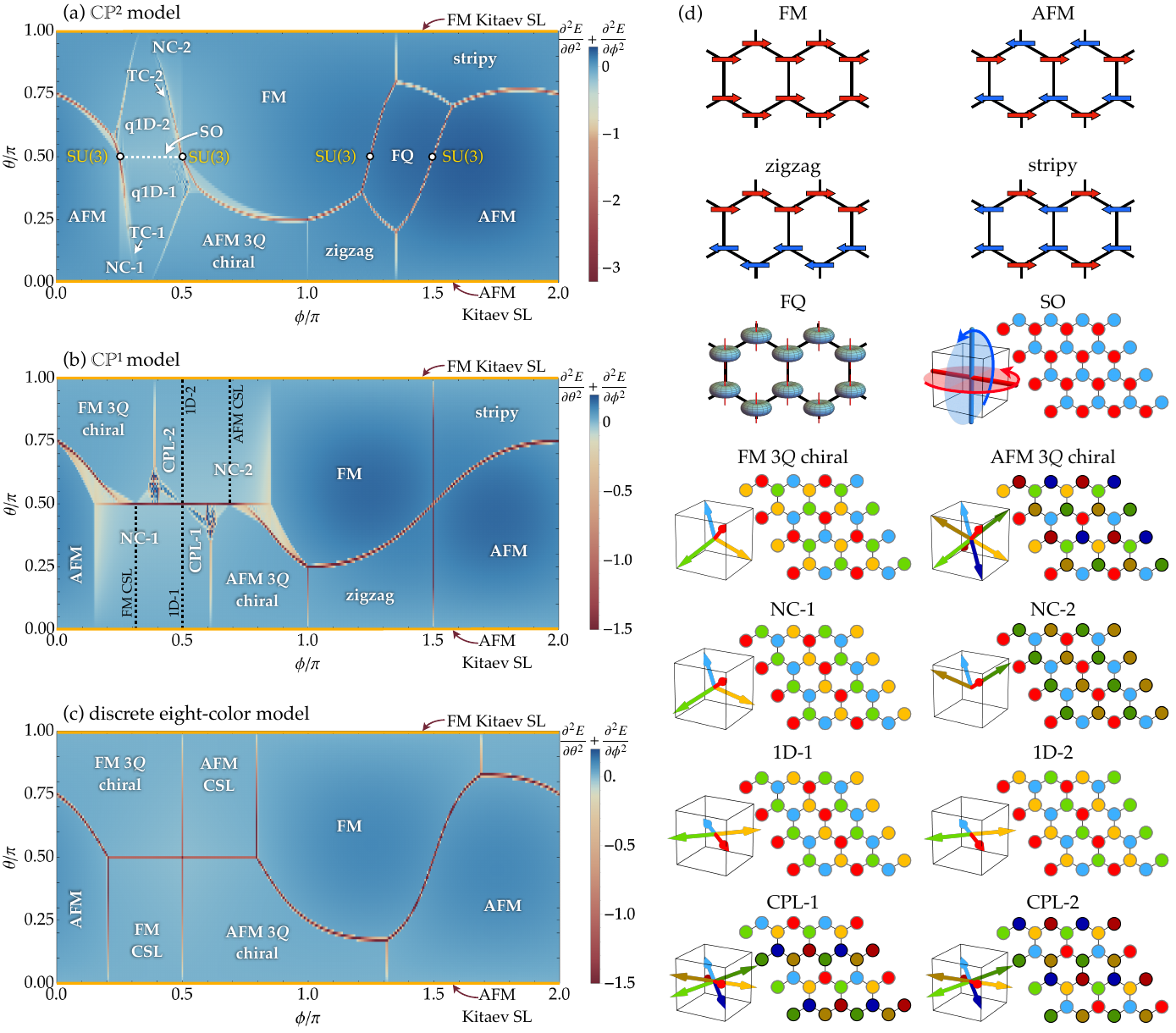} 	
	\caption{	
	Ground-state phase diagrams in the 2D equirectangular projection as function
	of $\theta$ and $\phi$ [see Eq.~\eqref{eq:parametrization.HBBQ-K}].
	We plot the second derivative of the internal energy 
	as contour color to make phase boundaries visible.
	(a) Solution of $\mathcal{H}^\mathcal{A}_{ {\sf BBQ-K} }$ [see Eq.~\eqref{eq:H.BBQ-K.A}] 
	in the spin space $\mathbb{CP}^{2}$, which accounts for all local degrees of freedom 
	of an $S=1$ moment, including dipolar and quadrupolar components. 
	(b) Solution of $\mathcal{H}^\mathcal{S}_{ {\sf BBQ-K} }$ [see Eq.~\eqref{eq:H.BBQ-K.S}] 
	in the spin space $\mathbb{CP}^{1}$, which considers only dipolar degree of 
	freedom of classical Heisenberg spins.
	(c) Solution of $\mathcal{H}^{ {\sf 8c} }_{ {\sf BBQ-K}}$ 
	for the discrete eight-color model [see Eq.~\eqref{eq:H.BBQ-K.8c}], where the dipolar 
	degree of freedom is discretized and allowed to point only at the corners of a unit 
	cube (see definition of spin states in Table~\ref{tab:8.color.spins}).
	(d) Real-space configurations of selected phases. 
	Ferromagnetic (FM), antiferromagnetic (AFM), and triple-$q$ 
	chiral magnetic orders (FM 3$Q$ chiral and AFM 3$Q$ chiral),
	as well as the semiclassical analog of the Kitaev spin liquid (SL), 
	are stabilized in all three phase diagrams. 
	Noncoplanar (NC), (quasi-)one-dimensional (q1D / 1D), canted planar (CPL), 
	and twisted conical (TC) phases are stabilized for the models with continuous 
	degrees of freedom in (a) and (b), while ferroquadrupolar (FQ), and 
	semiordered (SO) phases are purely quadrupolar 
	in nature and are only stabilized in (a).
	FM and AFM chiral spin liquid (CSL) states emerge as the ground state along 
	two singular lines in (b) and form a phase in (c).
	However, they do not appear in the ground state in (a), but instead 
	become a stable phase at finite temperatures
	(see discussion in Sec.~\ref{sec:SU3.model}).
	Results in (a) and (b) were obtained by variational energy minimization 
	for a finite-size cluster of $\ns = 28 \ 800$ spins under periodic boundary 
	conditions, while (c) was solved analytically for the thermodynamic limit.  
	}
	\label{fig:PD.all}
\end{figure*}
%%%%%%%%%%%%%%%%%%%%%%%%%%%%%%%%%%%%%%%
%

%% multipolar Kitaev systems and S=1 Kitaev, and our old results
%
Recently, higher-spin Kitaev systems have attracted substantial 
attention~\cite{Baskaran2008, Koga2018, Suzuki2018, Oitmaa2018, Dong2020, Hickey2020, Lee2020, Khait2021, Fukui2022, Rayyan2023, Ma2023, Liu2023, Georgiou2024},
as they offer a new avenue for studying quantum spin liquids 
with potentially multipolar character~\cite{Jin2022, Carvalho2023} and 
fractionalized excitations carrying multipolar moments~\cite{Rayyan2023}.
Theoretical proposals suggest that magnets with 
strong Hund's coupling and spin-orbit interactions could pave the way 
for experimental realizations~\cite{Stavropoulos2019}, with promising candidate 
materials such as NaNi$_2$BiO$_{6-\delta}$~\cite{Scheie2019} and KNiAsO$_4$~\cite{Taddei2023}. 
Furthermore, $S=1$ magnets naturally allow for higher-order interactions, such as 
biquadratic interactions, which effectively couple the quadrupole degree of freedom between 
individual spins. 

In this context, the recent work by the authors 
explored the 
interplay between spin nematics and spin liquids 
in the $S=1$ Kitaev model 
with bilinear-biquadratic (BBQ) interactions~\cite{Pohle2023}. 
The richness of this model unfolds with a plethora of phases, including spin nematic, and 
Kitaev spin liquid states, alongside 
triple-$q$ order with nonzero scalar spin chirality, ferro, antiferro, zigzag and stripy phases.
Particularly intriguing is the presence of exotic phases where dipole and quadrupole moments compete.
These include a quasi-one-dimensional 
coplanar phase, a twisted conical phase, and a noncoplanar ordered state.
A comprehensive phase diagram showcasing these diverse phases is shown in Fig.~\ref{fig:PD.all}(a).

Even though the model primarily stabilizes ordered phases in the ground state, it holds an 
intriguing surprise.
In the vicinity of the noncoplanar ordered state, the model hosts an unconventional 
classical chiral spin liquid (CSL). 
This CSL exhibits an extensive degeneracy of states and retains the Kitaev spin liquid (SL)
feature of $\mathbb{Z}_{2}$ flux order, along with a nonzero scalar spin chirality~\cite{Pohle2023}.
However, at a semiclassical level, this CSL does not appear in the ground state, as seen by its absence 
in Fig.~\ref{fig:PD.all}(a). 
Instead, it is stabilized at finite temperatures,  resisting magnetic dipolar order solely 
through thermal fluctuations, rather than quantum fluctuations or quantum entanglement.

%% Here we ... 
%
In this article, we expand on the theory discussed in our prior work, 
Ref.~[\onlinecite{Pohle2023}], with the goal of presenting a pedagogical 
explanation of the inherent nature of the CSL.
To achieve this, we start by representing local 
$S=1$ magnetic moments as SU(3) spin-coherent states in the spin space 
$\mathbb{CP}^{2}$.
While this approach is semiclassical and does not include entanglement 
between spins, it properly accounts for the onsite quadrupole
and dipole characteristics expected for $S=1$ moments~\cite{Remund2022}.
We proceed by systematically reducing the degrees of freedom for local spins
to the submanifold of SU(2) spin-coherent states in the spin space $\mathbb{CP}^{1}$.
This representation treats spins as classical dipole vectors pointing to 
the Bloch sphere, as done in conventional Monte Carlo (MC) simulations for 
classical Heisenberg spins.
Finally, we introduce a minimal model, the eight-color model, in which 
dipolar spins are restricted to eight discrete states, reminiscent of 
an eight-states clock model.
The corresponding phase diagrams for each of these cases are depicted in 
Figs.~\ref{fig:PD.all}(a), \ref{fig:PD.all}(b), and \ref{fig:PD.all}(c). 
The eight-color model, inspired by insights from simulations 
using  the $\mathbb{CP}^{1}$ and  $\mathbb{CP}^{2}$ models, effectively 
captures all physical properties of the CSL and enables us to understand 
its characteristics analytically.
The simplicity of the eight-color model allows us to identify local bond 
constraints that unequivocally determine all observed properties of 
this CSL.
It also allows us to implement a hexagon cluster update, which 
connects different states in the spin liquid manifold and dramatically 
reduces correlation times in numerical simulations. 
We complement our analysis with results for the dynamical structure factor,
unveiling a gapped continuum of excitations.

The remainder of this article is structured as follows. 
In Sec.~\ref{sec:model}, we introduce the $S=1$ BBQ-Kitaev model on the honeycomb 
lattice. 
We analyze the same model for different local spin degrees of freedom, 
namely, SU(3) spin-coherent states of $\mathbb{CP}^{2}$ (the $\mathbb{CP}^{2}$ model)
which contain onsite dipole and quadrupole moments,
SU(2) spin-coherent states of $\mathbb{CP}^{1}$ (the $\mathbb{CP}^{1}$ model)
which contain only onsite dipole moments,
and the eight-color model which allows for only eight discretized dipolar spin states.
In Sec.~\ref{sec:BBQ-K.results}, we present ground-state and finite-temperature results
of the BBQ-Kitaev model. 
In Sec.~\ref{sec:phase_diagrams}, we discuss the influence of the local spin degree 
of freedom by explicitly comparing the ground-state phase diagrams 
between the $\mathbb{CP}^{2}$,
the $\mathbb{CP}^{1}$, 
and the eight-color models.
In Sec.~\ref{sec:8c.model}, we discuss the eight-color model and provide an 
analytical explanation for all fundamental properties of the observed CSL, substantiated 
by comparisons to numerical simulations.
In Sec.~\ref{sec:O3.model}, we show numerical results for the $\mathbb{CP}^{1}$ 
model, and 
show that thermodynamic properties of the eight-color CSL remain intact even 
for continuous degrees of freedom.  
In Sec.~\ref{sec:SU3.model}, we show numerical results for the $\mathbb{CP}^{2}$
model, 
demonstrating that the CSL becomes an entropy-driven spin liquid at finite temperature. 
In Sec.~\ref{sec:Dynamics}, we present dynamical signatures of the CSL in the 
$\mathbb{CP}^{2}$
model, showing a largely gapped continuum of excitations. 
We conclude our work in Sec.~\ref{sec:Summary.Discussion} with a brief summary
and perspectives for possible future directions and experimental realizations.

%%%%%%%%%%%%%%%%%%%%%%%%%%%%%%%%%%%%%%%%%%
%
% 					METHODS and MODELS
%
%%%%%%%%%%%%%%%%%%%%%%%%%%%%%%%%%%%%%%%%%%
\section{BBQ-Kitaev model with different spin degrees of freedom}							
\label{sec:model}
%%%%%%%%%%%%%%%%%%%%%%%%%%%%%%%%%%%%%%%%%%
%

% The BBQ-K model 
%
We are interested in solving the bilinear-biquadratic (BBQ) model under influence 
of Kitaev interactions for $S=1$ magnetic moments on the honeycomb lattice.
The Hamiltonian is given by
%
%%%%%%%%%%%%%%%%%%%%%%
\begin{equation}
	\begin{aligned}
		\mathcal{H}^\mathcal{S}_{ {\sf BBQ-K} } 
			=  \sum_{\langle i,j \rangle} &\left[ \ J_1 \  {\bf S}_i \cdot{\bf S}_j \ 
				+ \ J_2 \ ( {\bf S}_i \cdot {\bf S}_j )^2 \ \right] 	\\
				&\ + 
				K  \sum_{\alpha = x,y,z} \  \sum_{ \langle ij \rangle_{\alpha}}  
				S_i^{\alpha} S_j^{\alpha} 	\, ,
		\label{eq:H.BBQ-K.S} 	
	\end{aligned}
\end{equation}
%%%%%%%%%%%%%%%%%%%%%%
%
where $J_1$, $J_2$, and $K$ respectively account for the Heisenberg (bilinear), biquadratic, and
Kitaev interaction strengths on nearest-neighbor bonds. 
The index \mbox{$\alpha = x,y,z$} selects the spin- and bond-anisotropic Kitaev interactions on the 
honeycomb lattice~\cite{Kitaev2006}. 
We normalize the total interaction strength as 
%
%%%%%%%%%%%%%%%%%%%%%%
\begin{equation}
	\left(J_1, J_2, K \right)  = \left( 
		\sin{\theta} \cos{\phi},  
		\sin{\theta} \sin{\phi},
		\cos{\theta} \right) 	\, ,
\label{eq:parametrization.HBBQ-K}
\end{equation}	
%%%%%%%%%%%%%%%%%%%%%%	
%
and use the angles $\theta$ and $\phi$ as new model parameters.
In this form, $\mathcal{H}^\mathcal{S}_{ {\sf BBQ-K} }$ recovers the well-known limits of 
the BBQ model for 
$\theta / \pi= 0.5$ ($K = 0$), 
the AFM Kitaev model at 
$\theta / \pi= 0$ ($K = 1$, $J_1 = J_2 = 0$), 
the FM Kitaev model at 
$\theta / \pi= 1$ ($K = -1$, $J_1 = J_2 = 0$),
and the Kitaev-Heisenberg model for $\phi / \pi= 0$ and 
$1$ ($J_2 = 0$).

% U(3) matrices  
%
An \mbox{$S=1$} moment contains
three magnetic states with \mbox{$S^z = \{ 1, 0, -1 \}$}, which 
are mathematically described by the $\mathfrak{su}(3)$ algebra 
and contain magnetic dipole and quadrupole components. 
To correctly describe all fluctuations of such an $S=1$ moment, we adopt the recently 
developed U(3) formalism~\cite{Remund2022}, an approach equivalent to the concept of SU(3) 
spin-coherent 
states~\cite{Papanicolaou1988, Perelomov1972, Gnutzmann1998, Nemoto2000, HaoZhang2021, Dahlbom2022a, Dahlbom2022b}, 
and suitable to access not only thermodynamic properties, but also dynamical properties of $S=1$
magnets at finite temperatures.
By embedding the underlying $\mathfrak{su}(3)$ algebra into the larger $\mathfrak{u}(3)$
algebra with an additional spin-length constraint, this formalism becomes capable  
of simultaneously evaluating dipole and quadrupole fluctuations at each site, despite the drawback
of losing quantum entanglement across the lattice.

In the U(3) formalism, each localized $S=1$ moment is represented
by a $3 \times 3$ Hermitian matrix
%
%%%%%%%%%%%%%%%%%%%%%%
\begin{equation}
		\mathcal{A}^{\alpha}_{\beta} = ({\rm d}^{\alpha})^* \  {\rm d}_{\beta} 	\, ,
		\label{eq:A-matrix_Director}
\end{equation}
%%%%%%%%%%%%%%%%%%%%%%
%
which we shall call henceforth the $\mathcal{A}$ matrix.
The $\mathcal{A}$ matrix is formally a tensor, as defined in Ref.~[\onlinecite{Remund2022}],  
where the superscript $\alpha$ and subscript $\beta$ denote the 
row and column of each index, respectively.
The $\mathcal{A}$ matrices can be written in terms of complex vectors ${\bf d}$ 
%
%%%%%%%%%%%%%%%%%%%%%%
\begin{equation}
		{\bf d} = \begin{pmatrix} 
			x_1 + i \ x_2		\\ 
			x_3 + i \ x_4		\\ 
			x_5 + i \ x_6		
		\end{pmatrix}  ,
\end{equation}
%%%%%%%%%%%%%%%%%%%%%%
%
also known as directors~\cite{Lauchli2006, Tsunetsugu2006}. 
We implicitly respect the spin-length constraint on the directors
%
%%%%%%%%%%%%%%%%%%%%%%
\begin{equation}
		{\bf d}^* {\bf d}	= 
		|{\bf d}|^2 		= 1	\, ,
\end{equation}
%%%%%%%%%%%%%%%%%%%%%%
%
by parametrizing its components as
%
%%%%%%%%%%%%%%%%%%%%%%
\begin{equation}
	\begin{aligned}
			x_1 &= \theta_2^{1/4} \ \theta_1^{1/2} \ \sin{\phi_1}		\ , \\ 
			x_2 &= \theta_2^{1/4} \ \theta_1^{1/2} \ \cos{\phi_1}		\ , \\ 
			x_3 &= \theta_2^{1/4} \ (1-\theta_1)^{1/2} \ \sin{\phi_2}		\ , \\ 
			x_4 &= \theta_2^{1/4} \ (1-\theta_1)^{1/2} \ \cos{\phi_2}	\ , \\ 
			x_5 &= (1-\theta_2^{1/2})^{1/2} \ \sin{\phi_3}			\ , \\ 
			x_6 &= (1-\theta_2^{1/2})^{1/2} \ \cos{\phi_3}			\ ,
	\end{aligned}
	\label{eq:sampling.d}
\end{equation}
%%%%%%%%%%%%%%%%%%%%%%
%
with $0 \leq \theta_1, \theta_2 \leq 1$ and $0 \leq \phi_1, \phi_2, \phi_3 < 2\pi$.
To fix the unphysical gauge 
we choose to make the $z$ component of 
${\bf d}$ purely real by setting $ \phi_3 = \pi/2$, providing in total  
four degrees of freedom with a restrained order-parameter space, formally known 
as the complex projective plane  $\mathbb{CP}^{2}$.
In this formalism, Eq.~(\ref{eq:H.BBQ-K.S}) can be rewritten in a bilinear 
form 
as
%
%
%%%%%%%%%%%%%%%%%%%%%%
\begin{equation}
	\begin{aligned}
		\mathcal{H}^{\mathcal{A}}_{ {\sf BBQ-K} } &=  \\
		\sum_{\langle ij \rangle} \Big[ J_1& \mathcal{A}^{\alpha}_{i  \beta} \mathcal{A}^{\beta}_{j  \alpha} 
		+ (J_2 - J_1) \mathcal{A}^{\alpha}_{i \beta} \mathcal{A}^{\alpha}_{j  \beta}  
		+ J_2 \mathcal{A}^{\alpha}_{i  \alpha} \mathcal{A}^{\beta}_{j \beta} \Big]   \\
		&- K  \sum_{ \langle ij \rangle_{\alpha} }  \epsilon^{\alpha \ \gamma}_{\ \beta}  
			\epsilon^{\alpha \ \eta}_{\ \delta} \mathcal{A}^{\beta}_{i \gamma} \mathcal{A}^{\delta}_{j \eta}  \, ,
	\label{eq:H.BBQ-K.A} 	
	\end{aligned}
\end{equation}
%%%%%%%%%%%%%%%%%%%%%%
%
where $\epsilon^{\alpha \ \gamma}_{\ \beta}$ is the Levi-Civita symbol.
Here and hereafter, we 
adopt the Einstein convention of summing over repeated indices. 
We shall refer to this form of the Hamiltonian in the following as ``$\mathbb{CP}^{2}$
model''.
The matrix $\mathcal{A}_{i}$ simultaneously 
incorporates the information of 
dipole and quadrupole moments 
at site $i$, which can be extracted respectively 
with
%
%%%%%%%%%%%%%%%%%%%%%%
\begin{align}
	S_{i}^{\alpha} &= -i \epsilon^{\alpha \ \gamma}_{\ \beta} \mathcal{A}^{ \beta}_{i  \gamma}	\, , \label{eq:dipole}	\\
	Q^{\alpha \beta}_{i} &= - \mathcal{A}^{\alpha}_{i \beta} - \mathcal{A}^{\beta}_{i  \alpha} 	
					+ \frac{2}{3}\delta^{\alpha \beta}  \mathcal{A}^{\gamma}_{i  \gamma}	\, ,  \label{eq:quadrupole}
\end{align}
%%%%%%%%%%%%%%%%%%%%%%
%
where $\delta^{\alpha \beta}$ is the Kronecker delta.
Here, we express spin quadrupoles 
in their symmetric and traceless rank-2 tensor form. 

To compute the averaged spin-dipole and spin-quadrupole norms
we respectively use
%
%%%%%%%%%%%%%%%%%%%%%%
\begin{align}
	|{\bf S}| & = \frac{1}{\ns} \sum^{\ns}_i |{\bf S}_i | \, , \label{eq:S.norm} \\
	|{\bf Q}| &= \frac{1}{\ns} \sum^{\ns}_i |{\bf Q}_i | \, ,  \label{eq:Q.norm}
\end{align}
%%%%%%%%%%%%%%%%%%%%%%
%
where vector components are expressed as 
%
%%%%%%%%%%%%%%%%%%%%%%
\begin{equation}
	{\bf S}_i = \begin{pmatrix}
	S_i^x \\ 
	S_i^y \\
	S_i^z
	\end{pmatrix} \, ,  
\end{equation}    
%%%%%%%%%%%%%%%%%%%%%%
%
and 
%
%%%%%%%%%%%%%%%%%%%%%%
\begin{equation}
	{\bf Q}_i = \begin{pmatrix}
	Q_i^{x^2-y^2}	\\
	Q_i^{3z^2-r^2}	\\
	Q_i^{xy}		\\
	Q_i^{xz}		\\
	Q_i^{yz}				
\end{pmatrix}  =  \begin{pmatrix}
	\frac{1}{2} \left(Q_i^{xx} - Q_i^{yy} \right) 	\\
	\frac{1}{\sqrt{3}} \left(Q_i^{zz} - \frac{1}{2}\left(Q_i^{xx} + Q_i^{yy}\right) \right) 	\\
	Q_i^{xy} 	\\
	Q_i^{xz} 	\\
	Q_i^{yz}	\\
\end{pmatrix}  \, .
\end{equation}
%%%%%%%%%%%%%%%%%%%%%%
%

% O(3) Heisenberg spins 
%
In Sec.~\ref{sec:phase_diagrams}, we find 
many nontrivial phases for the $\mathbb{CP}^{2}$
case, especially when the interactions 
frustrate the model.
To gain a better intuition of the underlying physics in such nontrivial phases,
we investigate Eq.~\eqref{eq:H.BBQ-K.S} also in the spin space 
$\mathbb{CP}^{1}$, by effectively 
excluding all quadrupole fluctuations and allowing only for fluctuations of local dipole 
moments. 
This simplification
corresponds to the treatment of classical Heisenberg spins, 
formally expressed as spin-coherent states of SU(2), 
which are parametrized by a point on the Bloch sphere
%
%%%%%%%%%%%%%%%%%%%%%%
\begin{equation}
		{\bf S} = \begin{pmatrix} 
			\sin{\theta_1} \cos{\phi_1} 	\\ 
			\sin{\theta_1} \sin{\phi_1}  		\\ 
			\cos{\theta_1}  	
		\end{pmatrix}  ,
\label{eq:sampling.S}
\end{equation}
%%%%%%%%%%%%%%%%%%%%%%
%
with $0 \leq \theta_1 \leq 1$ and $0 \leq \phi_1 < 2\pi$, the two local degrees of 
freedom for each spin on the lattice. 
In what follows, we refer to this model as ``$\mathbb{CP}^{1}$ model''.

% 8-color model 
%
%%%%%%%%%%%%%%%%%%%%%%%%%%%%%%%%%%%%%%%%%%
% Tab.    definition of spin direction in 8-color model 
%%%%%%%%%%%%%%%%%%%%%%%%%%%%%%%%%%%%%%%%%%
%
%%%%%%%%%%%%%%%%%%%%%%%%%%
\newcolumntype{C}{>{}c<{}} 
\begin{table}[t]
\def\arraystretch{2}
\centering
	\includegraphics[width=0.45\textwidth]{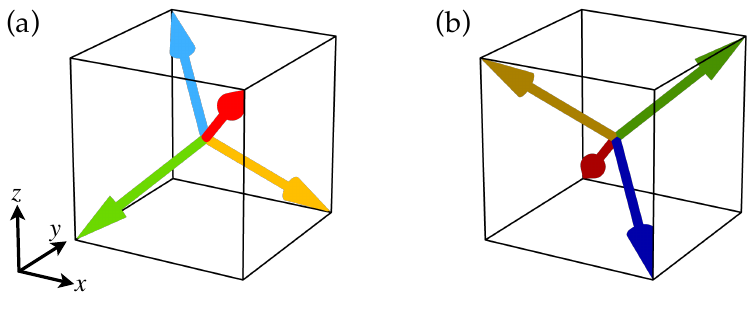}
     	\bigskip
      	\begin{tabular}{ C  C  C }
	\hhline{===}			
							& bright									\hspace{0.4cm}						& dark				\\
	\hhline{---}
	green	\hspace{0.3cm}		& $\boldsymbol \sigma_{\rm g} = \frac{1}{\sqrt{3}} \left\{ -1, -1, -1 \right \}$ 	\hspace{0.4cm}		& $\bar{\boldsymbol\sigma}_{\rm g} = \frac{1}{\sqrt{3}} \left\{ +1, +1, +1 \right \}$	\\
	yellow	\hspace{0.3cm}		& $\boldsymbol\sigma_{\rm y} = \frac{1}{\sqrt{3}} \left\{ +1, +1, -1 \right \}$ 	\hspace{0.4cm}		& $\bar{\boldsymbol\sigma}_{\rm y} = \frac{1}{\sqrt{3}} \left\{ -1, -1, +1 \right \}$		\\
	blue		\hspace{0.3cm}		& $\boldsymbol\sigma_{\rm b} = \frac{1}{\sqrt{3}} \left\{ -1, +1, +1 \right \}$ \hspace{0.4cm}		& $\bar{\boldsymbol\sigma}_{\rm b} = \frac{1}{\sqrt{3}} \left\{ +1, -1, -1 \right \}$		\\
	red		\hspace{0.3cm}		& $\boldsymbol\sigma_{\rm r} = \frac{1}{\sqrt{3}} \left\{ +1, -1, +1 \right \}$ 	\hspace{0.4cm}		& $\bar{\boldsymbol\sigma}_{\rm r} = \frac{1}{\sqrt{3}} \left\{ -1, +1, -1 \right \}$		\\
	\hhline{===}
	\end{tabular}
      	\caption{ 
	Definition of spin directions in the eight-color model of Eq.~\eqref{eq:H.BBQ-K.8c}.
	For pedagogical reasons, we assign four colors to spins, namely 
	``green'', ``yellow'', ``blue'', and ``red'', and distinguish between their bright and dark 
	contrast.
	Directions are explicitly shown in (a) for bright spins, and in (b) for dark spins.
	}   
	\label{tab:8.color.spins}
\end{table}
%%%%%%%%%%%%%%%%%%%%%%%%
%

In the $\mathbb{CP}^{2}$ model, a CSL phase is stabilized at finite temperatures
(see Sec.~\ref{sec:SU3.model}), while in the $\mathbb{CP}^{1}$ model, the CSL 
phase appears along singular lines in the ground state 
(see Sec.~\ref{sec:O3.model}).
In the CSL phase, we observe that spins align along well-defined, 
discrete directions, pointing to the eight corners of a unit cube
(see detailed discussion in Sec.~\ref{sec:8c.model}). 
This observation motivated us to further simplify Eq.~\eqref{eq:H.BBQ-K.S} by 
discretizing spins to only point to these eight corners.
We call this model the ``eight-color model'',  whose Hamiltonian is given by
% 
%
%%%%%%%%%%%%%%%%%%%%%%
\begin{equation}
	\begin{aligned}
		\mathcal{H}^{ {\sf 8c} }_{{\sf BBQ-K}}
			=  \sum_{\langle i,j \rangle} &\left[ \ J_1 \  \boldsymbol \sigma_i \cdot \boldsymbol \sigma_j \ 
				+ \ J_2 \ ( \boldsymbol \sigma_i \cdot \boldsymbol \sigma_j )^2 \ \right] 	\\
				&\ + K  \sum_{\alpha = x,y,z} \  \sum_{ \langle ij \rangle_{\alpha}}  
				\sigma_i^{\alpha} \sigma_j^{\alpha} 	\, .
		\label{eq:H.BBQ-K.8c} 
	\end{aligned}
\end{equation}
%%%%%%%%%%%%%%%%%%%%%%
%
%
As shown in Table~\ref{tab:8.color.spins}, we assign colors to the eight discrete, 
noncoplanar dipole spins $\boldsymbol \sigma$ such as ``green'', ``yellow'', ``blue'', and ``red'', and 
distinguish between ``bright'' spins,  
$\boldsymbol\sigma_{\rm g}$, 
$\boldsymbol\sigma_{\rm y}$, 
$\boldsymbol\sigma_{\rm b}$,  and 
$\boldsymbol\sigma_{\rm r}$, and ``dark'' spins,  
$\bar{\boldsymbol\sigma}_{\rm g}$, 
$\bar{\boldsymbol\sigma}_{\rm y}$, 
$\bar{\boldsymbol\sigma}_{\rm b}$, and 
$\bar{\boldsymbol\sigma}_{\rm r}$.
This simplified model is specifically designed to transparently understand the nature of 
the exotic chiral ordered and disordered phases.
%

%%%%%%%%%%%%%%%%%%%%%%%%%%%%%%%%%%%%%%%%%%
%
% 					THE BBQ-K MODEL FOR S=1
%
%%%%%%%%%%%%%%%%%%%%%%%%%%%%%%%%%%%%%%%%%%
\section{Results}									
\label{sec:BBQ-K.results}
%%%%%%%%%%%%%%%%%%%%%%%%%%%%%%%%%%%%%%%%%%
%
%

In our previous work in Ref.~[\onlinecite{Pohle2023}], we studied the
ground state and thermodynamic properties of 
$\mathcal{H}^{\mathcal{S}}_{ {\sf BBQ-K}}$ in Eq.~\eqref{eq:H.BBQ-K.S},
and revealed the existence of a chiral spin liquid (CSL) at finite temperatures.
In the present study, we build up an intuition for the microscopic origin 
and nature of this exotic CSL state by systematically restricting the 
local spin degree of freedom.
We begin with the SU(3) representation of Eq.~\eqref{eq:H.BBQ-K.S},
the $\mathbb{CP}^{2}$ model, given by $\mathcal{H}^{\mathcal{A}}_{ {\sf BBQ-K}}$ 
in Eq.~\eqref{eq:H.BBQ-K.A},
which captures all allowed components of an $S=1$ spin,
namely continuous dipolar and quadrupolar moments described in 
the spin space $\mathbb{CP}^{2}$.
Further, we solve the original model, $\mathcal{H}^{\mathcal{S}}_{ {\sf BBQ-K} }$
in Eq.~\eqref{eq:H.BBQ-K.S}, in the spin space $\mathbb{CP}^{1}$, which allows for 
dipole moments only [see Eq.~\eqref{eq:sampling.S}].
Finally, we consider the discrete eight-color model,
given by $\mathcal{H}^{ {\sf 8c} }_{{\sf BBQ-K}}$ in Eq.~\eqref{eq:H.BBQ-K.8c},
where dipolar spin moments are allowed to point only along 
eight discrete directions (see Table~\ref{tab:8.color.spins}).
All three models include the same BBQ-Kitaev interactions,
however for different degrees of freedom in spin space. 
While certain results have been previously presented in our 
earlier 
work~[\onlinecite{Pohle2023}], we will include some of them here 
to make the present paper self-contained.
%

%%%%%%%%%%%%%%%%%%%%%%%%%%%%%%%%%%%%%%%%%%
\subsection{Ground-state phase diagrams}								
\label{sec:phase_diagrams}
%%%%%%%%%%%%%%%%%%%%%%%%%%%%%%%%%%%%%%%%%%
%

% all 3 phase diagrams
%
In Figs.~\ref{fig:PD.all}(a)--\ref{fig:PD.all}(c) 
we respectively show the 
ground-state phase diagrams of 
$\mathcal{H}^\mathcal{A}_{ {\sf BBQ-K} }$, 
$\mathcal{H}^\mathcal{S}_{ {\sf BBQ-K} }$, 
and $\mathcal{H}^{ {\sf 8c} }_{ {\sf BBQ-K}}$,
solved in the spin space $\mathbb{CP}^{2}$, $\mathbb{CP}^{1}$, and 
for discrete eight-color spins. 
To obtain Figs.~\ref{fig:PD.all}(a) and \ref{fig:PD.all}(b), we performed 
variational energy minimization, using the gradient descent method based on the 
machine learning library JAX \cite{Jax2018, Optax2020} for sufficiently large 
finite-size clusters of $\ns = 2L^2= 28 \ 800$ 
spins (linear dimension $L=120$) under periodic boundary conditions
(see Appendix~\ref{app:method.details} for further details).
Figure~\ref{fig:PD.all}(c) was obtained in the thermodynamic limit
from analytic comparison of local bond energies.
We plot the phase diagrams 
in their 2D equirectangular projections parametrized by 
Eq.~\eqref{eq:parametrization.HBBQ-K}, and visualize phase boundaries 
by plotting $\partial^2 E/\partial \theta^2 + \partial^2 E/\partial \phi^2$, the sum of the 
second-order derivatives of the internal energy $E$ with respect to 
$\theta$ and $\phi$.
The corresponding real-space spin configurations of selected states are shown in 
Fig.~\ref{fig:PD.all}(d) with dipole moments depicted as arrows, and 
quadrupole moments in form of ``directors'' with ``doughnut-shaped'' 
spin-probability distributions~\cite{Lauchli2006, Tsunetsugu2006}.
It is worth noting 
that the phase diagrams for $\mathbb{CP}^{1}$ and eight-color 
models exhibit a self-duality 
around the high-symmetry points 
$(\phi/\pi ,\theta/\pi)=(0.5, 0.5)$ and 
$(\phi/\pi ,\theta/\pi)=(1.5, 0.5)$.
In contrast, the phase diagram for  $\mathbb{CP}^{2}$ lacks 
this self-dual symmetry, due to the presence of quadrupole moments.

%%%%%%%%%%%%%%%%%%%%%%%%%%%%%%%%%%%%%%%%%%
\subsubsection{Phase diagram for $\mathbb{CP}^{2}$ model} 
\label{sec:PD.CP2}									
%%%%%%%%%%%%%%%%%%%%%%%%%%%%%%%%%%%%%%%%%%
%

In Fig.~\ref{fig:PD.all}(a) we show the ground-state phase diagram of 
$\mathcal{H}^{\mathcal{A}}_{ {\sf BBQ-K} }$ in Eq.~\eqref{eq:H.BBQ-K.A},
by considering the full spin order-parameter space $\mathbb{CP}^{2}$.
When the biquadratic interaction is ferromagnetic, namely $J_2$  is 
negative \mbox{$(1.0 < \phi / \pi < 2.0)$}, 
the model offers a combination of phases which have previously been 
reported on the $S=1$ Kitaev-Heisenberg model~\cite{Stavropoulos2019, Dong2020} 
and  the BBQ model~\cite{Zhao2012}.
These comprise dipolar orders, in the form of ferromagnetic (FM), antiferromagnetic (AFM), 
zigzag, and stripy phases, together with a ferroquadrupolar (FQ) spin nematic
phase.

In contrast, when $J_2$ is positive \mbox{$(0.0 < \phi / \pi < 1.0)$}, 
the model provides a number of unconventional phases which emerge 
from the competition between frustrated interactions.
Adjacent to the zigzag and stripy phases, we find noncoplanar chiral ordered phases, 
also known as ``tetrahedral'' and ``cubic'' states~\cite{Martin2008, Akagi2010, Messio2011}, 
in which spins point to the corners of a unit cube [see Fig.~\ref{fig:PD.all}(d)].
These chiral configurations are represented by superpositions of three 
spiral states, and hence called triple-$q$ states.
The FM 3$Q$ chiral state is characterized by an extended 
eight-site magnetic unit cell, where four out of eight possible spin 
directions are selected and induce a net scalar spin chirality of 
\mbox{$|\kappa |^{\rm FM \ chiral}  = 8/ (3 \sqrt{3})$}
[see the definition of $\kappa$ in Eq.~\eqref{eq:kappa1} of 
Appendix~\ref{app:observables.details}].
The AFM 3$Q$ chiral state is essentially a reflection of the FM 3$Q$ chiral state, 
achieved by inverting spins on 
one of the two sublattices on the honeycomb lattice [see Fig.~\ref{fig:PD.all}(d)],
giving \mbox{$| \kappa |^{\rm AFM \ chiral} = 16/(3 \sqrt{3})$}. 
We note that the stripy and FM 3$Q$ chiral (zigzag and AFM 3$Q$ chiral) states are 
energetically degenerate at \mbox{$\phi/\pi=0$} (\mbox{$\phi/\pi = 1$})
along the vertical line for \mbox{$0.75 \leq \theta/\pi < 1.0$} (\mbox{$0.0 < \theta/\pi \leq 0.25$}).
However, this degeneracy is lifted by thermal fluctuations,
which select the stripy (zigzag) phase, as confirmed by our MC simulations (not shown).

% frustrated region of phase diagram 
%
% SO phase and 1D phase
Between FM, AFM, and 3$Q$ chiral ordered phases, dipole and quadrupole 
moments mix in nontrivial ways, and produce a range of exotic states.
In the limit of the BBQ model (\mbox{$\theta/\pi = 0.5$}), our semiclassical method 
realizes a semiordered (SO) (or semidisordered)~\cite{Papanicolaou1988, Niesen2017}, 
purely quadrupolar state, connecting the two SU(3) points at 
\mbox{$\phi/\pi = 0.25$} and \mbox{$\phi/\pi = 0.5$}~\footnote{
We notice several complicated phases around the two SU(3) points 
at $(\theta/\pi, \phi/\pi) = (0.5, 0.25)$ and $(0.5, 0.5)$, which will be discussed elsewhere.
}.
We note that this state is expected to be replaced by a plaquette valence bond crystal 
when quantum entanglement is fully taken into 
account~\cite{Corboz2013}.
Introducing nonzero 
Kitaev interactions (\mbox{$\theta/\pi \neq 0.5$}) immediately 
induces the formation of a nonzero dipole moment $|{\bf S}| \neq 0$. 
To minimize the dominant BBQ interactions, 
dipolar components of the spins align orthogonal to their nearest neighbors
within the $xy$, $yz$, or $zx$ plane.
In such a configuration, the Kitaev 
interaction energy is minimized on two
bonds, while leaving the third one to be zero.
Hence, dominant correlations prevail  along 
zigzag chains and stabilize coplanar, quasi-one-dimensional (q1D) 
states in extended regions away from the BBQ limit.
However, the presence of small quadrupolar correlations will induce a weak 
interchain coupling to form two-dimensional order.
%

% NC, TC, and others... 
Sandwiched between the q1D coplanar phases 
and the AFM/FM ordered phases, 
$\mathcal{H}^{\mathcal{A}}_{ {\sf BBQ-K} }$ minimizes its energy by forming 
spin textures with noncoplanar (NC) orientation and incommensurate twisted 
conical (TC) states.
These ordered states will give way to a finite-temperature 
chiral spin liquid (CSL) -- not shown in Fig.~\ref{fig:PD.all}(a) -- which shall 
be the major focus of this study.
We will discuss signatures of the CSL state explicitly for 
$\mathcal{H}^{\mathcal{A}}_{ {\sf BBQ-K} }$
by including all allowed spin fluctuations of $\mathbb{CP}^{2}$ in 
Sec.~\ref{sec:SU3.model}, while building up an intuitive understanding of its 
microscopic origin by 
systematically restricting the allowed spin space 
to $\mathbb{CP}^{1}$ and further to discretized states
in Secs.~\ref{sec:O3.model} and \ref{sec:8c.model}, respectively.

%%%%%%%%%%%%%%%%%%%%%%%%%%%%%%%%%%%%%%%%%%
\subsubsection{Phase diagram for $\mathbb{CP}^{1}$ model} 
\label{sec:PD.CP1}									
%%%%%%%%%%%%%%%%%%%%%%%%%%%%%%%%%%%%%%%%%%
%

Many phases in Fig.~\ref{fig:PD.all}(a) show dominant dipolar order 
mainly coming from the presence of bilinear Kitaev and Heisenberg 
interactions.
To build up an intuitive understanding of the microscopic origin of these 
phases, and to quantitatively understand the influence of local quantum fluctuations 
in our model, we are going to 
restrict the order parameter space from $\mathbb{CP}^{2}$ to $\mathbb{CP}^{1}$, 
by expressing an $S=1$ moment by a classical 
Heisenberg vector, as 
parametrized with 
Eq.~\eqref{eq:sampling.S}.
In this way we only allow for purely dipolar spin states and exclude 
onsite quadrupole moments.

In this framework, by minimizing the energy of $\mathcal{H}^{\mathcal{S}}_{ {\sf BBQ-K} }$
in Eq.~\eqref{eq:H.BBQ-K.S}, we obtain the ground-state phase diagram in 
Fig.~\ref{fig:PD.all}(b).
For negative $J_2$ $(1.0 < \phi / \pi < 2.0)$ the model 
captures the same dominant dipolar FM, AFM, zigzag, and stripy ordered phases, 
as in the $\mathbb{CP}^{2}$ case.
However, due to the exclusion of onsite quadrupolar degree of freedom, 
the FQ state is absent and has been replaced by a singular 
point at $\phi / \pi = 1.5$ and $\theta / \pi = 0.5$ where 
FM, AFM, zigzag, and stripy phases are energetically degenerate. 

In the region where $J_2$ is positive $(0.0 < \phi / \pi < 1.0)$, 
the model reproduces many phases as seen in Fig.~\ref{fig:PD.all}(a),
while the phase boundaries are rather strongly modified.
%
% noncoplanar phases
We find noncoplanar FM and AFM 3$Q$ chiral phases with same 
physical properties as in the $\mathbb{CP}^{2}$ case. 
Between AFM/FM ordered phases and AFM/FM 3$Q$ chiral phases the model 
stabilizes NC phases and canted planar (CPL) phases with 
spin configurations shown in Fig.~\ref{fig:PD.all}(d). 
Here, the NC phases are purely made of spin dipoles and exist in much 
wider regions compared to the $\mathbb{CP}^{2}$ case.
%
% 1D phase
At the phase boundary between the NC and CPL phases ($\phi / \pi = 0.5$), 
the model stabilizes coplanar, truly 1D phases, made of decoupled 
zigzag chains with a subextensive ground-state manifold 
[see Fig.~\ref{fig:PD.all}(d)].
While these \mbox{1D} phases appear on a singular line 
in the $\mathbb{CP}^{1}$ model, the presence of quadrupole moments in the 
$\mathbb{CP}^{2}$ case stabilizes \mbox{q1D} 
phases (not truly 1D due to presence of small but nonzero quadrupolar order), 
in much wider regions of the phase diagram [see Fig.~\ref{fig:PD.all}(a)].
%

% CSL phase 
Furthermore, the absence of quadrupole 
degrees of freedom allows the CSL state to exist in the ground state at 
the singular lines 
%
%%%%%%%%%%%%%%%%%%%%%%
\begin{equation}
	\frac{\phi}{\pi} = \frac{1}{\pi} \arctan \left(\frac32\right) \approx 0.312833 \, , \quad
	0.0 < \frac{\theta}{\pi} < 0.5	   \, ,  
	\label{eq:O3.CSL.AFM}
\end{equation}
%%%%%%%%%%%%%%%%%%%%%%
%
for the FM CSL and at 
%
%%%%%%%%%%%%%%%%%%%%%%
\begin{equation}
	\frac{\phi}{\pi} = 1- \frac{1}{\pi} \arctan \left(\frac32\right) \approx 0.687167  \, ,  \quad
	0.5< \frac{\theta}{\pi} <1.0 	\, ,  
	\label{eq:O3.CSL.FM}
\end{equation}
%%%%%%%%%%%%%%%%%%%%%%
%
for the AFM CSL, deep inside the NC phase.
The analytic derivation of these values is described in 
Appendix~\ref{app:PhaseBoundaries}.

In the CSL ground state, spin-dipole moments 
point exactly to the eight discrete corners of the unit cube, similar to the 
ordered AFM/FM 3$Q$ chiral states [see Fig.~\ref{fig:PD.all}(d)].
However, in strong contrast to the chiral ordered states, the CSL does not 
break translational symmetries of the lattice, and hence exhibits an extensive 
degeneracy of states.
This motivates us to further simplify the model by restricting the continuous spin 
space of $\mathbb{CP}^{1}$ to only eight discrete states (eight-color model), where dipolar spins 
are allowed to point only along the corners of a unit cube. 
We will explore this model in detail in the following subsections, and discuss ground-state 
and thermodynamic properties of the CSL 
explicitly for the $\mathbb{CP}^{1}$ model 
in Sec.~\ref{sec:O3.model}.

%%%%%%%%%%%%%%%%%%%%%%%%%%%%%%%%%%%%%%%%%%
\subsubsection{Phase diagram for eight-color model}
\label{sec:PD.8c}	
%%%%%%%%%%%%%%%%%%%%%%%%%%%%%%%%%%%%%%%%%%
%

To understand the intriguing nature of this CSL, we analyze the eight-color model described by 
$\mathcal{H}^{ {\sf 8c} }_{ {\sf BBQ-K}}$ in Eq.~\eqref{eq:H.BBQ-K.8c}, which enables
us to extract its most essential properties analytically. 
In this model we restrict the spin parameter space by allowing for only eight 
discrete dipolar spin directions, pointing along the corners of a unit cube, as defined in 
Table~\ref{tab:8.color.spins}. 

As shown in the ground-state phase diagram in Fig.~\ref{fig:PD.all}(c), this model 
stabilizes trivial magnetically-ordered states such as FM and 
AFM states in the region where the biquadratic interaction 
$J_2$ is mostly negative $(1.0 < \phi / \pi < 2.0)$.
For predominantly positive  $J_2$  $(0.0 < \phi / \pi < 1.0)$,
the model stabilizes nontrivial chiral ordered, triple-$q$ states, as also 
observed in the $\mathbb{CP}^{2}$ and $\mathbb{CP}^{1}$
models in Figs.~\ref{fig:PD.all}(a) and \ref{fig:PD.all}(b),
respectively.
Since coplanar spin configurations are not allowed in this model, zigzag, stripy, 1D, 
and other states with canting angles away from the (111) axes are absent. 

Importantly, between trivial magnetically-ordered and triple-$q$ states, the 
model stabilizes the CSL states in wide regions of the phase diagram.
The FM CSL phase exists for
%
%%%%%%%%%%%%%%%%%%%%%%
\begin{equation}	
	\frac{1}{\pi} \ \arctan{\left( \frac{3}{4} \right) } < \frac{\phi}{\pi} < 0.5	\, ,	\quad  0.0 < \frac{\theta}{\pi} <0.5 \, ,  
\label{eq:FM.chiral.PD}
\end{equation}
%%%%%%%%%%%%%%%%%%%%%%
%
while the AFM CSL phase appears for 
%
%%%%%%%%%%%%%%%%%%%%%%
\begin{equation} 
	0.5 < \frac{\phi}{\pi} < 1 - \frac{1}{\pi} \arctan{ \left(\frac{3}{4} \right) } \, ,	\quad   0.5 < \frac{\theta}{\pi} <1.0	\, .   
	\label{eq:AFM.chiral.PD}
\end{equation}
%%%%%%%%%%%%%%%%%%%%%%
%
The phase boundaries can be obtained analytically
by simple energy comparison, as described in Appendix~\ref{app:PhaseBoundaries}.

Both states are classical spin liquids with a nonzero scalar spin chirality 
and residual entropy at zero temperature, which is associated with their 
extensive number of degenerate ground states.
We note that the phase boundary lines between chiral order and disordered phases 
at $\phi / \pi = 0.5$ and $\theta / \pi = 0.5$ also host
classical spin liquids, which, however, has different properties than the CSL
discussed here.

By comparison between measured physical observables, we argue that this
CSL with all its properties survives in the $\mathbb{CP}^{1}$ and $\mathbb{CP}^{2}$ models,
while additional degrees of freedom in spin space only weaken its stability 
in the phase diagram.
For pedagogical reasons, we will start our journey by investigating the CSL 
on the eight-color model.

%%%%%%%%%%%%%%%%%%%%%%%%%%%%%%%%%%%%%%%%%%
%
% 			The 8-color model
%
%%%%%%%%%%%%%%%%%%%%%%%%%%%%%%%%%%%%%%%%%%
\subsection{Analysis of eight-color model}	
\label{sec:8c.model}
%%%%%%%%%%%%%%%%%%%%%%%%%%%%%%%%%%%%%%%%%%
%
%

In Fig.~\ref{fig:PD.all}(c), we see that the eight-color model
stabilizes the CSL ground state in wide regions of the phase diagram 
[see Eqs.~\eqref{eq:FM.chiral.PD} and \eqref{eq:AFM.chiral.PD}].
In Sec.~\ref{sec:CSL.properties}, we use the simplicity of the eight-color model to 
analytically obtain bond-dependent spin constraints which determine
the ground-state properties of this CSL. 
In Sec.~\ref{sec:8c.model.thermo}, we validate our analytical results by numerical MC 
simulations and compare thermodynamic properties between the CSL and 
chiral ordered states.

%%%%%%%%%%%%%%%%%%%%%%%%%%%%%%%%%%%%%%%%%%
\subsubsection{Eight-color chiral spin liquid }	
\label{sec:CSL.properties}
%%%%%%%%%%%%%%%%%%%%%%%%%%%%%%%%%%%%%%%%%%
%
%

% 
The eight-color model given by $\mathcal{H}^{ {\sf 8c} }_{{\sf BBQ-K}}$ 
in Eq.~\eqref{eq:H.BBQ-K.8c} 
allows us to understand the nature of the CSL analytically from 
simple comparison of local bond energies. 
In the following, we consider the FM CSL
for parameters outlined in 
Eq.~\eqref{eq:FM.chiral.PD}, where all couplings are positive, with 
$0 < J_1 < 0.8$,  $J_2 > 0.6$, and $K > 0$.
Since $\mathcal{H}^{ {\sf 8c} }_{{\sf BBQ-K}}$ is symmetric in parameter space 
the following conclusions will also hold for the 
AFM CSL after inversion of the spin direction on one of the two sublattices 
on the honeycomb lattice.

We first compare energies in absence of Kitaev interactions at $\theta / \pi = 0.5$.
The nearest-neighbor energy levels, exemplified 
for only ``bright green'' spins  $\boldsymbol \sigma_{\rm g}$ (see Table~\ref{tab:8.color.spins}),
are given as
%
%%%%%%%%%%%%%%%%%%%%%%
\begin{align}
	&(\boldsymbol\sigma_{\rm g}, \boldsymbol\sigma_{\rm g}) : \ 	E = J_1 + J_2	\, ,	\\
	&(\boldsymbol\sigma_{\rm g} , \bar{\boldsymbol\sigma}_{\rm g}): \ E = -J_1 + J_2	\, ,	\\
	&(\boldsymbol\sigma_{\rm g} , \bar{\boldsymbol\sigma}_{\rm r}), \
	(\boldsymbol\sigma_{\rm g} , \bar{\boldsymbol\sigma}_{\rm y}), \
	(\boldsymbol\sigma_{\rm g} , \bar{\boldsymbol\sigma}_{\rm b}) : \ E = \frac{1}{3}J_1 + \frac{1}{9}J_2	\, ,	
	\label{eq:8c.lowEnergy.3}\\
	&(\boldsymbol\sigma_{\rm g} , \boldsymbol\sigma_{\rm r}), \
	(\boldsymbol\sigma_{\rm g} , \boldsymbol\sigma_{\rm y}), \
	(\boldsymbol\sigma_{\rm g} , \boldsymbol\sigma_{\rm b}): \ E = -\frac{1}{3} J_1 + \frac{1}{9}J_2	\, , 
	\label{eq:8c.lowEnergy.4} 
\end{align}
%%%%%%%%%%%%%%%%%%%%%%
%
%%%%%%%%%%%%%%%%%%%%%%
%
where the lowest-energy configurations in Eq.~\eqref{eq:8c.lowEnergy.4} 
exclude nearest neighbors of same color 
(e.g., ``bright green''--``bright green'') and different 
contrast (e.g., ``bright green''--``dark red'').

The degeneracy of the lowest-energy states
can be lifted by introducing the bond-dependent Kitaev interactions, giving the 
ground state energy 
%
%%%%%%%%%%%%%%%%%%%%%%
\begin{align}
	E_{\rm GS} = -\frac{1}{3} J_1 + \frac{1}{9}J_2 - \frac{1}{3}K  \, .
\label{eq:8c.GS.energy}
\end{align}
%%%%%%%%%%%%%%%%%%%%%%
%
Intriguingly, while $K<0$ gives only one solution for Eq.~\eqref{eq:8c.GS.energy}
inducing FM 3$Q$ chiral order,
the case of $K>0$ allows for multiple spin pairs 
\mbox{$(\boldsymbol\sigma_{\rm g} , \boldsymbol\sigma_{\rm y})$} and 
\mbox{$(\boldsymbol\sigma_{\rm g} , \boldsymbol\sigma_{\rm r})$}  
on the $x$ bonds,
\mbox{$(\boldsymbol\sigma_{\rm g}  , \boldsymbol\sigma_{\rm y})$} and  
\mbox{$(\boldsymbol\sigma_{\rm g}  , \boldsymbol\sigma_{\rm b})$} 
on the $y$ bonds, and
\mbox{$(\boldsymbol\sigma_{\rm g} , \boldsymbol\sigma_{\rm b})$} and  
\mbox{$(\boldsymbol\sigma_{\rm g} , \boldsymbol\sigma_{\rm r})$} 
on the $z$ bonds. 
Since the Kitaev interactions couple spin components on each bond differently, 
the allowed color pairs in the ground state will be bond dependent.

%
%%%%%%%%%%%%%%%%%%%%%%%%%%%%%%%%%%%%%
%. Fig. -- Ground state: bond constraints and hexagon updates
%%%%%%%%%%%%%%%%%%%%%%%%%%%%%%%%%%%%%
\begin{figure}[t]
	\centering
	\includegraphics[width=0.45\textwidth]{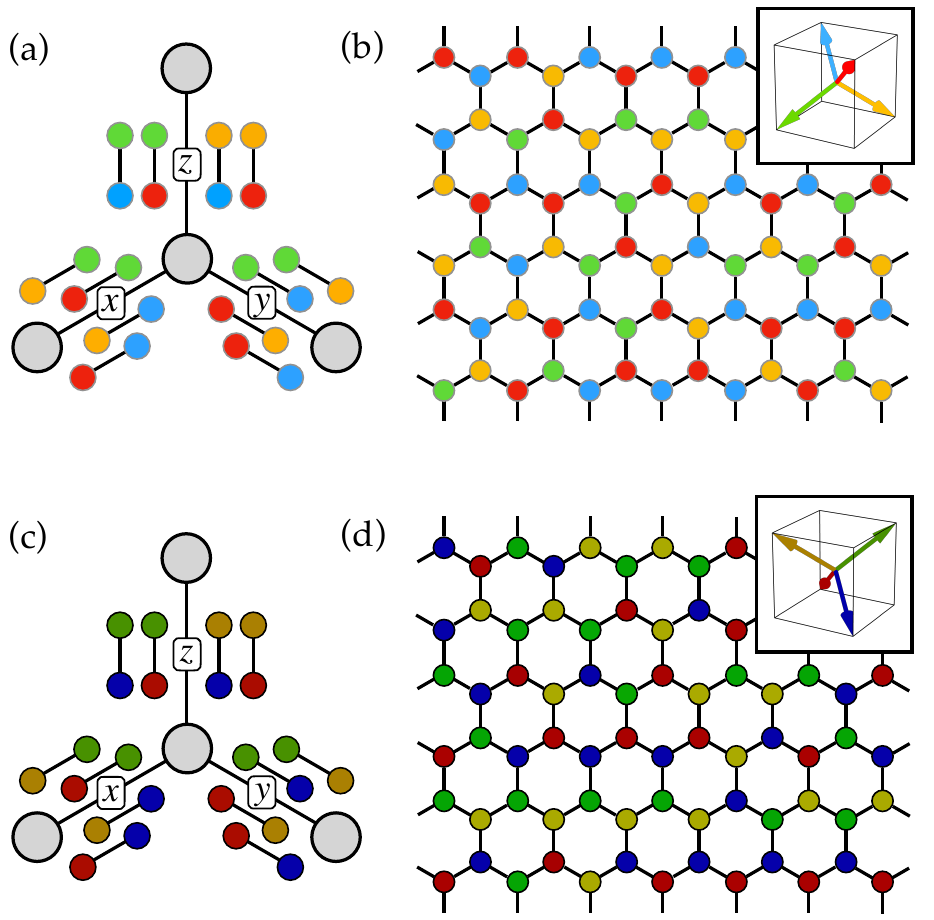}
	\caption{
	Macroscopic degeneracy in the FM chiral spin liquid (FM CSL).
	(a) Constraints on allowed color pairs on bonds
	for ``bright'' spins and (b) a corresponding 
	representative spin configuration in the FM CSL ground-state manifold.
	Corresponding figures for ``dark'' spins are shown in (c) and (d).
	The definitions of spin orientations and colors are shown in the insets
	of (b) and (d) (see also Table~\ref{tab:8.color.spins}).
	}
	\label{fig:8cCSL.bond.constraints}
\end{figure}
%%%%%%%%%%%%%%%%%%%%%%%%%%%%%%%%%%%%%%%
%

By doing the same energy analysis for the other ``bright'' spins 
($\boldsymbol\sigma_{\rm y}$, $\boldsymbol\sigma_{\rm b}$, $\boldsymbol\sigma_{\rm r}$),
we obtain the complete set of constraints on allowed color pairs on bonds:
%
%%%%%%%%%%%%%%%%%%%%%%
\begin{align}
	( \boldsymbol\sigma_{\rm g}, \boldsymbol\sigma_{\rm y} ) , \
	( \boldsymbol\sigma_{\rm g}, \boldsymbol\sigma_{\rm r} ) , \
	 ( \boldsymbol\sigma_{\rm b}, \boldsymbol\sigma_{\rm y} ) , \ 
	 (\boldsymbol \sigma_{\rm b}, \boldsymbol\sigma_{\rm r} ) , 
\label{eq:8c.constraints.x}
\end{align}
%%%%%%%%%%%%%%%%%%%%%%
%
on the $x$ bonds,
 %
%%%%%%%%%%%%%%%%%%%%%%
\begin{align}
	(\boldsymbol \sigma_{\rm g}, \boldsymbol\sigma_{\rm y} ) , \
	( \boldsymbol\sigma_{\rm g}, \boldsymbol\sigma_{\rm b} ) , \
	(\boldsymbol \sigma_{\rm r}, \boldsymbol\sigma_{\rm y} ) , \
	 ( \boldsymbol\sigma_{\rm r}, \boldsymbol\sigma_{\rm b} ) , 
\label{eq:8c.constraints.y}
\end{align}
%%%%%%%%%%%%%%%%%%%%%%
%
on the $y$ bonds, and
%
%%%%%%%%%%%%%%%%%%%%%%
\begin{align}
	( \boldsymbol\sigma_{\rm g}, \boldsymbol\sigma_{\rm b} ) , \
	( \boldsymbol\sigma_{\rm g},  \boldsymbol\sigma_{\rm r } ) , \
	 (\boldsymbol \sigma_{\rm y}, \boldsymbol \sigma_{\rm b} ) , \
	 ( \boldsymbol\sigma_{\rm y},  \boldsymbol\sigma_{\rm r}  ) , 
\label{eq:8c.constraints.z}
\end{align}
%%%%%%%%%%%%%%%%%%%%%%
%
on the $z$ bonds.
The same constraints apply for all ``dark'' colored spins.
We visualize these bond constraints 
for ``bright'' and ``dark'' spins in Figs.~\ref{fig:8cCSL.bond.constraints}(a) and 
\ref{fig:8cCSL.bond.constraints}(c), respectively.

Every spin in the CSL ground state adheres to these bond constraints, as 
one can verify by examining Figs.~\ref{fig:8cCSL.bond.constraints}(b) 
and \ref{fig:8cCSL.bond.constraints}(d), which depict representative spin 
configurations in the FM CSL ground-state manifold within "bright" and 
"dark" spin sectors, respectively. 
However, since every spin can choose between two different colored 
spins as their neighbors, the bond constraints are not strong enough 
to enforce long-range order.

%%%%%%%%%%%%%%%%%%%%%%%%%%%%%%%%%%%%%
%  Fig. - finite scalar chirality
%%%%%%%%%%%%%%%%%%%%%%%%%%%%%%%%%%%%%
%
\begin{figure}[t]
	\centering
  	\includegraphics[width=0.49\textwidth]{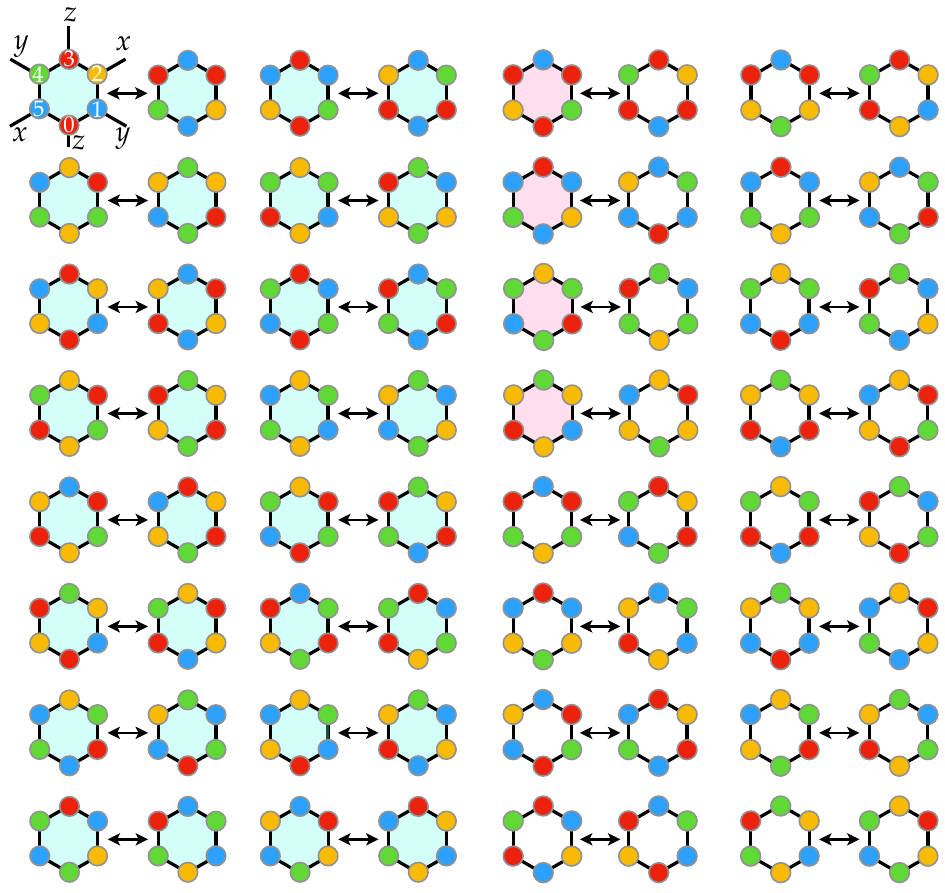}	
	\caption{ 
	All allowed arrangements of ``bright'' spins on a single hexagon in the 
	FM CSL state,
	with definitions for colors given in Table~\ref{tab:8.color.spins}.
	Configurations are arranged in pairs which are related to each other by 
	a six-site hexagon cluster update following Eq.~\eqref{eq:hex.update.S}.
	Definitions of site indices and labels for the Kitaev bonds are shown on the top 
	left hexagon.
	Hexagons are shaded in light cyan, pink, and white, indicating the 
	value of scalar spin chirality 
	$\kappa_p = -8/(3\sqrt{3})$, $16/(3\sqrt{3})$, and $0$, respectively.
	All states exhibit $W_p = 1$ for the semiclassical analog of the 
	$\mathbb{Z}_{2}$-flux operator.
	Definitions of $\kappa_p$ and $W_p$ are given in  
	Appendix~\ref{app:observables.details}.
	}
	\label{fig:chiralities}
\end{figure}
%%%%%%%%%%%%%%%%%%%%%%%%%%%%%%%%%%%%%
%

%% entropy
The resulting extensive ground state manifold induces a nonzero residual 
entropy in the CSL, 
which can be understood from the fact that two spin configurations 
at one hexagon can always be transformed into each other by a local hexagon cluster update:
%
%%%%%%%%%%%%%%%%%%%%%%
\begin{equation}
	{\bf S}^{\rm  new}_{p, j} = R^{\alpha}_j \ {\bf S}_{p, j} \, ,
\label{eq:hex.update.S}
\end{equation}
%%%%%%%%%%%%%%%%%%%%%%
%
where $R^{\alpha}_j$ is the rotation matrix of angle $\pi$ about the axis \mbox{$\alpha = x, y, z$};
$\alpha$ is the site-dependent label for the Kitaev bond pointing outwards
the hexagon $p$ at site $j = 0, \cdots, 5$ (see definition on site labels and 
$\alpha$ at top left in Fig.~\ref{fig:chiralities}).
In Fig.~\ref{fig:chiralities} we show all allowed spin configurations 
on a single hexagon in the ``bright'' manifold of the FM CSL,
which are related by such a six-site hexagon update.
Since the CSL allows always for two spin configurations per hexagon,
we obtain the normalized residual entropy in the ground state 
%
%%%%%%%%%%%%%%%%%%%%%%
\begin{equation}
	\frac{S(T \to 0)}{\ns} 
		= \frac{1}{\ns} \log{2^{N_{\rm h}}} 
		= \frac{1}{2} \log{2} \, ,
\label{eq:8c.entropy}
\end{equation}
%%%%%%%%%%%%%%%%%%%%%%
%
where $N_{\rm h} = \ns/2$ is the number of hexagons on the whole lattice.
This is $1/6$ of the total entropy per site, $\log{8}$.

%% scalar spin chirality
%
Interestingly, the set of all possible states on a hexagon
in Fig.~\ref{fig:chiralities} homogeneously exhibits a nonzero value for the 
octupole moment, 
$\sum_i S_i^x S_i^y S_i^z$, which induces broken time-reversal symmetry.
This time-reversal symmetry breaking consequently generates a nonzero 
scalar spin chirality on each hexagon, $\kappa_p$, as we defined in 
Eq.~\eqref{eq:kappa2} of  Appendix~\ref{app:observables.details}.
The $64$ states per hexagon 
form three groups with different values of 
$\kappa_p = -8/(3\sqrt{3})$ (cyan), $16/(3\sqrt{3})$ (pink), and $0$ (white) for $32$, $4$, and $28$
states respectively, which results in the averaged nonzero value of
%
%%%%%%%%%%%%%%%%%%%%%%
\begin{equation}
	\begin{aligned}
	\kappa 	&= \frac{1}{64} \sum_p \kappa_p  \\
			&= \frac{1}{64} \left( -\frac{8}{3\sqrt{3}}\times 32 
	     		  + \frac{16}{3\sqrt{3}}\times 4 + 0\times 28 \right)  \\
			&= - \frac{1}{\sqrt{3}} \approx -0.577		\, .
	\end{aligned}
\label{eq:8c.kappa}
\end{equation}
%%%%%%%%%%%%%%%%%%%%%%
%
Furthermore, we find that the semiclassical analog of the $\mathbb{Z}_{2}$-flux operator, $W_p$, 
defined in Eq.~\eqref{eq:Wp}, gives \mbox{$W_p = 1$} for every spin configuration in 
Fig.~\ref{fig:chiralities}.
We note that \mbox{$W_p = -1$} for the AFM CSL, offering a way to distinguish 
the two spin liquids from each other.

Besides the residual entropy and nonzero scalar spin chirality, 
the bond-dependent spin constraints in the CSL ground state show
extremely short-ranged spin correlations.
As visualized in Fig.~\ref{fig:8cCSL.bond.constraints}, nearest-neighbor color 
pairs are always of same contrast  (``bright''--``bright'' or ``dark''--``dark''), but never of 
the same color.
By comparison of spin components, as given in 
Table~\ref{tab:8.color.spins}, the correlations for all allowed 
nearest neighbors (distance $\ell = 1$) in the ground-state 
are therefore always exactly
%
%%%%%%%%%%%%%%%%%%%%%%
\begin{equation}
	D(\ell = 1) = \boldsymbol\sigma_{i} 
			\cdot \boldsymbol\sigma_{j} 
			= - \frac{1}{3} \, .
\label{eq:8c.Dr1}
\end{equation}
%%%%%%%%%%%%%%%%%%%%%%
%
Yet again, by considering the same color constraints, one finds that spin configurations 
for next-nearest neighbors and beyond are all equally allowed and give on average
%
%%%%%%%%%%%%%%%%%%%%%%
\begin{equation}
	D(\ell > 1) = \boldsymbol\sigma_{i} 
				\cdot \boldsymbol\sigma_{k} 
	= 3 \times \left(- \frac{1}{3} \right) + 1 \times 1  = 0 \, ,
\label{eq:8c.Dr2}
\end{equation}
%%%%%%%%%%%%%%%%%%%%%%
%
where the contribution $1\times1$ comes from a pair
of the same color, which is forbidden for nearest neighbors, but allowed for 
further neighbors. 
Interestingly, such extremely short-range correlations coincide with analytical predictions 
for the $S=1$ Kitaev SL in Ref.~[\onlinecite{Baskaran2008}].

%%%%%%%%%%%%%%%%%%%%%%%%%%%%%%%%%%%%%%%%%%
\subsubsection{Finite-temperature properties of eight-color model}
\label{sec:8c.model.thermo}
%%%%%%%%%%%%%%%%%%%%%%%%%%%%%%%%%%%%%%%%%%
%
%

%%%%%%%%%%%%%%%%%%%%%%%%%%%%%%%%%%%%%
%. Fig. -- Thermodynamics: chiral order & CLS 
%%%%%%%%%%%%%%%%%%%%%%%%%%%%%%%%%%%%%
\begin{figure*}[t]
	\centering
	\captionsetup[subfloat]{labelformat=empty}		
	\subfloat[chiral order]{
		\includegraphics[width=0.45\textwidth]{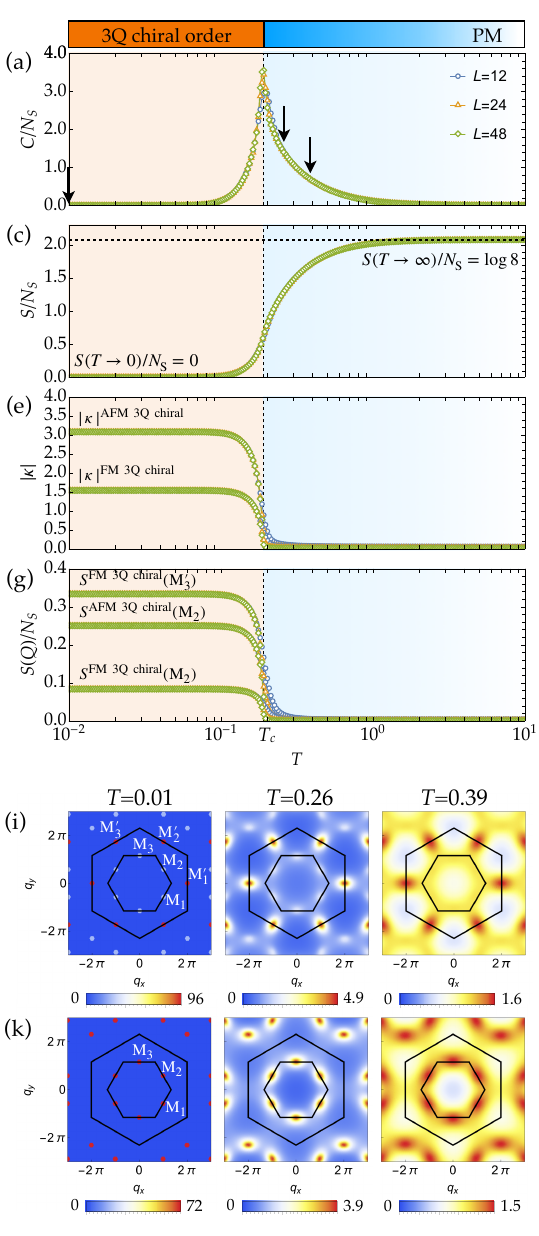} }	\hspace{1cm}
	\subfloat[chiral spin liquid (CSL) ]{
		\includegraphics[width=0.45\textwidth]{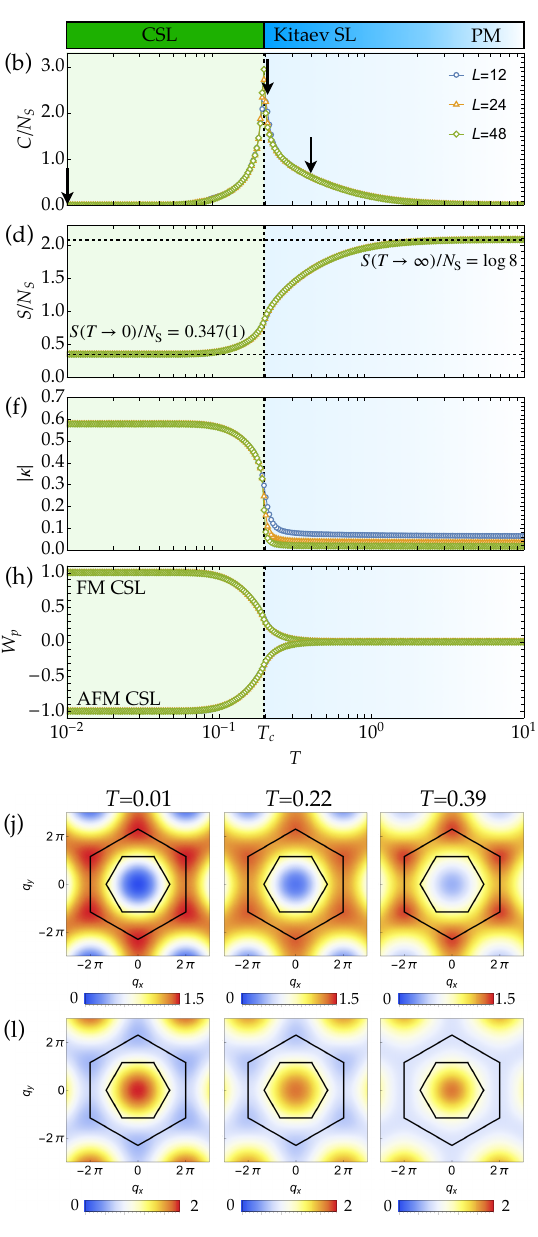} }	
	\caption{	
	Comparison of thermodynamic quantities between 
	the FM 3$Q$ chiral (\mbox{$\phi/\pi = 0.2$}, \mbox{$\theta/\pi = 0.9$}) and 
	AFM 3$Q$ chiral (\mbox{$\phi/\pi = 0.8$}, \mbox{$\theta/\pi = 0.1$}) 
	ordered phases (left column), and the
	FM CSL (\mbox{$\phi/\pi = 0.35$}, \mbox{$\theta/\pi = 0.25$}) and 
	AFM CSL (\mbox{$\phi/\pi = 0.65$}, \mbox{$\theta/\pi = 0.75$}) 
	phases (right column) for the eight-color model given by 
	$\mathcal{H}^{ {\sf 8c} }_{{\sf BBQ-K}}$ in Eq.~\eqref{eq:H.BBQ-K.8c}
	[see the phase diagram in Fig.~\ref{fig:PD.all}(c)].
	Shown are the temperature dependencies of 
	(a)--(b) the specific heat per site, $C/\ns$ [Eq.~\eqref{eq:spec.heat}], 
	(c)--(d) the entropy per site, $S/\ns$ [Eq.~\eqref{eq:entropy}], and
	(e)--(f) the absolute value of the scalar spin chirality, $|\kappa|$ [Eq.~\eqref{eq:kappa1}]. 
	(g) shows the spin structure factor $S_{\rm S}({\bf Q})$ 
	[Eqs.~\eqref{eq:Sq.MC} and \eqref{eq:defn.m.S2}]
	for the chiral ordered phases at characteristic ordering vectors indicated in (i) and (k). 
	(h) shows the semiclassical analog of the $\mathbb{Z}_{2}$-flux operator 
	[Eq.~\eqref{eq:Wp}] for the CSL phases. 
	The momentum resolved spin structure factors, $S_{\rm S}({\bf q})$, are shown for 
	(i) the FM 3$Q$ chiral ordered phase, 
	(j) the FM CSL phase, 
	(k) the AFM 3$Q$ chiral ordered phase, and 
	(l) the AFM CSL phase, for three different temperatures indicated by  
	black arrows in (a) and (b).
	Data were obtained from classical MC simulations 
	on clusters of size $L = 12, 24$, and $48$ ($\ns = 288, 1152$, and $4608$).
	$S_{\rm S}({\bf q})$ are shown for $L = 48$, except for (i) and (k) at $T=0.01$,
	where $L=12$.
	}
	\label{fig:8c.chiral.order.CSL.thermo}
\end{figure*}
%%%%%%%%%%%%%%%%%%%%%%%%%%%%%%%%%%%%%%%
%

In Fig.~\ref{fig:8c.chiral.order.CSL.thermo} we show finite-temperature results from
classical MC simulations of $\mathcal{H}^{ {\sf 8c} }_{{\sf BBQ-K}}$ in 
Eq.~\eqref{eq:H.BBQ-K.8c}, for chiral ordered phases 
(FM 3$Q$ chiral at $\phi / \pi = 0.2$ and $\theta / \pi = 0.9$, and 
AFM 3$Q$ chiral at $\phi / \pi = 0.8$ and $\theta / \pi = 0.1$), 
in the left column, and chiral liquid phases 
(FM CSL at $\phi / \pi = 0.35$ and $\theta / \pi = 0.25$, and
AFM CSL at $\phi / \pi = 0.65$ and $\theta / \pi = 0.75$) in the right column
for finite-size systems of linear dimension $L = 12, 24$, and $48$.
Since the Hamiltonian is symmetric in parameter space, 
the FM and AFM states are related to each other by inversion of the spins on one 
sublattice of the honeycomb lattice. 
Therefore, energy related observables, such as the 
specific heat $C$ [Eq.~\eqref{eq:spec.heat}] and the 
entropy $S$ [Eq.~\eqref{eq:entropy}] are identical for the FM and AFM 
states (here shown only once), while magnetic observables, like the 
scalar spin chirality $\kappa$ [Eq.~\eqref{eq:kappa1}], the 
spin structure factor $S_{\rm{S}}({\bf q})$ [Eqs.~\eqref{eq:Sq.MC} and \eqref{eq:defn.m.S2}], 
and the semiclassical analog of the $W_p$ flux operator [Eq.~\eqref{eq:Wp}] 
are distinct.

%% specific heat and entropy 
%
The specific heat, $C$ [Eq.~\eqref{eq:spec.heat}], in the first row of Fig.~\ref{fig:8c.chiral.order.CSL.thermo}, exhibits 
a sharp singularity at (a) \mbox{$T_c = 0.186(5)$} and (b) \mbox{$T_c = 0.192(5)$}~\footnote{
Values in parentheses represent the statistical error in the last digit, determined
by the standard deviation among five independent simulation runs.}.
At these transition temperatures, the system spontaneously breaks its chirality
by selecting a configuration where spins are either all bright or all dark 
in the FM 3$Q$ chiral ordered and FM CSL states
[see Eq.~\eqref{eq:8c.lowEnergy.4} and discussion in Sec.~\ref{sec:CSL.properties}].
Due to the symmetry of the Hamiltonian in parameter space, 
the AFM counterparts of these states similarly break chirality. 
However, in the AFM versions, there is a mixture of bright and dark spins due to the 
inversion of spins on one sublattice of the honeycomb lattice.
We confirmed the transition into the CSL to be a second-order phase transition, 
following the $\mathbb{Z}_{2}$ Ising universality class. 
The selection of either bright or dark spins generates a nonzero octupolar  
moment, $\sum_i S_i^x S_i^y S_i^z$, which acts as an Ising order parameter. 
This parameter is time-reversal odd, and consequently leads to the development 
of a nonzero scalar spin chirality.
We leave more detailed analyses for future work.

%% entropy 
%
Even though, the spontaneous breaking of chirality 
seems identical, the left column of Fig.~\ref{fig:8c.chiral.order.CSL.thermo} 
shows a transition into a magnetically  
ordered state, while the right column 
corresponds to a transition into the disordered CSL state. 
This can be directly confirmed in the second row of Fig.~\ref{fig:8c.chiral.order.CSL.thermo},
which shows the entropy, $S$ [see Eq.~\eqref{eq:entropy}], as obtained from numerical 
integration of the specific heat over the available temperature range $10^{-2} \leq  T \leq 10$.
The total entropy of the system, $\log{8}$, is released by passing through the phase 
transition at $T_c$, resulting in zero entropy $S(T \to 0) / \ns  = 0$
in the chiral ordered states as shown in Fig.~\ref{fig:8c.chiral.order.CSL.thermo}(c). 
In contrast, the CSLs in Fig.~\ref{fig:8c.chiral.order.CSL.thermo}(d) show a 
residual entropy of 
%
%%%%%%%%%%%%%%%%%%%%%%
\begin{equation}
	\frac{S(T\to0)}{\ns} 
			= 0.347(1) \approx \frac{1}{2} \log{2} \, ,   \\
\end{equation}
%%%%%%%%%%%%%%%%%%%%%%
%
which matches with the analytically predicted value in Eq.~\eqref{eq:8c.entropy},
and is direct numerical evidence of a disordered ground state  
with extensive degeneracy.
%

%% scalar spin chirality 
%
Furthermore, in Figs.~\ref{fig:8c.chiral.order.CSL.thermo}(e) and 
\ref{fig:8c.chiral.order.CSL.thermo}(f) we show the absolute value of the scalar 
spin chirality, $|\kappa|$ [see Eq.~\eqref{eq:kappa1}],
for the ordered and disordered chiral states. 
In both cases, $|\kappa|$ scales to zero above $T_c$, and becomes 
nonzero below $T_c$.
At low temperatures, we measure for the FM and AFM 3$Q$ chiral ordered 
states the following values 
%
%%%%%%%%%%%%%%%%%%%%%%
\begin{align}
	|\kappa|^{\rm FM \ 3{\it Q} \ chiral}  &= 1.5396(1) \approx \frac{8}{3 \sqrt{3}} \, ,	\\
	|\kappa|^{\rm AFM \ 3{\it Q}  \ chiral} &= 3.0792(1) \approx \frac{16}{3 \sqrt{3}} \,  ,
\end{align}
%%%%%%%%%%%%%%%%%%%%%%
%
and for the FM and AFM CSL states commonly  
%
%%%%%%%%%%%%%%%%%%%%%%
\begin{equation}
	|\kappa|^{\rm CSL} = 0.5774(1) \approx \frac{1}{\sqrt{3}}\, .
\label{eq:kappa.CSL}
\end{equation}
%%%%%%%%%%%%%%%%%%%%%%
%
Yet again, the result for the CSL matches very well with 
our analytical estimate 
in Eq.~\eqref{eq:8c.kappa}.
Furthermore, as shown in Fig.~\ref{fig:8c.chiral.order.CSL.thermo}(h),
the semiclassical analog of the $\mathbb{Z}_{2}$-flux operator reaches 
perfectly \mbox{$W_p =1$} in the FM CSL, and \mbox{$W_p = -1$} in the AFM CSL,
as expected analytically.

%% S(q) - chiral order 
%
Characteristic magnetic signatures, which clearly distinguish ordered from disordered states,
can be found in the dipolar equal-time structure factor $S_{\rm{S}}({\bf q})$ 
[see Eqs.~\eqref{eq:Sq.MC} and \eqref{eq:defn.m.S2}].
We show the $S_{\rm{S}}({\bf q})$ at three different temperatures indicated 
by black arrows in Fig.~\ref{fig:8c.chiral.order.CSL.thermo}(a), for the 
FM and AFM 3$Q$ chiral ordered phases in Figs.~\ref{fig:8c.chiral.order.CSL.thermo}(i) 
and \ref{fig:8c.chiral.order.CSL.thermo}(k), respectively.
By reducing the temperature, the spectral weight accumulates around 
the M points in the Brillouin zone. 
Below $T_c$, Bragg peaks develop at the M$_1$, M$_2$, and M$_3$ points 
in the AFM 3$Q$ chiral state.
Additionally, Bragg peaks appear at the
M$'_1$, M$'_2$, and M$'_3$ points in the FM 3$Q$ chiral state. 
These clearly identify both phases as triple-$q$ ordered states.
The intensities of selected Bragg peaks are explicitly shown as functions of 
temperature in Fig.~\ref{fig:8c.chiral.order.CSL.thermo}(g). 
%

%% S(q) - chiral spin liquid 
%
Meanwhile, the $S_{\rm{S}}({\bf q})$ for the CSL phases behave very differently. 
In Figs.~\ref{fig:8c.chiral.order.CSL.thermo}(j) and 
\ref{fig:8c.chiral.order.CSL.thermo}(l) we show the $S_{\rm{S}}({\bf q})$ for the 
FM and AFM CSL states, respectively,
at three different temperatures indicated by black arrows in 
Fig.~\ref{fig:8c.chiral.order.CSL.thermo}(b).
The signal is very diffuse and does not show any qualitative difference 
below and above $T_c$.
In fact, the scattering function is reminiscent of the $S_{\rm{S}}({\bf q})$ for the 
Kitaev SL~\cite{Samarakoon2017}, showing very diffuse scattering around the 
Brillouin zone edge 
in the FM CSL, and accumulation of intensity around the $\Gamma$ point in 
the AFM CSL. 
This diffuse scattering signature and the absence of any Bragg peaks are the direct 
evidence of a magnetically disordered state below $T_c$.
%

%% spin-spin correlations in chiral spin liquid 
%
%
%%%%%%%%%%%%%%%%%%%%%%%%%%%%%%%%%%%%%
%. Fig. -- CSL - spin-spin correlations
%%%%%%%%%%%%%%%%%%%%%%%%%%%%%%%%%%%%%
\begin{figure}[t]
	\centering
	\includegraphics[width=0.45\textwidth]{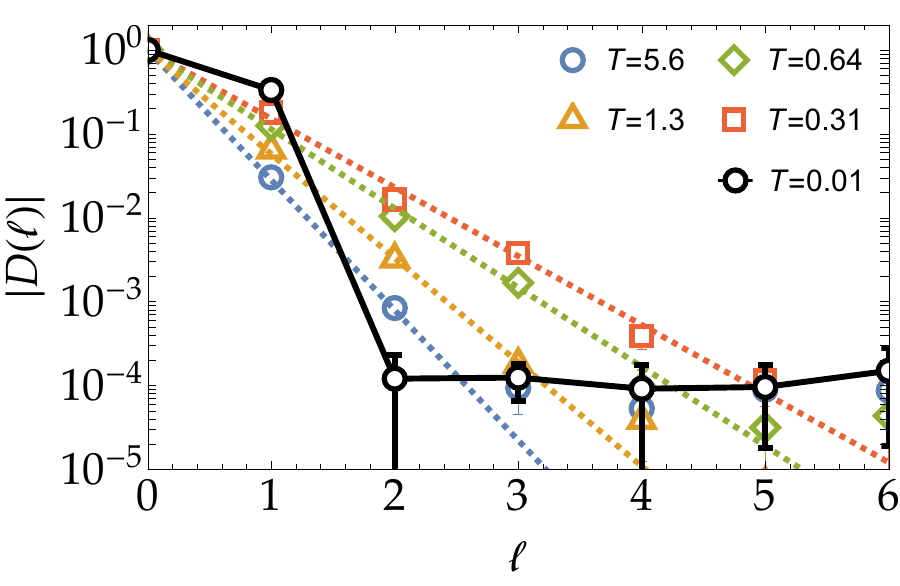}
	\caption{	
	Semilogarithmic plot of spin-spin correlations $|D(\ell)|$ 
	[see Eq.~\eqref{eq:Dr}] 
	for the FM CSL case at five different temperatures. 
	Correlations show an exponential decay above the critical temperature 
	$T_c = 0.192(5)$ [peak in Fig.~\ref{fig:8c.chiral.order.CSL.thermo}(b)].
	Below $T_{\rm c}$, correlations show a large value for nearest-neighbor 
	contributions, $|D(\ell = 1)| = 0.3333(1)$, and a negligibly small value 
	beyond nearest neighbors, $|D(\ell>1)| \lesssim 10^{-4}$.
	Data were obtained from classical MC simulations of the 
	eight-color model given by $\mathcal{H}^{ {\sf 8c} }_{{\sf BBQ-K}}$ 
	in Eq.~\eqref{eq:H.BBQ-K.8c}, at model parameters 
	$\phi / \pi = 0.35$ and $\theta / \pi = 0.25$, for a cluster of size 
	$L = 48$ ($\ns = 4608$).
	Correlations are shown as function of the Manhatten distance $\ell$,
	and were obtained after averaging over symmetrically equivalent 
	paths along zigzag chains on the honeycomb lattice. 
	}
	\label{fig:8c.CSL.SS}
\end{figure}
%%%%%%%%%%%%%%%%%%%%%%%%%%%%%%%%%%%%%%%
%

Since momentum-resolved spin-spin correlations of the CSL  
seem very similar to those in the Kitaev SL, we 
investigate correlations also in real space. 
Figure~\ref{fig:8c.CSL.SS} shows the MC data of the spin-spin 
correlations 
%
%%%%%%%%%%%%%%%%%%%%%%
\begin{equation}
	D(\ell) = \langle \boldsymbol\sigma_0 \cdot \boldsymbol\sigma_{\ell} \rangle \, ,
\label{eq:Dr}
\end{equation}
%%%%%%%%%%%%%%%%%%%%%%
%
in the FM CSL at $\phi / \pi = 0.35$ and $\theta / \pi = 0.25$ 
measured at five different temperatures. 
The correlations are averaged over symmetrically equivalent paths
and are plotted semilogarithmically, where 
$\ell$ corresponds to the Manhatten distance along zigzag chains 
on the honeycomb lattice.
For temperatures above $T_c$, correlations decay exponentially, with 
increasing correlation length upon reduction of $T$. 
Below $T_c$, the behavior changes dramatically and 
exhibits strong nearest-neighbor 
correlations with $|D(\ell = 1)| = 0.3333(1)$ 
at the lowest measured temperatures $T=0.01$, 
while correlations beyond the nearest neighbors become negligibly 
small with $|D(\ell>1)| \lesssim 10^{-4}$.
These results confirm our analytical prediction in Eqs.~\eqref{eq:8c.Dr1} 
and \eqref{eq:8c.Dr2} of extremely short-range correlations in the CSL.

In short summary of this subsection, our investigation focused on 
a simplified model, the eight-color model, $\mathcal{H}^{ {\sf 8c} }_{{\sf BBQ-K}}$
in Eq.~\eqref{eq:H.BBQ-K.8c}, with the aim of unraveling the unique 
nature and nontrivial characteristics of the eight-color CSL.
We discovered that the CSL is characterized 
by bond-dependent spin constraints, as 
visualized in Fig.~\ref{fig:8cCSL.bond.constraints}, giving rise to 
exotic properties such as residual entropy, nonzero scalar spin 
chirality, nonzero $\mathbb{Z}_{2}$-flux order parameter, and 
extremely short-range spin correlations.
We have successfully verified the existence of all these properties through 
unbiased numerical MC simulations.
%

%%%%%%%%%%%%%%%%%%%%%%%%%%%%%%%%%%%%%%%%%%
%
% 			The O(3) model
%
%%%%%%%%%%%%%%%%%%%%%%%%%%%%%%%%%%%%%%%%%%
\subsection{Analysis of $\mathbb{CP}^{1}$ model}	
\label{sec:O3.model}
%%%%%%%%%%%%%%%%%%%%%%%%%%%%%%%%%%%%%%%%%%
%
%

In the previous section, we investigated the eight-color model, as a simplification of the 
original BBQ-Kitaev model, $\mathcal{H}^\mathcal{S}_{ {\sf BBQ-K} }$ in Eq.~\eqref{eq:H.BBQ-K.S},
by discretising the continuous spin space of an $S=1$ moment to only eight allowed, noncoplanar 
states. 
This helped us to transparently understand the nature and origin of the exotic 
CSL phase in a simple analytical way. 

In this section, we are going to partially relax this constraint, and allow dipolar spin components 
to take continuous values, namely being a classical Heisenberg vector
described in the order parameter space $\mathbb{CP}^{1}$.
Compared to the eight-color model, the $\mathbb{CP}^{1}$ model, with its enlarged
spin space, offers a variety of additional and nontrivial phases.
Understanding these phases will help us in bridging our comprehension 
from the simplified eight-color model to the more complex $\mathbb{CP}^{2}$ model.

%%%%%%%%%%%%%%%%%%%%%%%%%%%%%%%%%%%%%%%%%%
\subsubsection{Ground-state properties of $\mathbb{CP}^{1}$ model}	
\label{sec:O3.model.GS}
%%%%%%%%%%%%%%%%%%%%%%%%%%%%%%%%%%%%%%%%%%
%

%
%%%%%%%%%%%%%%%%%%%%%%%%%%%%%%%%%%%%%
%. Fig. --Energy comparison along \theta = 0.06
%%%%%%%%%%%%%%%%%%%%%%%%%%%%%%%%%%%%%
\begin{figure}[thbp!]
	\centering
	\includegraphics[width=0.45\textwidth]{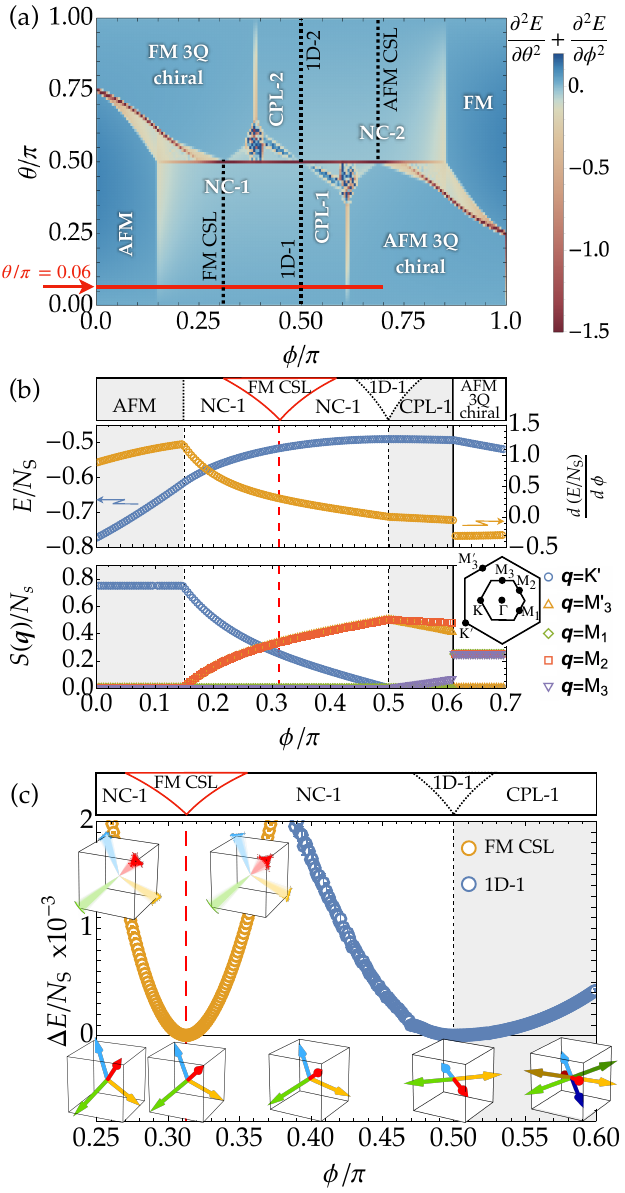}
	\caption{	
	Competing phases of $\mathcal{H}^\mathcal{S}_{ {\sf BBQ-K} }$ in 
	Eq.~\eqref{eq:H.BBQ-K.S} for classical Heisenberg spins ($\mathbb{CP}^{1}$ model), 
	obtained by variational energy 
	minimization for a finite-size cluster of linear dimension $L=120$ 
	($\ns = 28 \ 800$ spins) under periodic boundary conditions.
	(a) Zoom-in of the ground-state phase diagram in Fig.~\ref{fig:PD.all}(b) 
	to the region for $0.0 \leq \phi/\pi \leq 1.0$.
	(b) Normalized energy and its derivative, $E/\ns$ and $\partial (E/\ns) / \partial \phi$,
	(top), and the structure factor for dipole moments, $S_{\rm S}({\bf q})$
	[Eqs.~\eqref{eq:Sq.MC} and \eqref{eq:defn.m.S2}], 
	at high-symmetry momenta (bottom) at $\theta/\pi = 0.06$ [red line in (a)].
	The solid and dashed lines represent first- and second-order 
	phase transitions, respectively.
	(c) Energy difference, $\Delta E/\ns$, measured from the ground state to the 
	metastable FM CSL and 1D-1 phases at $\theta/\pi = 0.06$. 
	The lower insets show spin configurations in each ordered phase, while the upper ones represent 
	metastable states of the FM CSL. 
	}
	\label{fig:O3.energy.comparison}
\end{figure}
%%%%%%%%%%%%%%%%%%%%%%%%%%%%%%%%%%%%%%%
%

%% general overview of phases
%
In Fig.~\ref{fig:O3.energy.comparison}(a) we show a zoom-in of the ground-state phase 
diagram in Fig.~\ref{fig:PD.all}(b) for positive $J_2$ $(0.0 < \phi / \pi < 1.0)$, where the 
model is mostly frustrated.
Between FM/AFM ordered and FM/AFM 3$Q$ chiral phases the model stabilizes
NC phases, and canted planar (CPL) phases over a wide range of model parameters. 
The canting angle of spins in those phases depends on the model parameters. 
Notably, specific choices of $\phi$ allow us to stabilize disordered CSLs and 
dimensionally-reduced 1D phases in the ground state, as indicated by the 
black dotted lines in Fig.~\ref{fig:O3.energy.comparison}(a)
\footnote{
We observed several unconventional phases near the two SU(3) points 
at $(\theta/\pi, \phi/\pi) = (0.5, 0.25)$ and $(0.5, 0.5)$, which we reserve 
for a more in-depth discussion in the future.}.

%% evolution between AFM - NC - CSL - NC - 1D - X - AFM chiral 
%
To see these phase competitions in more detail, in Fig.~\ref{fig:O3.energy.comparison}(b) 
we concentrate on the lower part of the phase diagram at $\theta / \pi =0.06$ 
[red line in Fig.~\ref{fig:O3.energy.comparison}(a)], 
and explicitly show the normalized energy, $E/\ns$, and 
its derivative, $\partial (E/\ns) / \partial \phi$ (top), together with the $S_{\rm S}({\bf q})$ 
[Eqs.~\eqref{eq:Sq.MC} and \eqref{eq:defn.m.S2}] at 
relevant ordering vectors K$'$, M$_1$, M$_2$, M$_3$, and M$'_3$ (bottom).
The transition from CPL-1 to AFM 3$Q$ chiral phase at $\phi / \pi \approx 0.611$ is of first order 
(solid black line), while transitions from AFM to NC-1 at $\phi / \pi \approx 0.148$ and 
NC-1 to CPL-1 at $\phi / \pi = 0.5$ are of second order (dashed black lines),
as seen by a jump and kinks in $\partial (E/\ns) / \partial \phi$, respectively. 
The Bragg peaks at the K$'$ points 
monotonically reduce their intensity 
in the \mbox{NC-1} phase as $\phi$ increases, while 
additional ordering vectors at the M$_2$ and M$'_3$ points increase.
In the CPL-1 phase, intensities at the M$_3$ and M$_1$ points become also 
nonzero, while the AFM 3$Q$ chiral state shows an equal weight at the M$_1$, 
M$_2$, and M$_3$ points, forming the triple-$q$ order, as also obtained in the 
eight-color model in Fig.~\ref{fig:8c.chiral.order.CSL.thermo}(k).
%

%% CSL
%
In Fig.~\ref{fig:O3.energy.comparison}(c) we show the energy difference $\Delta E/\ns$ 
between the FM CSL (yellow circles) and the 1D-1 state (blue circles) relative to 
the ground state for $\theta/\pi = 0.06$.
The lower insets illustrate the spin configurations within each ordered phase, while the upper 
insets represent metastable states of the FM CSL. 
Upon increasing $\phi / \pi$ above $\approx 0.148$, spins in the ground state 
start to cant away from the collinear AFM arrangement to establish the NC-1 order, 
with a four-site magnetic unit cell [see Fig.~\ref{fig:PD.all}(d)].
The canting angle of spins in the NC-1 phase is $\phi$ dependent, and the spontaneous 
selection of a particular direction for this canting angle induces a single-$q$ order 
with Bragg peaks at the M$_2$ points in the first Brillouin zone, supplemented by other peaks 
in the extended Brillouin zone [see Fig.~\ref{fig:O3.energy.comparison}(b)].

Upon tuning the model to $\phi / \pi = (1/\pi )\arctan(3/2)$ [see Eq.~\eqref{eq:O3.CSL.AFM}], 
the spins achieve isotropic alignment across the lattice, by pointing precisely 
towards the corners of a unit cube.
At this critical point, the system allows
an extensive number of spin configurations to form the CSL ground state,
characterized by all properties 
as discussed on the eight-color model in Sec.~\ref{sec:CSL.properties}.
In the vicinity of this critical point, we find a CSL like state as a metastable state. 
Spins in this metastable CSL are slightly canted away from the (111) directions 
as shown in the insets of Fig.~\ref{fig:O3.energy.comparison}(c), while still retaining 
the underlying nature of the CSL with macroscopic degeneracy.
%

%% 1D phase 
%
Upon further increasing $\phi$, spins in the NC-1 ground state continue to 
cant until they achieve complete coplanarity, precisely at the boundary between the NC-1 
and CPL-1 phases, at $\phi / \pi = 0.5$. 
At this specific point, nearest-neighbor spins are mutually orthogonal to each other, and  
effectively cancel nearest-neighbor bilinear and biquadratic interactions.
Coplanar spins globally select either the $xy$, $yz$, or $zx$ plane 
[the inset of Fig.~\ref{fig:O3.energy.comparison}(c) shows the case of the $xy$ plane],
by setting one of their spin components to zero.
Therefore the dominating Kitaev interactions for 
one particular bond-type will be zero, and effectively decouple individual zigzag chains 
throughout the whole lattice. 
The system effectively forms a dimensionally-reduced 1D phase.
Increasing $\phi$ beyond this point will again impose canting of spins, but this time with 
an enlarged eight-site magnetic unit cell forming the magnetic order of the CPL-1 state
[see Fig.~\ref{fig:PD.all}(d)].

%%%%%%%%%%%%%%%%%%%%%%%%%%%%%%%%%%%%%%%%%%
\subsubsection{Finite-temperature properties of $\mathbb{CP}^{1}$ model}
\label{sec:O3.model.thermo}
%%%%%%%%%%%%%%%%%%%%%%%%%%%%%%%%%%%%%%%%%%
%
%

%%%%%%%%%%%%%%%%%%%%%%%%%%%%%%%%%%%%%
%. Fig. -- Thermodynamics:CLS & 1D
%%%%%%%%%%%%%%%%%%%%%%%%%%%%%%%%%%%%%
\begin{figure}[t]
	\centering
	\includegraphics[width=0.45\textwidth]{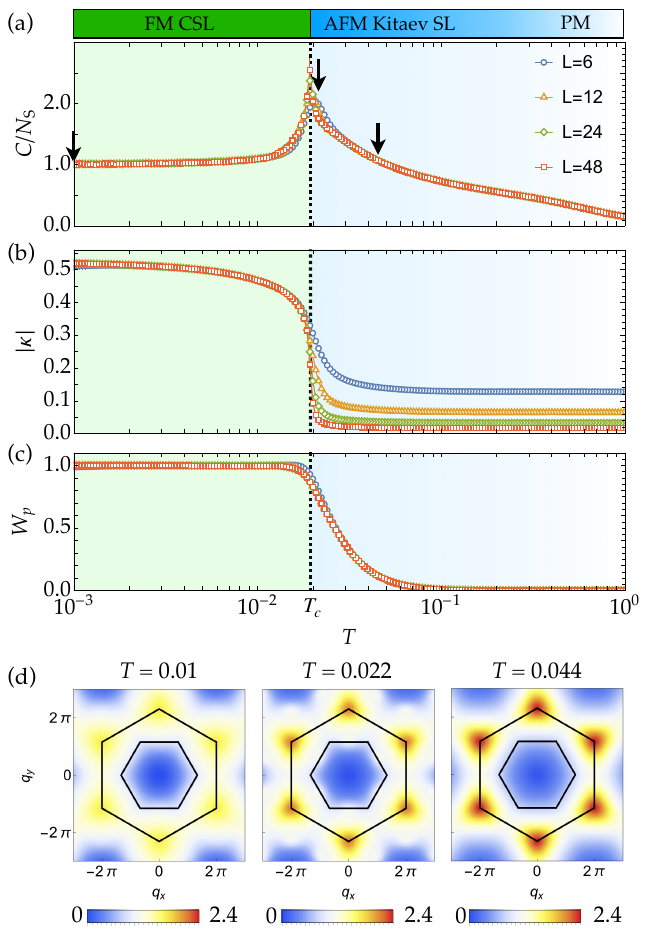}	
	\caption{	
	Finite-temperature properties of the FM CSL phase 
	for the $\mathbb{CP}^{1}$ model
	given by $\mathcal{H}^\mathcal{S}_{ {\sf BBQ-K} }$ in 
	Eq.~\eqref{eq:H.BBQ-K.S} at \mbox{$\phi / \pi = (1/\pi) \arctan (3/2)$}
	and \mbox{$\theta/\pi = 0.06$}.
	Shown are temperature dependencies of 
	(a) the specific heat per site, $C/\ns$ [Eq.~\eqref{eq:spec.heat}],	
	(b) the absolute value of the scalar spin chirality $|\kappa|$ [Eq.~\eqref{eq:kappa1}], and 
	(c) the semiclassical analog of the $\mathbb{Z}_{2}$-flux operator, $W_p$
	[Eq.~\eqref{eq:Wp}].
	MC simulations were performed for $L=6,12,24$, and 
	$48$ ($\ns = 72, 288, 1152$, and $4608$).
	(d) shows the spin structure factor $S_{\rm S}({\bf q})$ 
	[Eqs.~\eqref{eq:Sq.MC} and \eqref{eq:defn.m.S2}] at temperatures indicated
	by black arrows in (a) for $L=24$.
	}
	\label{fig:O3.thermo.CSL}
\end{figure}
%%%%%%%%%%%%%%%%%%%%%%%%%%%%%%%%%%%%%%%
%

Let us now turn to the thermodynamic properties of the CSL in the 
$\mathbb{CP}^{1}$ model.  
In Fig.~\ref{fig:O3.thermo.CSL}, we show results from classical MC simulations of
$\mathcal{H}^\mathcal{S}_{ {\sf BBQ-K} }$ in Eq~\eqref{eq:H.BBQ-K.S},
at model parameters $\theta/\pi = 0.06$ and $\phi / \pi = (1/\pi) \arctan (3/2)$,
which stabilize the FM CSL in the ground state [see Eq.~\eqref{eq:O3.CSL.AFM}].

In Figs.~\ref{fig:O3.thermo.CSL}(a)--\ref{fig:O3.thermo.CSL}(c) we respectively 
show the specific heat $C/\ns$ [Eq.~\eqref{eq:spec.heat}], the scalar spin chirality 
$|\kappa|$ [Eq.~\eqref{eq:kappa1}], and the semiclassical analog of 
the $\mathbb{Z}_{2}$-flux operator, $W_p$ [Eq.~\eqref{eq:Wp}], over a
wide range of temperatures. 
$C/\ns$ shows a clear singularity at $T_c = 0.0193(6)$, which scales weakly with system size,
and has a shape which looks somewhat similar to the observed singularity in the eight-color model
in Fig.~\ref{fig:8c.chiral.order.CSL.thermo}(b).

For $T\to 0$, we observe $C/\ns \to 1$, as expected for classical Heisenberg spins 
in the absence of soft mode excitations.
We note that the acceptance ratio for single-spin flip updates at such low temperatures 
becomes strongly suppressed (see Appendix~\ref{app:acc.SS.hex}), suggesting that 
continuous local spin motion is not able to adiabatically connect different ground-state configurations.
However, where the single-spin flip fails, the hexagon update becomes successful, 
demonstrating that in the CSL phase decorrelation primarily occurs through cluster 
updates [see Eq.~\eqref{eq:hex.update.S}].

The scalar spin chirality $|\kappa|$ scales to zero in the paramagnetic regime, and takes 
nonzero values below $T_c$, as observed in the eight-color model in 
Fig.~\ref{fig:8c.chiral.order.CSL.thermo}(f).
However, in comparison, $|\kappa|$ does not rapidly saturate, but rather 
monotonically reaches a value slightly smaller
than in the eight-color model [see Eq.~\eqref{eq:kappa.CSL}].
We note that the asymptotic value of $|\kappa|$ becomes 
$\theta$ dependent (not shown here).
However, variational energy minimization confirms that at $T=0$, spins precisely align 
along the (111) direction, as also illustrated in the lower insets of
Fig.~\ref{fig:O3.energy.comparison}(c).
This seemingly contradictory observation implies that different configurations 
of the CSL ground state manifold in Fig.~\ref{fig:chiralities} are selected with 
varying weights in the $\mathbb{CP}^{1}$ model, resulting in a reduction 
of $\kappa$ at finite $T$.
At the lowest temperatures, the $\mathbb{Z}_{2}$-flux operator takes a 
value \mbox{$W_p \approx 1$}, as observed in the eight-color model.

In Fig.~\ref{fig:O3.thermo.CSL}(d) we show the spin structure factor $S_{\rm S}({\bf q})$ 
at three different temperatures 
indicated by black arrows in Fig.~\ref{fig:O3.thermo.CSL}(a).
The signal is diffuse across the entire range of temperatures and closely resembles the 
spin structure factor of the eight-color FM CSL [see Fig.~\ref{fig:8c.chiral.order.CSL.thermo}(j)]
with an enlarged concentration of intensity around the K' points 
[see inset in Fig.~\ref{fig:O3.energy.comparison}(b)]
and, almost invisible,
some additional weak intensity accumulation around the M points 
in the first Brillouin zone at $T=0.022$. 

Thermodynamic properties for the CSL in both the eight-color model and the 
$\mathbb{CP}^{1}$ model match remarkably well. 
This alignment provides strong evidence that the eight-color CSL, with all of its physical 
properties analyzed in Sec.~\ref{sec:CSL.properties}, remains preserved even after 
relaxing the spin degree of freedom to $\mathbb{CP}^{1}$. 
Given that the CSL persists as a metastable state in the vicinity of its optimal model parameters
given by Eq.~\eqref{eq:O3.CSL.AFM} [see Fig.~\ref{fig:O3.energy.comparison}(c)], we anticipate 
its survival for an enlarged region of model parameters at finite temperatures 
due to the substantial entropy associated with its macroscopic degeneracy.

%%%%%%%%%%%%%%%%%%%%%%%%%%%%%%%%%%%%%%%%%%
%
% 			The CP2 model
%
%%%%%%%%%%%%%%%%%%%%%%%%%%%%%%%%%%%%%%%%%%
\subsection{Analysis of $\mathbb{CP}^{2}$ model}
\label{sec:SU3.model}
%%%%%%%%%%%%%%%%%%%%%%%%%%%%%%%%%%%%%%%%%%
%
%

The original BBQ-Kitaev model allows for both dipole and quadrupole components 
of $S=1$ magnetic moments, which are correctly described in the spin  
space $\mathbb{CP}^{2}$. 
In the last two subsections we simplified this model by restricting the spin 
degree of freedom to extract the essential properties of the CSL state.
In the present subsection we come back to the original model by reintroducing all 
allowed degrees of freedom for $S=1$ moments.  
We will not only see that the CSL survives at finite temperatures even in the 
presence of quadrupolar degree of freedom, but also quantitatively examine the impact 
of local quantum fluctuations on the ground-state phase diagram.
Some results in this subsection have already been presented in our earlier 
work in Ref. [\onlinecite{Pohle2023}], but we include them here to make the discussion 
and comparison to $\mathbb{CP}^{1}$
and eight-color models self-contained.

%%%%%%%%%%%%%%%%%%%%%%%%%%%%%%%%%%%%%%%%%%
\subsubsection{Ground-state properties of $\mathbb{CP}^{2}$ model}	
\label{sec:U3.model.GS}
%%%%%%%%%%%%%%%%%%%%%%%%%%%%%%%%%%%%%%%%%%
%
%

%
%%%%%%%%%%%%%%%%%%%%%%%%%%%%%%%%%%%%%
%. Fig. --Energy comparison along \theta = 0.06
%%%%%%%%%%%%%%%%%%%%%%%%%%%%%%%%%%%%%
\begin{figure}[htbp!]
	\centering
	\includegraphics[width=0.45\textwidth]{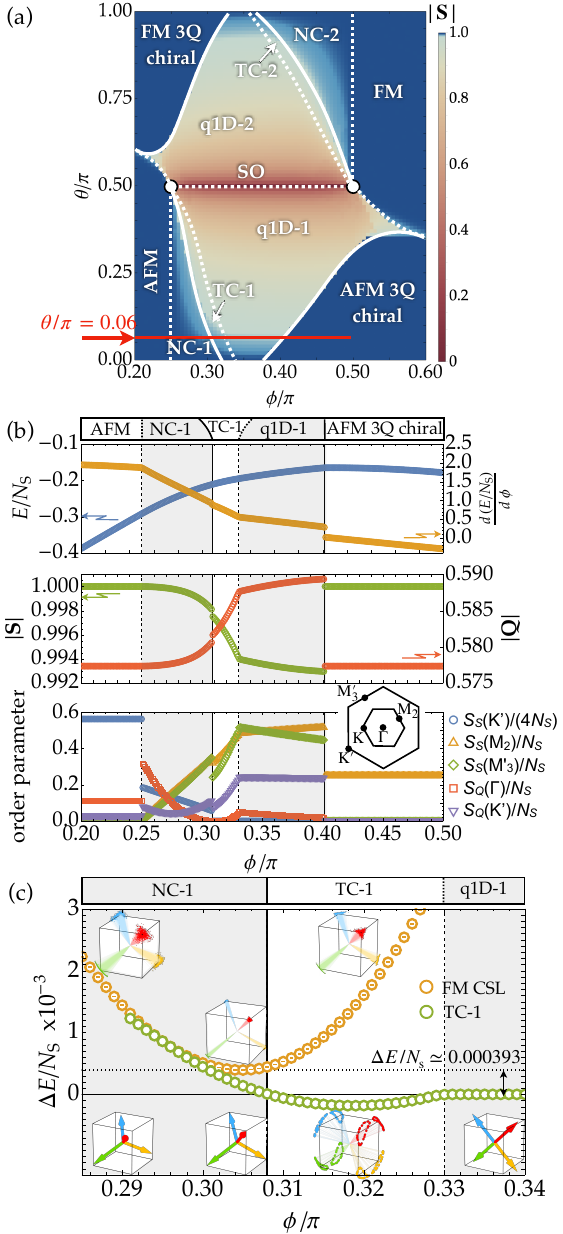}
	\caption{	
	Competing phases of $\mathcal{H}^\mathcal{A}_{ {\sf BBQ-K} }$ in 
	Eq.~\eqref{eq:H.BBQ-K.A} for $S=1$ magnetic moments  ($\mathbb{CP}^{2}$ model), 
	obtained by variational energy 
	minimization for a finite-size cluster of linear dimension $L=120$ 
	($\ns = 28 \ 800$ spins) under periodic boundary conditions.
	(a) Averaged spin norm $|{\bf S}|$ [see Eq.~\eqref{eq:S.norm}] in the region 
	$0.2 \leq \phi / \pi \leq 0.5$. 
	The solid and dashed lines represent first- and second-order 
	phase transitions,
	respectively.
	(b) The normalized energy $E/\ns$ and its derivative $d (E/\ns) / d \phi$ 
	(top panel), norms for dipole moment, $|{\bf S}|$, and quadrupole moment, 
	$|{\bf Q}|$ [see Eq.~\eqref{eq:Q.norm}] (middle panel), 
	and the structure factors for dipole moments, $S_{\rm{S}}({\bf q})$
	[Eqs.~\eqref{eq:Sq.MC} and \eqref{eq:defn.m.S}],
	and quadrupole moments, $S_{\rm{Q}}({\bf q})$ [Eqs.~\eqref{eq:Sq.MC} and \eqref{eq:defn.m.Q}] 
	(bottom panel), along $\theta/\pi = 0.06$ [red line in (a)].	
	(c) Energy difference $\Delta E/\ns$, measured from the NC-1 state.
	The lower insets show spin configurations 
	in each ordered phase, while the upper ones resemble metastable states
	of the FM CSL. 
	}
	\label{fig:U3.energy.comparison}
\end{figure}
%%%%%%%%%%%%%%%%%%%%%%%%%%%%%%%%%%%%%%%
%

%
In Fig.~\ref{fig:U3.energy.comparison}(a), we show the spin-dipole norm, $|{\bf S}|$
[see Eq.~\eqref{eq:S.norm}],  
in the frustrated region of the phase diagram in Fig.~\ref{fig:PD.all}(a).
In most of the phases, the spin norm is $|{\bf S}| \neq 1$, indicating the presence of 
nonzero onsite quadrupole moments. 
The onsite dipole and quadrupole moments mix in nontrivial ways to 
alleviate frustration, leading to quantitative changes in phases and their 
boundaries, in contrast to the $\mathbb{CP}^{1}$ model
shown in 
Fig.~\ref{fig:O3.energy.comparison}(a).

The purely quadrupolar SO phase at \mbox{$\theta / \pi=0.5$} 
[see Fig.~\ref{fig:PD.all}(d)], immediately changes into a 
q1D ordered state in presence of finite Kitaev interactions.
For both $\theta/\pi>0.5$ and $\theta/\pi<0.5$, the length of dipolar spin 
moments monotonically increases and behaves essentially the same as described for 
the 1D phase in the $\mathbb{CP}^{1}$ model in Sec.~\ref{sec:O3.model.GS},
where spins form dominant correlations along isolated 1D zigzag chains.
The key distinction from the $\mathbb{CP}^{1}$ model
lies in the presence of quadrupole correlations,
which introduce weak interactions between the chains at very low temperatures.
This leads to the formation 
of a 2D ordered state exhibiting one-dimensional character over a wide region 
at low and intermediate temperatures.
This characteristic is the reason why we refer to the 
state as a quasi-1D state.
Remarkably, quadrupole correlations stabilize the
q1D phases in a wide region within the phase diagram, which is in stark contrast 
to the $\mathbb{CP}^{1}$ model, where the 1D phases remain stable only 
along a singular line.
However, to this expense, the NC phases are notably suppressed, along 
with the neighboring TC phases.

In Fig.~\ref{fig:U3.energy.comparison}(b), we focus along the line at $\theta / \pi = 0.06$ 
[red line in Fig.~\ref{fig:U3.energy.comparison}(a)], and plot 
$E/\ns$ and $\partial(E/\ns)/ \partial \phi$ (top panel), the spin dipole norm $|{\bf S}|$ 
[see Eq.~\eqref{eq:S.norm}] and 
quadrupole norm $|{\bf Q}|$  [see Eq.~\eqref{eq:Q.norm}] (middle panel), and the structure factors 
for dipoles, $S_{\rm S}({\bf q})$ [Eqs.~\eqref{eq:Sq.MC} and \eqref{eq:defn.m.S}], and
for quadrupoles, $S_{\rm Q}({\bf q})$ [Eqs.~\eqref{eq:Sq.MC} and \eqref{eq:defn.m.Q}], 
at relevant ordering vectors (bottom panel).
We identify first-order phase transitions (solid lines) from NC-1 to TC-1 at 
\mbox{$\phi / \pi = 0.3085(5)$}, and from q1D-1 to AFM 3$Q$ chiral at \mbox{$\phi / \pi = 0.4015(1)$}. 
Additionally, we find second-order transitions (dashed lines) from 
AFM to NC-1 at \mbox{$\phi / \pi = 0.250(1)$}, and from \mbox{TC-1} to \mbox{q1D-1} at 
\mbox{$\phi / \pi = 0.3305(5)$}.
Notably, our measurements reveal a reduction in the spin dipole norm $|{\bf S}| < 1$ 
within the \mbox{NC-1}, \mbox{TC-1}, and q1D-1 phases. 
It is worth noting 
that the reduction of $|{\bf S}|$ is $\theta$ dependent; however, in this 
parameter region near the Kitaev limit, it is relatively small, with less than $1$~\%.
This highlights the significance of dipole moments and justifies our comparison to the simplified 
$\mathbb{CP}^{1}$ model
for dominant Kitaev interactions, as done in Sec.~\ref{sec:O3.model}.

In Fig.~\ref{fig:U3.energy.comparison}(c), we present the energy difference between the 
metastable FM CSL and TC-1 states to the NC-1 ordered state.
The presence of local quadrupole moments induces a small, yet finite energy difference to 
the FM CSL bounded by
%
%%%%%%%%%%%%%%%%%%%%%%
\begin{equation}
	\frac{\Delta E}{\ns} = 0.000393(4) \, ,
\label{eq:energy.gap}
\end{equation}
%%%%%%%%%%%%%%%%%%%%%%
%
for $\phi / \pi = 0.305(1)$. 
Consequently, the CSL becomes an excited state rather than the ground state, with 
spin configurations no longer perfectly aligned with the corners of the unit cube, but 
still retaining the essential properties of the CSL as a metastable state.
In fact, we find that the FM CSL is a metastable state with small energy gap in a 
wide region of the phase diagram, and smoothly merges with the AFM Kitaev SL 
for $\theta \to 0$ (see the Supplemental Material in Ref.~\cite{Pohle2023}).
The evolution of spin canting angles, illustrated in the bottom insets, closely resembles the 
general behavior discussed in the $\mathbb{CP}^{1}$ model 
[see Fig.~\ref{fig:O3.energy.comparison}(c)]. 
We note, however, a noticeable difference: At $\phi / \pi = 0.3085(5)$, the TC-1 state 
[see spin configurations in lower inset of Fig.~\ref{fig:U3.energy.comparison}(c)]
becomes the ground state, a scenario which seems absent in the $\mathbb{CP}^{1}$ model.
Furthermore, the q1D-1 state maintains its status as the ground state for $\phi / \pi > 0.3305(5)$, 
in contrast to the $\mathbb{CP}^{1}$ model 
where it remains the ground state only along a singular line.

%%%%%%%%%%%%%%%%%%%%%%%%%%%%%%%%%%%%%%%%%%
\subsubsection{Finite-temperature properties of $\mathbb{CP}^{2}$ model}
\label{sec:U3.model.thermo}
%%%%%%%%%%%%%%%%%%%%%%%%%%%%%%%%%%%%%%%%%%
%
%

%
%%%%%%%%%%%%%%%%%%%%%%%%%%%%%%%%%%%%%
%. Fig. -- Thermodynamics: CLS
%%%%%%%%%%%%%%%%%%%%%%%%%%%%%%%%%%%%%
\begin{figure}[htbp!]
	\centering
	\includegraphics[width=0.45\textwidth]{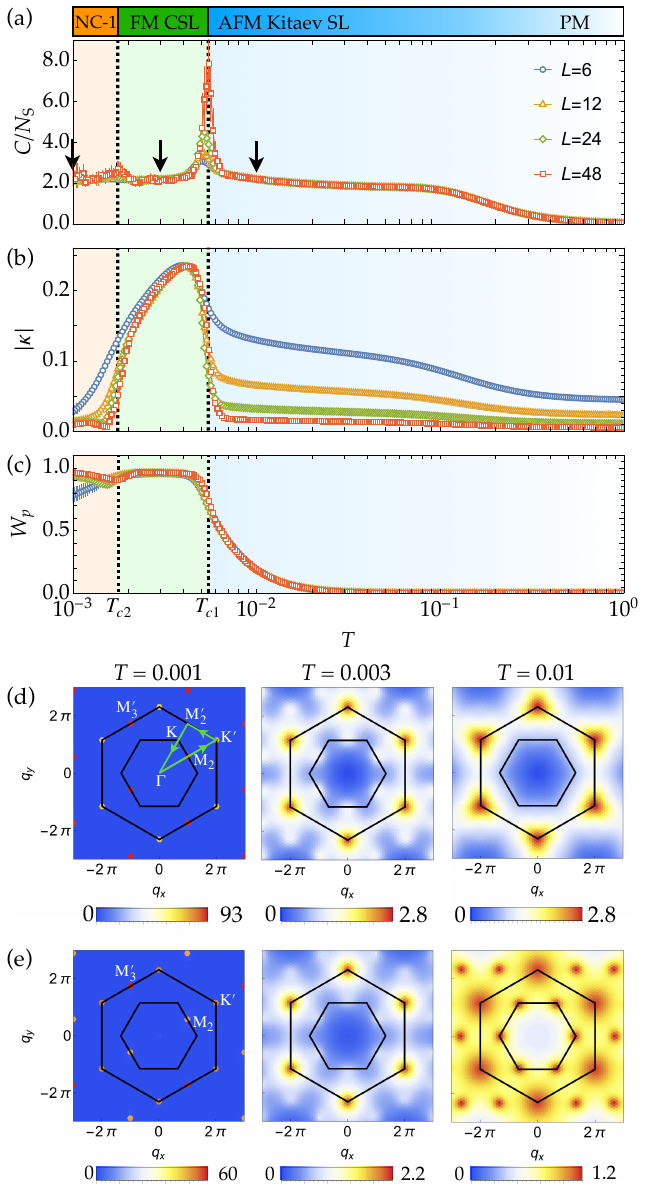}	
	\caption{		
	MC simulations of 
	$\mathcal{H}^{\mathcal{A}}_{ {\sf BBQ-K} }$ in Eq.~\eqref{eq:H.BBQ-K.A},
	for the $\mathbb{CP}^{2}$ model 
	at $\phi / \pi = 0.305$ and $\theta / \pi = 0.06$ reveal the presence of the FM CSL 
	at finite temperatures.
	Shown are temperature dependencies of 
	(a) the specific heat per site, $C/\ns$  [Eq.~\eqref{eq:spec.heat}],
	(b) the absolute value of the scaler spin chirality $|\kappa|$ [Eq.~\eqref{eq:kappa1}], and 
	(c) the semiclassical analog of the $\mathbb{Z}_{2}$-flux operator, $W_p$
	[Eq.~\eqref{eq:Wp}].
	Simulations were done for $L=6,12, 24$, and $48$ ($\ns =72,288,1152,$ and $4608$).
	(d) and (e) show the spin structure factor for dipoles, $S_{\rm S}({\bf q})$, and 
	quadrupoles $S_{\rm Q}({\bf q})$ [Eqs.~\eqref{eq:Sq.MC}--\eqref{eq:defn.m.Q}], 
	respectively, at temperatures indicated by black 
	arrows in (a) for $L = 24$ at $T =0.01$ and $T =0.003$, while for $L=12$
	at $T =0.001$.
	The green path in the left panel of (d) corresponds to the path used 
	for the dynamical structure factor in Fig~\ref{fig:Sqw}.
	}
	\label{fig:U3.thermo.CSL}
\end{figure}
%%%%%%%%%%%%%%%%%%%%%%%%%%%%%%%%%%%%%%%
%

The extensive degeneracy 
in the FM CSL and the small but finite energy gap to the 
ground state suggest that the system undergoes an entropy-driven phase 
transition from NC-1 order to the FM CSL at finite temperature.
To confirm this, we show finite-temperature MC simulations for
$\mathcal{H}^{\mathcal{A}}_{ {\sf BBQ-K} }$ [Eq.~\eqref{eq:H.BBQ-K.A}] in 
Fig.~\ref{fig:U3.thermo.CSL}, for model parameters where 
the FM CSL shows the smallest energy difference to the ground state 
(\mbox{$\theta / \pi = 0.06$} and \mbox{$\phi / \pi = 0.305$}).
Figures~\ref{fig:U3.thermo.CSL}(a)--\ref{fig:U3.thermo.CSL}(d) are basically the same 
as those in Ref.~[\onlinecite{Pohle2023}], but we additionally show 
the static structure factors for 
quadrupoles $S_{\rm Q}({\bf q})$ 
in Fig.~\ref{fig:U3.thermo.CSL}(e).

The specific heat $C/\ns$ [Eq.~\eqref{eq:spec.heat}] in Fig.~\ref{fig:U3.thermo.CSL}(a) shows two 
singularities at $T_{c1}=0.0055(2)$ and $T_{c2}=0.0017(2)$, 
which scale weakly with system size.
For $T\to0$, we observe $C/\ns \to 2$, as expected for $S=1$ magnetic moments described 
within the spin space of $\mathbb{CP}^{2}$ 
in the absence of soft-mode fluctuations \cite{Remund2022}.
Below $T_{c2}$, characteristic Bragg peaks emerge in the structure factors at the
momentum points, corresponding to the NC-1 order.
Therefore, we associate $T_{c2}$ with a symmetry-breaking transition into the NC-1 ordered phase
\cite{Pohle2023}.

The intermediate phase between $T_{c1}$ and $T_{c2}$ shows all properties of the FM 
CSL discussed in the previous subsections, namely a nonzero, size-independent 
scalar spin chirality $|\kappa|$ [Eq.~\eqref{eq:kappa1}], a value of $W_p$ [Eq.~\eqref{eq:Wp}]
which is almost $+1$, and a very diffuse structure factor for spin dipoles, $S_{\rm S}({\bf q})$.
The values of both quantities, $|\kappa|$ and $W_p$, are somewhat reduced compared 
to the ideal eight-color CSL discussed in Sec.~\ref{sec:8c.model}. 
This reduction stems from the decrease in spin lengths $|{\bf S}|$ caused by the presence 
of small, but finite, quadrupole moments [see Fig.~\ref{fig:U3.energy.comparison}(b)] and 
temperature fluctuations.

The diffuse $S_{\rm S}({\bf q})$ is comparable to the structure factors found in the 
eight-color and $\mathbb{CP}^{1}$ models 
[see Figs.~\ref{fig:8c.chiral.order.CSL.thermo}(j)
and \ref{fig:O3.thermo.CSL}(d), respectively], albeit some additional weak intensity 
features around the M points in the first Brillouin zone.
The quadrupole structure factor, $S_{\rm Q}({\bf q})$, also suggests the 
absence of quadrupolar order, with scattering patterns closely resembling those for dipoles. 
This similarity arises because spin moments are primarily dipolar in nature, and 
dipole characteristics are implicitly reflected in the quadrupole components.
Taken these observations, we associate $T_{c1}$ with a discrete 
chiral symmetry breaking into the FM CSL, as observed in the eight-color model
and $\mathbb{CP}^{1}$ model cases [see Figs.~\ref{fig:8c.chiral.order.CSL.thermo}(b) and 
\ref{fig:O3.thermo.CSL}(a), respectively].
It is worth noting that the intermediate-temperature CSL persists over an extended region 
around $\phi / \pi \approx 0.3$ and $\theta / \pi \approx 0.2$, as shown in 
the Supplemental Material of Ref.~[\onlinecite{Pohle2023}].

The scattering in $S_{\rm S}({\bf q})$ at $T=0.01$, above $T_{c1}$, exhibits 
characteristics reminiscent of the semiclassical 
AFM Kitaev SL, as our chosen model parameters contain strong Kitaev 
interactions. 
The intensity exhibits diffuse accumulation around the Brillouin zone edge,
similar to the results for the ``pure'' Kitaev model in Appendix~\ref{app:dynamics.Kitaev},
however, with a stronger accumulation of diffuse intensity around the K' points.
Meanwhile, the structure factor $S_{\rm Q}({\bf q})$ shows a more uniform intensity 
distribution with dominant scattering intensity around the K and K' points.

All physical observables characterizing the CSL in the $\mathbb{CP}^{2}$ model 
closely resemble those 
measured in the eight-color model (Sec.~\ref{sec:8c.model}) and the $\mathbb{CP}^{1}$ model
(Sec.~\ref{sec:O3.model}). 
This remarkable consistency strongly suggests that the dominant properties of the CSL, 
as discussed in Sec.~\ref{sec:CSL.properties}, remains robust even after considering all allowed
local degrees of freedom for an $S=1$ moment in the spin space $\mathbb{CP}^{2}$.
However, it is important to emphasize that the CSL no longer represents the ground state
but emerges as an entropically driven state at finite temperature.

%%%%%%%%%%%%%%%%%%%%%%%%%%%%%%%%%%%%%%%%%%
%
% 			Dynamics
%
%%%%%%%%%%%%%%%%%%%%%%%%%%%%%%%%%%%%%%%%%%
\subsection{Dynamical properties of the chiral spin liquid }	
\label{sec:Dynamics}
%%%%%%%%%%%%%%%%%%%%%%%%%%%%%%%%%%%%%%%%%%
%
%

In this section we analyze dynamical properties of the CSL in the $\mathbb{CP}^{2}$ model.
We show that the excitations corresponding to the CSL, in fact, are gapped, consistent 
with our observation of extreme short-range correlations in 
Fig.~\ref{fig:8c.CSL.SS}.

%%%%%%%%%%%%%%%%%%%%%%%%%%%%%%%%%%%%%%%%%%
\subsubsection{Excitation spectrum}	
%%%%%%%%%%%%%%%%%%%%%%%%%%%%%%%%%%%%%%%%%%
%

%%%%%%%%%%%%%%%%%%%%%%%%%%%%%%%%%%%%%%%%%%%
%. Fig. -- S(q,w) for AFM CSL in SU(3) model 
%%%%%%%%%%%%%%%%%%%%%%%%%%%%%%%%%%%%%%%%%%%
\begin{figure*}[tbp]
	\centering
	\subfloat[][ NC-1, ${\tilde S}_{\rm{S}}({\bf q}, \omega), T=0.001$ ]{
		\includegraphics[width=0.33\textwidth]{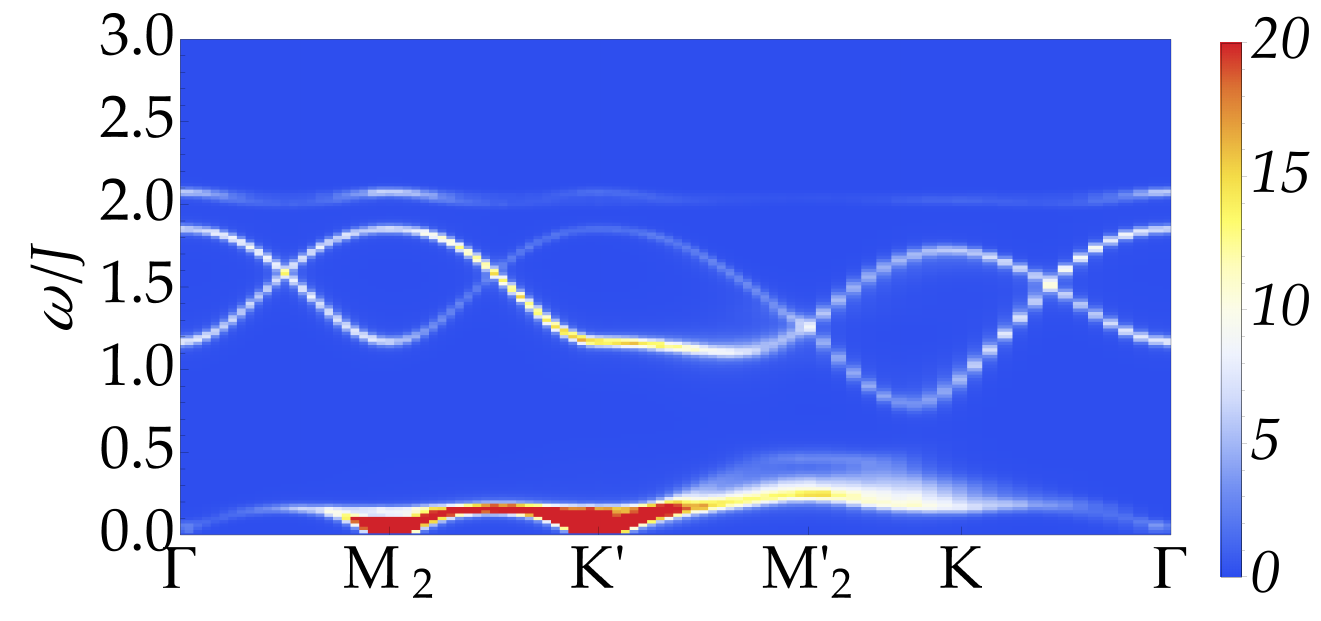}} 
	\subfloat[][ FM CSL, ${\tilde S}_{\rm{S}}({\bf q}, \omega), T=0.003$]{
		\includegraphics[width=0.33\textwidth]{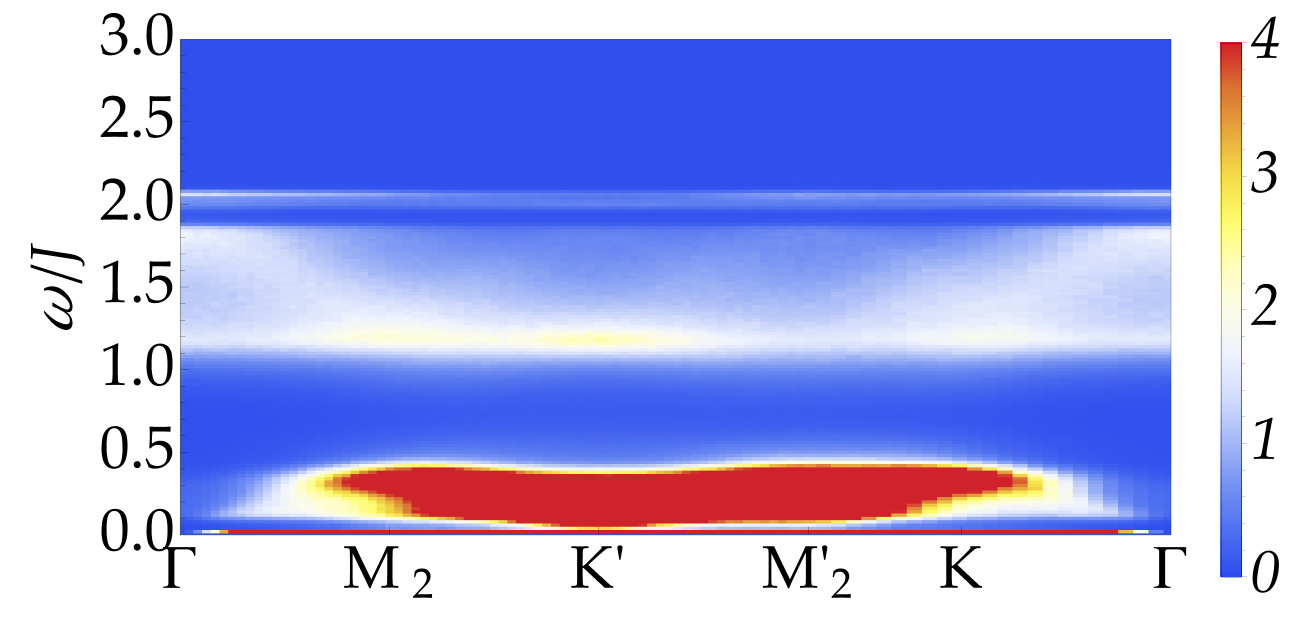}} 	
	\subfloat[][ AFM Kitaev SL, ${\tilde S}_{\rm{S}}({\bf q}, \omega), T=0.01$]{
		\includegraphics[width=0.33\textwidth]{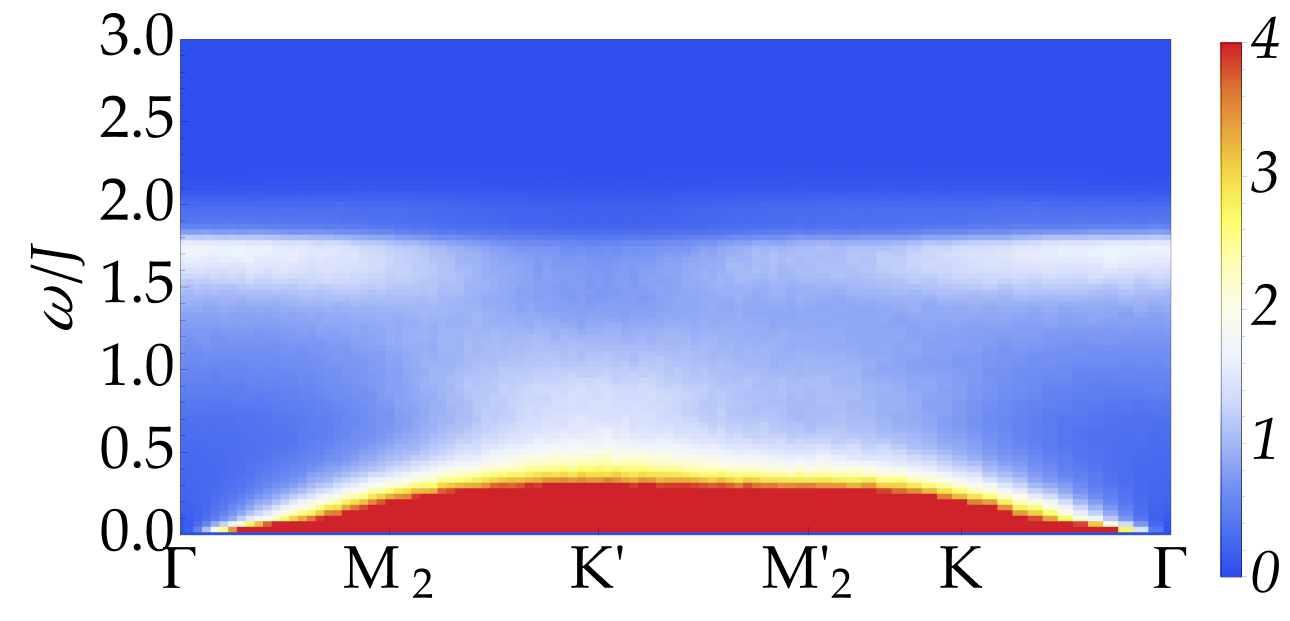}} 	\\
	\subfloat[][ NC-1, ${\tilde S}_{\rm{Q}}({\bf q}, \omega), T=0.001$]{
		\includegraphics[width=0.33\textwidth]{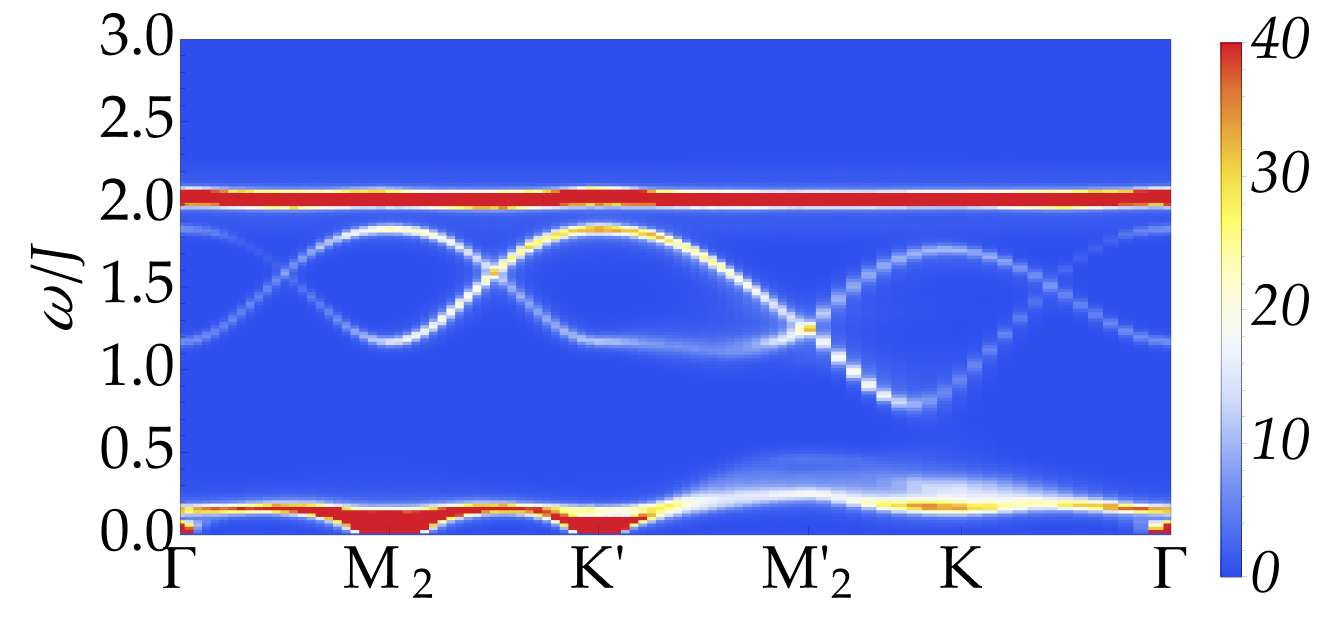}} 
	\subfloat[][ FM CSL, ${\tilde S}_{\rm{Q}}({\bf q}, \omega), T=0.003$]{
		\includegraphics[width=0.33\textwidth]{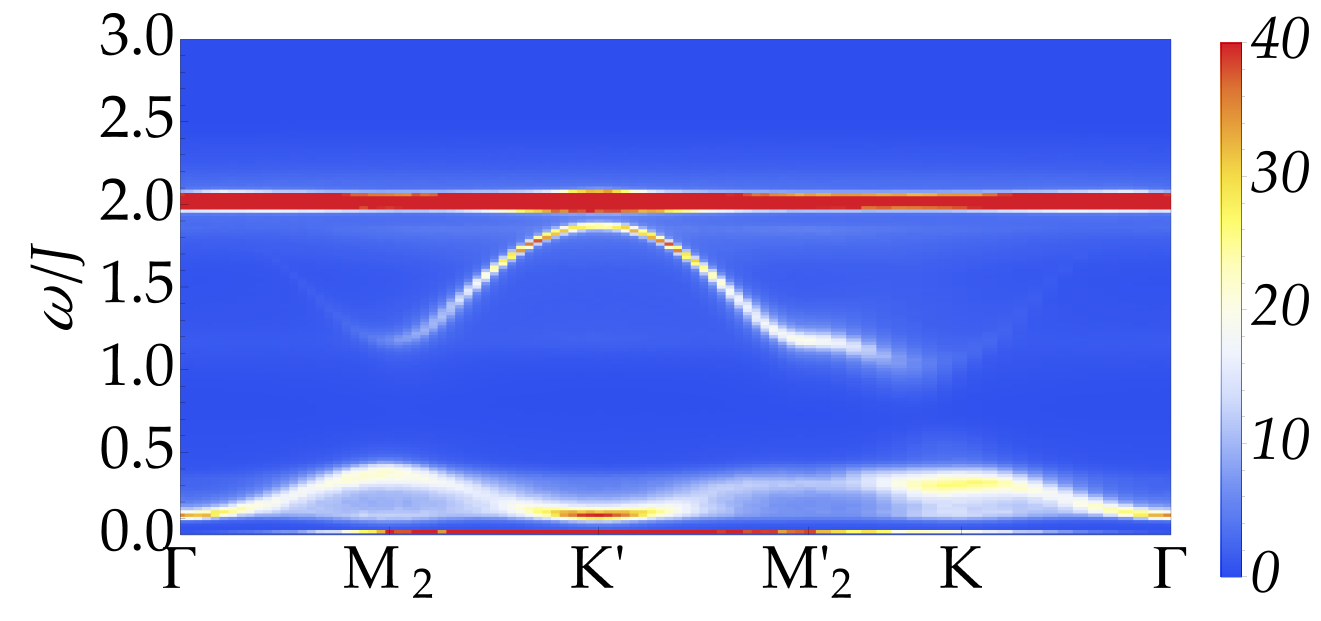}} 	
	\subfloat[][ AFM Kitaev SL, ${\tilde S}_{\rm{Q}}({\bf q}, \omega), T=0.01$]{
		\includegraphics[width=0.33\textwidth]{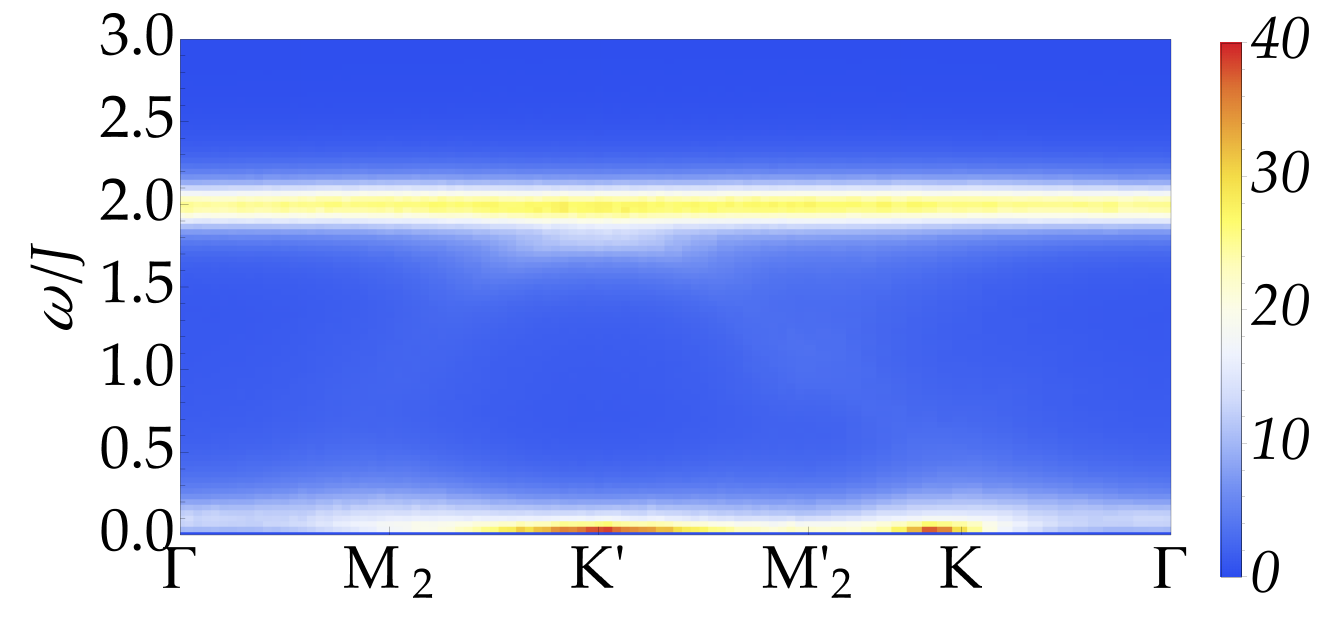}} 	
	\caption{
	Dynamical structure factors [see Eqs.~\eqref{eq:Sqw}--\eqref{eq:Dyn.correct}] 
	for spin-dipole moments, ${\tilde S}_{\rm{S}}({\bf q}, \omega)$ (top panels), 
	and spin-quadrupole moments, $\tilde{S}_{\rm{Q}}({\bf q}, \omega)$ 
	(bottom panels) of $\mathcal{H}^{\mathcal{A}}_{ {\sf BBQ-K} }$ in Eq.~\eqref{eq:H.BBQ-K.A},
	for the $\mathbb{CP}^{2}$ model, at $\phi / \pi = 0.305$ and $\theta / \pi = 0.06$.
	Results are shown for (a) and (d) the NC-1 ordered phase at $T=0.001$,
	(b) and (e) the FM CSL at $T=0.003$, and 
	(c) and (f) the semiclassical analog of the AFM Kitaev SL at $T=0.01$.
	Temperatures are indicated by black arrows in Fig.~\ref{fig:U3.thermo.CSL}(a).
	Data were obtained from molecular dynamics simulations (details in 
	Appendix~\ref{app:method.details}) for finite-size clusters of $L=48$ $(N=4608)$.
	Results are plotted along the green path in momentum space, as 
	indicated in the left panel of Fig.~\ref{fig:U3.thermo.CSL}(d).
	}
	\label{fig:Sqw}
\end{figure*}
%%%%%%%%%%%%%%%%%%%%%%%%%%%%%%%%%%%%%%%%%%%
%

In Fig.~\ref{fig:Sqw}, we show the dynamical structure factors for 
spin-dipole moments, $\tilde{S}_{\rm{S}}({\bf q}, \omega)$ (top panels), 
and spin-quadrupole moments, $\tilde{S}_{\rm{Q}}({\bf q}, \omega)$ (bottom panels),  
which are obtained from molecular dynamics simulations 
(see details in Appendix~\ref{app:method.details})
at the same model parameters as chosen 
in Fig.~\ref{fig:U3.thermo.CSL}.
We show the spectrum along the path in momentum space, as indicated by green lines 
in the left panel of Fig.~\ref{fig:U3.thermo.CSL}(d) and compare dynamics 
between the \mbox{NC-1} ordered phase at $T = 0.001$, the FM CSL at $T = 0.003$, and the 
semiclassical analog of the AFM Kitaev SL 
at $T=0.01$.
Their corresponding equal-time structure factors are plotted in 
Figs.~\ref{fig:U3.thermo.CSL}(d) and \ref{fig:U3.thermo.CSL}(e).

% dynamics for NC-1 phase: short
%
Figures~\ref{fig:Sqw}(a) and \ref{fig:Sqw}(d) show 
the dynamical structure factors for the NC-1 ordered phase.
For model parameters used here, the energy cost of a single-spin flip is primarily controlled by the 
Kitaev bond energies, limiting the bandwidth of excitations to $\omega / J \approx 2$.
Given the magnetic order in this phase, well-defined spin waves and quadrupole waves  emerge 
in the dipole and quadrupole channels, respectively. 
The signal reveals a nontrivial dispersion, which can be separated into three energy regimes.
In the low-energy regime for $\omega/J \lesssim 0.5$, linearly dispersing 
Goldstone modes exist at the magnetic ordering vectors $\Gamma$, M$_2$, and 
K' (M'$_3$ is not shown here), in accordance to the Bragg-peaks 
seen in the left column of 
Figs.~\ref{fig:U3.thermo.CSL}(d) and \ref{fig:U3.thermo.CSL}(e).
Moving to higher energy between \mbox{$0.6 \lesssim \omega/J \lesssim 1.9$}, two well-defined 
gapped bands exist, equally present in spin-dipole as well as spin-quadrupole channels.
At the energy maximum, the spectrum features an almost 
flat band, corresponding to quadrupole excitations captured by strong intensity in 
$\tilde{S}_{\rm{Q}}({\bf q}, \omega)$.
The band is almost flat, as the biquadratic interactions $J_2$
are small compared to the dominant Kitaev interactions $K$.
A weak-intensity ``shadow'' of this excitation is visible in the dipole channel, coming from 
mixing between dipole and quadrupole excitations due to spin-anisotropic
interactions.

% dynamics for AFM CSL
%
When the system turns into the FM CSL by increasing the temperature above $T_{c2}$, noticeable 
changes occur in the dynamical signatures, as shown in Figs.~\ref{fig:Sqw}(b) and \ref{fig:Sqw}(e).
The once well-defined spin waves at higher energy in the NC-1 ordered phase 
form a broad continuum of diffuse excitations with a gap of approximately $1.2J$ and a band 
maximum of around $1.9J$.
We associate this continuum with the gapped excitations of the FM CSL.
Considering the simplified eight-color model in Sec.~\ref{sec:8c.model}, an elementary 
excitation out of the CSL ground-state manifold is comprised of a single-spin flip, violating the 
ground-state constraints in Fig.~\ref{fig:8cCSL.bond.constraints}.
However, such a single-spin flip will violate the ground-state constraints not only on one 
but on two bonds, with an energy cost on each bond of $\frac{2}{3} K$ 
[see Eq.~\eqref{eq:8c.GS.energy}].
By using $K = \cos{(0.06 \pi)}$, for the model parameters used in this calculation
(see Fig.~\ref{fig:U3.thermo.CSL}), we estimate a minimum 
excitation energy of 
%
%%%%%%%%%%%%%%%%%%%%%%
\begin{equation}
	\Delta \omega^{\rm CSL}_{\rm min} = 2 \times \frac{2}{3} K \approx 1.31 J	\, .
\label{eq:gap.CSL}
\end{equation}
%%%%%%%%%%%%%%%%%%%%%%
%
This estimate aligns well with the energy gap observed in Fig.~\ref{fig:Sqw}(b). 
Above this energy region, quadrupole excitations do not form a 
continuum but instead exhibit well-defined quadrupole waves, 
with a remnant of one of the two branches seen in the NC-1 phase in 
Fig.~\ref{fig:Sqw}(d).
This is rather intriguing, since the energy-integrated correlations 
in Figs.~\ref{fig:U3.thermo.CSL}(d) and \ref{fig:U3.thermo.CSL}(e) show a 
diffuse structure which is very similar for both $S_{\rm S}({\bf q})$ and $S_{\rm Q}({\bf q})$.
This suggests that excitations in the CSL continuum induce coherent quadrupole dynamics 
at finite frequencies.

Furthermore, we observe that the well-defined Goldstone modes, corresponding to the NC-1 order 
change into a strong zero-energy flat band 
and a diffuse continuum of excitations with a small energy gap of $\sim 0.1J$ 
and bandwidth of $\sim 0.5J$.
As the zero-energy mode is challenging to distinguish from the diffuse low-energy continuum, 
we provide the dynamics for the FM CSL in the $\mathbb{CP}^{1}$ model in 
Appendix~\ref{app:dynamics.CSL.O3}.
Here, dynamics exhibit more pronounced and sharper bands, allowing us to validate their 
genuine physical presence and ensuring that they are not artifacts of the simulations.

% dynamics for AFM Kitaev SL
%
In Figs.~\ref{fig:Sqw}(c) and \ref{fig:Sqw}(f), we show results above $T_{c1}$ in the 
semiclassical analog of the AFM Kitaev SL. 
It appears that the three energy regimes observed in the FM CSL phase merge 
to form a spin structure factor that spans the full energy range up to 
$\omega / J \approx 2$, with strong intensity below $\sim 0.5 J$. 
The quadrupole channel is dominated by an almost flat band at 
$\omega / J \approx 2$, which corresponds to the same 
excitations of quadrupolar origin as discussed in Figs.~\ref{fig:Sqw}(d) and \ref{fig:Sqw}(e).
The observed dynamical correlations in this finite-temperature phase strongly resemble 
the dynamics of the semiclassical Kitaev SL, as shown in the comparison in 
Appendix~\ref{app:dynamics.Kitaev}.
This similarity supports our interpretation of a phase transition from 
the semiclassical Kitaev SL to the CSL state, since 
chosen model parameters ensure that Kitaev interactions strongly 
favor Kitaev SL physics above $T_{c1}$.

%%%%%%%%%%%%%%%%%%%%%%%%%%%%%%%%%%%%%%%%%%
\subsubsection{Energy cross sections}	
%%%%%%%%%%%%%%%%%%%%%%%%%%%%%%%%%%%%%%%%%%
%

% energy cross sections: AFM CSL 
%
%%%%%%%%%%%%%%%%%%%%%%%%%%%%%%%%%%%%%%%%%%%
%. Fig. -- S(q,w) line cuts for AFM CSL 
%%%%%%%%%%%%%%%%%%%%%%%%%%%%%%%%%%%%%%%%%%%
\begin{figure}[t]
	\centering
	\includegraphics[width=0.49\textwidth]{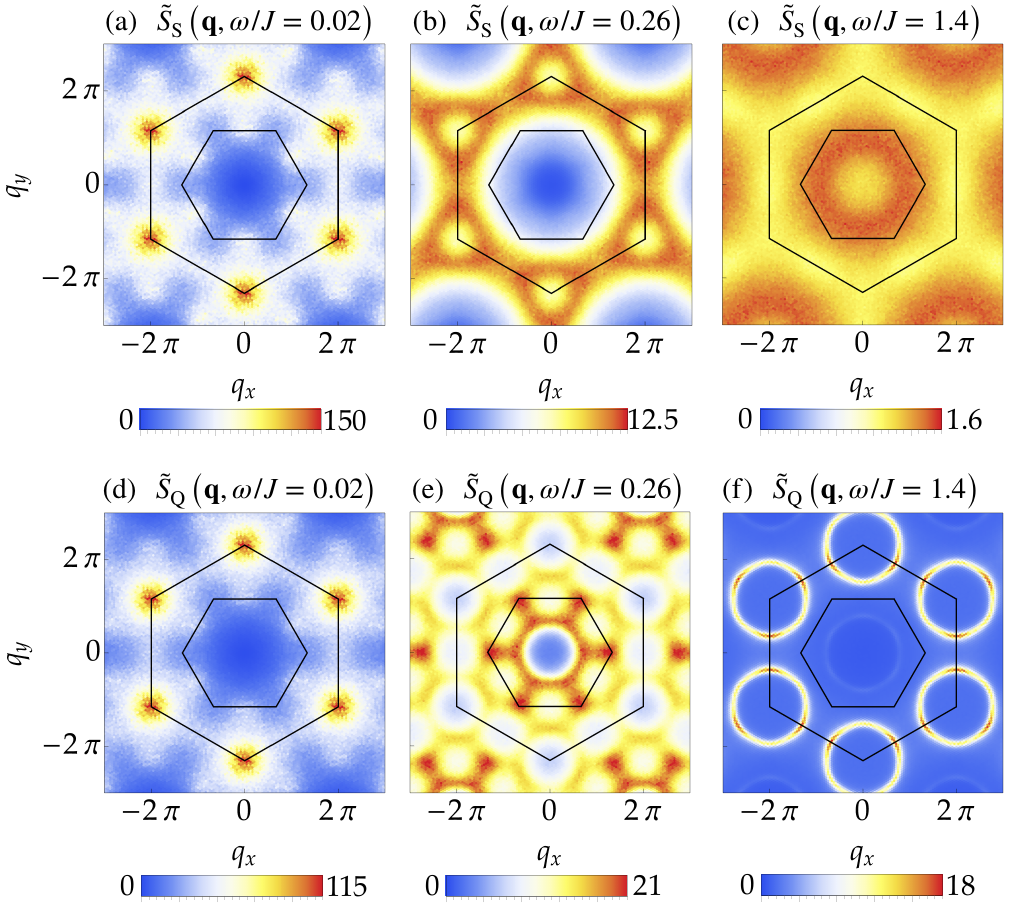}	
	\caption{
	Energy cross sections of the dynamical structure factors 
	[see Eqs.~\eqref{eq:Sqw}--\eqref{eq:Dyn.correct}] for the FM CSL phase in
	the $\mathbb{CP}^{2}$ model of $\mathcal{H}^{\mathcal{A}}_{ {\sf BBQ-K} }$ 
	[see Eq.~\eqref{eq:H.BBQ-K.A}], 
	as shown in Figs.~\ref{fig:Sqw}(b) and \ref{fig:Sqw}(e).
	Results are shown in 
	(a)--(c) for dipoles, $\tilde{S}_{\rm{S}}({\bf q}, \omega)$, while in
	(d)--(f) for quadrupoles, $\tilde{S}_{\rm{Q}}({\bf q}, \omega)$.
	}
	\label{fig:Sqw.lineCuts}
\end{figure}
%%%%%%%%%%%%%%%%%%%%%%%%%%%%%%%%%%%%%%%%%%%
%

In the following, we examine the dynamical signatures in the 
FM CSL, shown in Figs.~\ref{fig:Sqw}(b) and \ref{fig:Sqw}(e), at $T = 0.003$, in more detail 
by analyzing energy cross sections.
In Fig.~\ref{fig:Sqw.lineCuts} we show dynamical structure factors 
[see Eqs.~\eqref{eq:Sqw}--\eqref{eq:Dyn.correct}] for spin-dipole moments, 
$\tilde{S}_{\rm{S}}({\bf q}, \omega)$ (top panels) and spin-quadrupole moments, 
$\tilde{S}_{\rm{Q}}({\bf q}, \omega)$ (bottom panels) at particular
energy values.
%
% zero-energy intensity 
%
The cross section at low energy, $\omega / J = 0.02$, in 
Figs.~\ref{fig:Sqw.lineCuts}(a) and \ref{fig:Sqw.lineCuts}(d) shows a characteristic 
high-intensity, diffuse scattering signal with dominant weight around the K' points 
and some accumulation of weak spectral weight around the M points.
These signals are almost completely captured by the equal-time structure factors 
$S_{\rm S}({\bf q})$ and $S_{\rm Q}({\bf q})$, shown in Figs.~\ref{fig:U3.thermo.CSL}(d)
and \ref{fig:U3.thermo.CSL}(e) (middle column). 
%	

% intermediate energies
%
At small, but finite energies, the spectrum shows a diffuse signal extending up to 
$\omega / J \approx 0.5$.
An energy cross section within this energy range at $\omega / J = 0.26$ reveals 
complicated scattering features with finite intensity across the entire 
Brillouin zone.
We observe dominant weight around the edge of the extended Brillouin zone 
for dipoles in Fig.~\ref{fig:Sqw.lineCuts}(b), and predominant scattering 
within the first Brillouin zone for quadrupoles in Fig.~\ref{fig:Sqw.lineCuts}(e).

The energy cross section within the diffuse and almost 
featureless continuum for spin dipoles at $\omega / J = 1.4$ reveals an almost 
uniform distribution of intensity over the entire Brillouin zone in Fig.~\ref{fig:Sqw.lineCuts}(c).
Within this energy range, where dipoles exhibit a continuum, quadrupole 
excitations display well-defined energy bands, as visible in Fig.~\ref{fig:Sqw}(e).
Shown in Fig.~\ref{fig:Sqw.lineCuts}(f), these bands manifest as rings around the 
K' points in momentum space.

% our understanding is correct, since O(3) model shows same dynamics ! 
%
In Appendix~\ref{app:dynamics.CSL.O3}, we explicitly compare the dynamical 
properties of the CSL in the $\mathbb{CP}^{2}$ model
with the results obtained for the $\mathbb{CP}^{1}$ model.
Our findings reveal a qualitative agreement between the two models, providing 
additional support for our interpretation that the spin liquid physics of the CSL 
is primarily governed by spin dipoles.

%%%%%%%%%%%%%%%%%%%%%%%%%%%%%%%%%%%%%%%%%%
%
% 			SUMMARY and DISCUSSION
%
%%%%%%%%%%%%%%%%%%%%%%%%%%%%%%%%%%%%%%%%%%
\section{Summary and perspectives}	
\label{sec:Summary.Discussion}
%%%%%%%%%%%%%%%%%%%%%%%%%%%%%%%%%%%%%%%%%%
%
%

%%%%%%%%%%%%%%%%%%%%%%%%%%%%%%%%%%%%%%%%%%
%\subsection{Summary of results}	
%%%%%%%%%%%%%%%%%%%%%%%%%%%%%%%%%%%%%%%%%%
%
%summarize key points and results
%
In this paper, we explored the $S=1$ Kitaev model with bilinear-biquadratic (BBQ) interactions 
on the honeycomb lattice 
for SU(3) spin-coherent states in the spin space $\mathbb{CP}^{2}$, a model which 
hosts a diversity of exotic phases, including 
multiple-$q$ states with nonzero scalar spin chirality, 
quasi-one-dimensional coplanar phases, twisted conical phases, and noncoplanar ordered states.
Building upon the theory presented in our prior work, Ref.~\cite{Pohle2023}, our goal
was to provide a pedagogical explanation of the nature and unique properties of a 
finite-temperature chiral spin liquid (CSL), apparent in this model. 
To achieve this, we utilized variational energy minimization and classical Monte Carlo (MC)
with molecular dynamics simulations, using the recently developed U(3) formalism 
specifically designed for simulating $S=1$ magnets~\cite{Remund2022}.
%

% essence of main results: advantage of eight-color model and dynamics
%
By progressively restricting the spin degree of freedom from $\mathbb{CP}^{2}$ to 
$\mathbb{CP}^{1}$ and eventually to a discrete eight-color model, we unveiled the significant 
impact of the local Hilbert space dimension on the ground state and its excitation properties.
The eight-color model, designed to capture the most intriguing aspects of the CSL, offered 
an intuitive and analytical understanding of its physical properties. 
This model not only uncovered the origin of the extensive degeneracy in the CSL, but also 
clarified the reasons behind the nonzero scalar spin chirality, extreme short-ranged correlations,
and the presence of a Kitaev spin liquid (SL) feature of $\mathbb{Z}_{2}$-flux order.
We derived bond-dependent spin constraints that unequivocally determine the nature 
of this eight-color CSL and validated our findings through unbiased classical MC 
simulations.
Our results convincingly demonstrated that the dominant properties of the eight-color CSL 
persist even in the more complex $\mathbb{CP}^{1}$ and $\mathbb{CP}^{2}$ model cases.
Importantly, the enlarged spin degree of freedom does not alter the fundamental properties of 
the CSL but rather restricts its stability in the ground state. 
Consequently, the CSL survives as an entropy-driven, robust phase at finite temperatures 
in the original $\mathbb{CP}^{2}$ model.
This stability in the more complex $\mathbb{CP}^{2}$ model enabled us to simulate 
the dynamical structure factor, which revealed a nontrivial excitation spectrum characterized 
by a high-intensity zero-energy mode with a diffuse continuum of excitations in the 
low-frequency region. 
At higher frequencies, a broad continuum of excitations was observed, which we attributed
to the excitations of the discrete eight-color CSL.

%%%%%%%%%%%%%%%%%%%%%%%%%%%%%%%%%%%%%%%%%%
%\subsection{Future perspectives}	
%%%%%%%%%%%%%%%%%%%%%%%%%%%%%%%%%%%%%%%%%%
%
% analysis of CSL as classical topological liquid
% (i) nature of the phase transitions to CSL
%
CSLs represent a well-established branch of SLs, wherein the SL 
state persists even after the spontaneous breaking of time-reversal 
symmetry~\cite{Kalmeyer1987, Wen1989}.
While examples of CSLs with exactly known quantum-ground states 
exist~\cite{Schroeter2007, Yao2007, Fu2019, Peri2020}, studies across various models 
have indicated that the spontaneous breaking of discrete time-reversal symmetry typically 
occurs at finite temperatures~\cite{Momoi1997, Kato2010, Nasu2015, Kato2017, Mishchenko2020}.
The CSL investigated in our present study is a classical spin liquid that also spontaneously breaks 
time-reversal symmetry at finite temperature.
We confirmed that this transition is of second order, following 
the $\mathbb{Z}_{2}$ Ising universality class.
Remarkably, the correlations above and below this phase transition
are strikingly similar
[see Figs.~\ref{fig:8c.chiral.order.CSL.thermo}(j) and \ref{fig:8c.chiral.order.CSL.thermo}(l)],
suggesting a potential deeper connection to correlations seen in Kitaev SLs~\cite{Samarakoon2017}.
It would be intriguing to explore whether the nature of the phase transition in the
$\mathbb{CP}^{1}$ and $\mathbb{CP}^{2}$ models exhibits similarities to the phase transition observed 
in the eight-color model, and whether such behavior could also be found in other 
lattice models which host classical CSLs~\cite{Gomez2024}.
Furthermore, the defining bond-dependent spin constraints, as illustrated in Fig.~\ref{fig:8cCSL.bond.constraints}, 
can be expressed in terms of a suitable lattice gauge theory~\cite{YanPohle}.
This directly leads to the intriguing question of whether the CSL can also exist on other
three-coordinated lattices with different geometries and whether there is a potential relevance 
to compass models~\cite{Nussinov2015}.
We leave these intriguing questions for future work.
%

% importance of treating right local Hilbert space
%  (ii) extension to other interactions
%
We strongly advocate for simulating systems with higher-order interactions within their 
corresponding local Hilbert space dimension. 
While it may be theoretically straightforward to formulate and solve models with any type of spin 
interactions, it is crucial to ensure that the local Hilbert space respects the underlying 
Lie algebra, which physically allows for such interactions~\cite{Dahlbom2022b, Dahlbom2023}.
For instance, biquadratic interactions, $J_2$, are absent for spin moments described 
by Pauli matrices in the $\mathfrak{su}(2)$ Lie algebra.  
To address such interactions in a physically realistic setting, it is necessary to formulate the problem 
in SU(3), with generators in form of Gell-Mann matrices. 
To illustrate the profound impact of changing the local spin degree 
of freedom, our results in Fig.~\ref{fig:PD.all} provide a compelling demonstration using a specific example. 
Also other examples, such as the BBQ model on the triangular lattice, reveal very 
different properties for spins living in $\mathbb{CP}^{1}$ 
(classical Heisenberg spins)~\cite{Kawamura2007, Korshunov2012}, 
compared to spins in $\mathbb{CP}^{2}$~\cite{Lauchli2006, Smerald2013, Remund2022}.
In particular, realistic Mott insulators, in their effective low-energy description, can accommodate 
hybridized multipole moments, which can be efficiently studied within their corresponding SU($N$) 
representation~\cite{Iwazaki2023, Mellado2023}.

%possible future direction with methods treating long-range entanglement
% (iii) quantum effects
%
The U(3) formalism~\cite{Remund2022} has been used in the present
paper as a semiclassical approximation 
that neglects quantum entanglement between different sites.
A very promising and important direction for future studies would involve the treatment 
of the same model with methods which can account for long-range entanglement.
In such a case, we expect that quantum fluctuations will broaden the Kitaev SL phases, 
which, in our semiclassical simulations, are confined to the singular Kitaev points located 
at the north and south poles of the phase diagram in Fig.~\ref{fig:PD.all}.
In fact, this expectation has already been realized by recent iPEPS studies in
Ref.~[\onlinecite{Mashiko2024}], which examined the unfrustrated regime of 
the $S=1$  BBQ-Kitaev model for $1 < \phi/\pi < 2$ [cf. Fig.~\ref{fig:PD.all}(a)].
Moreover, recent work utilizing DMRG techniques on the $S=1$ Kitaev model has identified 
the emergence of chiral ordered phases in the presence of symmetric off-diagonal 
interactions~\cite{Luo2024} and vortex states in the presence of single-ion anisotropy~\cite{Singhania2023}. 
These findings provide intriguing results that underscore the richness of the 
quantum problem, offering promising avenues for further exploration.
Furthermore, given that the CSL phases in the $\mathbb{CP}^2$ model persist at finite temperatures 
in close proximity to the Kitaev points, it might be possible that quantum fluctuations 
will stabilize the CSL phases even in the ground state.
This raises the possibility of an intriguing connection to Schwinger boson mean-field theory, which 
proposes the existence of a gapped chiral QSL~\cite{Ralko2024}.
Additionally, it is known that the SO phase in the BBQ model will transform into a plaquette 
valence bond solid (pVBS), as observed from iPEPS simulations~\cite{Corboz2013}. 
It would be interesting to investiagte how far the pVBS extends in the presence of Kitaev 
interactions and whether the one-dimensional phase (1D/q1D) survives in this scenario or not. 
If such a phase persists in a quantum simulation, it could potentially stabilize exotic phases 
like the Haldane phase~\cite{Lauchli2006b}, quantum loop states~\cite{Savary2021}, or 
other phases with dimensionally reduced characteristics, as observed in 
anisotropic $S=1/2$ Kitaev models~\cite{Gohlke2022, Feng2023}.

% (iv) implications to experiments
%
The $S=1$ Kitaev model with BBQ interactions offers a diverse range of 
interesting and exotic phases.
In the Kitaev limit we recovered the semiclassical analog of the $S = 1$ Kitaev SL, 
as suggested for honeycomb materials with Ni$^{2+}$ ions~\cite{Stavropoulos2019}.
However, when considering the experimental realization of higher-$S$ Kitaev materials, it
becomes apparent that achieving ideal interactions poses significant challenges.
$S=1$ magnets naturally introduce higher-order biquadratic interactions, placing potential
candidate materials likely in the vicinity of the Kitaev limit, but with finite 
$J_1$ and $J_2$ interactions.
In such cases, the surrounding 
phases become relevant for characterizing and classifying 
synthesized materials. 
A ferroquadrupolar spin nematic state is expected for negative biquadratic interactions, $J_2$, 
which can be expected in materials with spin-phonon 
coupling~\cite{Tchernyshyov2011}.
States, such as triple-$q$, q1D, and CSL states appear for positive biquadratic interactions
$J_2$, which are expected in materials with orbital 
degeneracy~\cite{Yoshimori1981, Hoffmann2020, Soni2022}.
Phases with scalar spin chirality, such as the FM/AFM CSL or triple-$q$ ordered states found in this 
study, are of particular interest, as they may lead to unconventional phenomena, such as the
magnon Hall effect~\cite{Katsura2010, Onose2010, Matsumoto2011}.
In fact, triple-$q$ states were also found in the Kitaev-Heisenberg model under magnetic 
field~\cite{Janssen2016} and play a crucial role in understanding the physics of materials such as
the honeycomb cobaltites
Na$_2$Co$_2$TeO$_6$~\cite{Kruger2023, Francini2024}, and 
Na$_3$Co$_2$SbO$_6$~\cite{Gu2024}.
Despite the presence of non-Kitaev interactions, the CSL presented in our study is stable at finite 
temperatures in the vicinity of the Kitaev limit. 
This observation suggests the intriguing possibility of stabilizing a similar state in real materials, 
potentially leading to the emergence of a novel Kitaev-like SL that breaks time-reversal 
symmetry.

%%%%%%%%%%%%%%%%%%%%%%%%%%%%%%%%%%%%%
% 				Acknowledgements
%%%%%%%%%%%%%%%%%%%%%%%%%%%%%%%%%%%%%
\begin{acknowledgments}
%%%%%%%%%%%%%%%%%%%%%%%%%%%%%%%%%%%%%

The authors are pleased to acknowledge helpful conversations with 
Cristian Batista, Matthias Gohlke, Yasuyuki Kato, Geet Rakala, Kotaro Shimizu, 
and Han Yan.
This work was supported by the Quantum Liquid Crystals JSPS KAKENHI 
Grants \mbox{No. JP19H05822} and \mbox{No. JP19H05825}, 
MEXT as "Program for Promoting Research on the Supercomputer Fugaku"
(Grant \mbox{No. JPMXP1020230411}), and 
the Theory of Quantum Matter Unit, OIST.
Numerical calculations were carried out using HPC facilities provided by 
the Supercomputer Center of the Institute for Solid State Physics, 
the University of Tokyo.

%%%%%%%%%%%%%%%%%%%%%%%%%%%%%%%%%%%%%
\end{acknowledgments}
%%%%%%%%%%%%%%%%%%%%%%%%%%%%%%%%%%%%%

%%%%%%%%%%%%%%%%%%%%%%%%%%%%%%%%%%%%%%%%%
%
%					APPENDIX
%
%%%%%%%%%%%%%%%%%%%%%%%%%%%%%%%%%%%%%%%%%
\appendix									
%%%%%%%%%%%%%%%%%%%%%%%%%%%%%%%%%%%%%%%%%

%%%%%%%%%%%%%%%%%%%%%%%%%%%%%%%%%%%%%%%%%%
\section{Details of methods}	
\label{app:method.details}	
%%%%%%%%%%%%%%%%%%%%%%%%%%%%%%%%%%%%%%%%%%
%

In this Appendix, we describe the numerical methods used in this study. 
Simulations are performed for different spin degrees of freedom,
namely, SU(3) spin-coherent states on the complex-projective plane $\mathbb{CP}^{2}$,
SU(2) spin-coherent states on $\mathbb{CP}^{1}$, and discretized spins in the 
eight-color model.
Energies in each case are evaluated at the level of classical approximations 
on local bonds, and hence 
do not allow us to treat long-range quantum entanglement beyond a single site.

%%%%%%%%%%%%%%%%%%%%%%%%%%%%%%%%%%%%%%%%%%
\subsection{Variational energy minimization}	
\label{app:JAX}	
%%%%%%%%%%%%%%%%%%%%%%%%%%%%%%%%%%%%%%%%%%
%

To determine the ground-state phase diagrams, as shown in Fig.~\ref{fig:PD.all}, we 
perform large-scale variational energy minimization using the machine learning 
library JAX~\cite{Jax2018}.
In Fig.~\ref{fig:PD.all}(a) we minimize $\mathcal{H}^{\mathcal{A}}_{ {\sf BBQ-K} }$
in Eq.~\eqref{eq:H.BBQ-K.A} by optimizing $\theta_1$, $\theta_2$, $\phi_1$, $\phi_2$ 
and $\phi_3$ in Eq.~\eqref{eq:sampling.d} as independent variational parameters at each 
site on the lattice.
Similarly, in Fig.~\ref{fig:PD.all}(b) we minimize $\mathcal{H}^{\mathcal{S}}_{ {\sf BBQ-K} }$ in 
Eq.~\eqref{eq:H.BBQ-K.S} using local variational parameters $\theta_1$ and $\phi_1$ in 
Eq.~\eqref{eq:sampling.S}.
We employ the gradient processing and optimization 
library ``Optax'' with the optimizer ``Adam''~\cite{Optax2020}.
The initial conditions considered include relevant ordered and disordered states, 
as depicted in Fig.~\ref{fig:PD.all}(d), as well as random initial states.
The optimization process involves $5000$-$10000$ steps per model parameter.

%%%%%%%%%%%%%%%%%%%%%%%%%%%%%%%%%%%%%
\subsection{Monte Carlo simulation}
\label{app:u3MC}
%%%%%%%%%%%%%%%%%%%%%%%%%%%%%%%%%%%%%

To obtain thermodynamic properties we perform classical MC simulations 
where spins are locally updated at every site on the physical lattice of size $\ns$.
In the $\mathbb{CP}^{2}$ spin space, we sample $\mathcal{A}$ matrices by 
choosing parameters $0 \leq \theta_1, \theta_2 \leq 1$ 
and $0 \leq \phi_1, \phi_2 < 2\pi$ in Eq.~(\ref{eq:sampling.d}) at random,
while in the $\mathbb{CP}^{1}$ spin space we sample classical vector spins with
$-1 \leq \cos{\theta_1} \leq 1$ 
and $0 \leq \phi_1< 2\pi$ in Eq.~(\ref{eq:sampling.S}).
For a spin flip in the eight-color model we randomly select, with equal weight,
one out of all eight possible spin states, as defined in Table~\ref{tab:8.color.spins},
Following the standard single-spin flip Metropolis algorithm~\cite{Metropolis1953} we 
accept or reject a new spin state at site $i$, after evaluating the local 
bond energies.
In the actual calculations, a single MC step consists of $\ns$ local spin-flip attempts 
on randomly chosen sites.
Simulations are carried out in parallel for replicas at 
different temperatures, using the replica-exchange method, 
initiated by the parallel tempering algorithm~\cite{Swendsen1986, Hukushima1996, Earl2005} 
every 100 MC steps.
Results for thermodynamic quantities are averaged over $5 \times 10^5$ statistically 
independent samples, after initial $5 \times 10^5$ MC steps 
for slowly heating from the ground state, as obtained by the variational
energy minimization   
and further $5 \times 10^5$ MC steps for thermalization.
Error bars are estimated by averaging over 10 independent simulation runs.

MC simulations in the frustrated parameter region, where
the eight-color CSL appears, suffer from severe slowing down, due to the failure of 
single-spin flip updates.
As explained in Sec~\ref{sec:CSL.properties}, the CSL is expected to have an extensive number of 
degenerate states, which are connected via simultaneous flips
of a group of spins covering at least one hexagon (see Fig.~\ref{fig:chiralities}).
To efficiently sample over an extensive manifold of states, we combine single-spin flip updates 
with an additional cluster update of six spins.
For the eight-color model and the $\mathbb{CP}^{1}$ model
we adopt Eq.~\eqref{eq:hex.update.S}, while for the 
$\mathbb{CP}^{2}$ model,
where spins are represented by $3 \times 3$ Hermitian $\mathcal{A}$-matrices 
[see Eqs.~\eqref{eq:A-matrix_Director}--\eqref{eq:sampling.d}], we 
transform Eq.~\eqref{eq:hex.update.S} into the matrix form 
%
%%%%%%%%%%%%%%%%%%%%%%
\begin{align}
	\mathcal{A}_{p,j}^{\rm new} = (R^{\alpha}_{j})^{-1} \mathcal{A}_{p,j} \ R^{\alpha}_{j} \, . 
\label{eq:hex.update.A}
\end{align}
%%%%%%%%%%%%%%%%%%%%%%
%
By comparing energies before and after the cluster update, we accept or reject the 
new state following the Metropolis argument~\cite{Metropolis1953}.
Within one MC step we implement $N_{\rm h} = \ns/2$ cluster-update attempts 
for randomly chosen hexagons on the whole lattice.
The efficiency of this update within the CSL phase strongly depends on the degrees 
of freedom of spins.
As discussed for the acceptance ratio in Appendix~\ref{app:acc.SS.hex}, 
this cluster update becomes 
rejection free in the eight-color model, while showing 
a reduced efficiency for 
$\mathbb{CP}^{1}$ and $\mathbb{CP}^{2}$ model 
calculations. 
%

%%%%%%%%%%%%%%%%%%%%%%%%%%%%%%%%%%%%%
\subsection{Molecular dynamics simulation}
\label{app:u3MD}
%%%%%%%%%%%%%%%%%%%%%%%%%%%%%%%%%%%%%

Dynamics are present in models with continuous spin degree of freedom, 
here in our case for the $\mathbb{CP}^{2}$ and $\mathbb{CP}^{1}$ models.
The equations of motion for spins respecting the spin space $\mathbb{CP}^{2}$
[spin coherent states of SU(3)] are expressed within our formalism in terms of 
$\mathcal{A}$ matrices (see Sec.~\ref{sec:model}).
This allows us to rewrite the original BBQ-Kitaev Hamiltonian in Eq.~\eqref{eq:H.BBQ-K.S} 
into the bilinear form of Eq.~\eqref{eq:H.BBQ-K.A}, 
%
%%%%%%%%%%%%%%%%%%%%%%
\begin{equation}
	\mathcal{H}^{\mathcal{A}}_{ {\sf BBQ-K} } =  
	\mathcal{H}^{\mathcal{A}}_{\sf BBQ} + \mathcal{H}^{\mathcal{A}}_{\sf K[{\it x}]}
	+ \mathcal{H}^{\mathcal{A}}_{\sf K[{\it y}]} + \mathcal{H}^{\mathcal{A}}_{\sf K[{\it z}]} 
\end{equation}
%%%%%%%%%%%%%%%%%%%%%%
%	
with
%
%%%%%%%%%%%%%%%%%%%%%%
\begin{align}	
	\mathcal{H}^{\mathcal{A}}_{\sf BBQ}  &= 
	\sum_{\langle ij \rangle} \Big[ J_1 \mathcal{A}^{\alpha}_{i  \beta} \mathcal{A}^{\beta}_{j  \alpha} 
	+ (J_2 - J_1) \mathcal{A}^{\alpha}_{i \beta} \mathcal{A}^{\alpha}_{j  \beta}  
	+ J_2 \mathcal{A}^{\alpha}_{i  \alpha} \mathcal{A}^{\beta}_{j \beta} \Big]   \, , \\
	 \mathcal{H}^{\mathcal{A}}_{\sf K[{\it x}]} &=  - K_x  \sum_{ \langle ij \rangle_x}  
			\left( \mathcal{A}^{y}_{i z} - \mathcal{A}^{z}_{i y} \right) 
			 \left( \mathcal{A}^{y}_{j z} - \mathcal{A}^{z}_{j y} \right)	 \, ,	\\
	 \mathcal{H}^{\mathcal{A}}_{\sf K[{\it y}]} &=  - K_y  \sum_{ \langle ij \rangle_y}  
			\left( \mathcal{A}^{x}_{i z} - \mathcal{A}^{z}_{i x} \right) 
			 \left( \mathcal{A}^{x}_{j z} - \mathcal{A}^{z}_{j x} \right)	 \, ,	\\
	 \mathcal{H}^{\mathcal{A}}_{\sf K[{\it z}]}  &=  - K_z  \sum_{ \langle ij \rangle_z}  
			\left( \mathcal{A}^{x}_{i y} - \mathcal{A}^{y}_{i x} \right) 
			 \left( \mathcal{A}^{x}_{j y} - \mathcal{A}^{y}_{j x} \right)	 \, ,
\end{align}
%%%%%%%%%%%%%%%%%%%%%%
%
where $\mathcal{A}$ matrices respect the commutation relations~\cite{Remund2022}:
%
%%%%%%%%%%%%%%%%%%%%%%
\begin{equation}
	[\mathcal{A}^{\alpha}_{i \beta}, \mathcal{A}^{\gamma}_{i \eta}] = 
	\delta^{\gamma}_{\beta} \mathcal{A}^{\alpha}_{i \eta} - 
	\delta^{\alpha}_{\eta} \mathcal{A}^{\gamma}_{i \beta}		\, ,   \quad
	[\mathcal{A}^{\alpha}_{i \beta}, \mathcal{A}^{\gamma}_{j \eta}] = 0	\, .
\label{eq:A.com.rel}
\end{equation}
%%%%%%%%%%%%%%%%%%%%%%
%
By using Eq.~\eqref{eq:A.com.rel}, one can solve the Heisenberg equations of motion, 
written in terms of $\mathcal{A}$ matrices
%
%%%%%%%%%%%%%%%%%%%%%%
\begin{equation}
	\frac{d}{dt}  \mathcal{A}^{\alpha}_{i \beta} = 
	\frac{d}{dt}  \mathcal{A}^{\alpha}_{i \beta}\Big|_{\sf BBQ} + \frac{d}{dt}  \mathcal{A}^{\alpha}_{i \beta}\Big|_{\sf K[{\it x}]}  
	+ \frac{d}{dt}  \mathcal{A}^{\alpha}_{i \beta}\Big|_{\sf K[{\it y}]} + \frac{d}{dt}  \mathcal{A}^{\alpha}_{i \beta}\Big|_{\sf K[{\it z}]}  \, ,
\label{eq:EOM1}
\end{equation}
%%%%%%%%%%%%%%%%%%%%%%
%
which can be explicitly solved for contributions from the BBQ interactions
%
%%%%%%%%%%%%%%%%%%%%%%
\begin{equation}
	\begin{aligned}
		\frac{d}{dt}  \mathcal{A}^{\alpha}_{i \beta}\Big|_{\sf BBQ}
		= &-i \left[\mathcal{A}^{\alpha}_{i \beta}, \mathcal{H}^{\mathcal{A}}_{ {\sf BBQ}} \right] \\
		= &-i \sum_{\delta} \Big[ J_1 \left(\mathcal{A}^{\alpha}_{i \gamma}  \mathcal{A}^{\gamma}_{i+{\delta}, \beta}
						-    \mathcal{A}^{\gamma}_{i \beta}  \mathcal{A}^{\alpha}_{i+{\delta}, \gamma} \right)	\\
		&+ (J_2 - J_1) \left(\mathcal{A}^{\alpha}_{i \gamma}  \mathcal{A}^{\beta}_{i+{\delta}, \gamma}
						-    \mathcal{A}^{\gamma}_{i \beta}  \mathcal{A}^{\gamma}_{i+{\delta}, \alpha} \right) \Big] \, ,
	\label{eq:EOM2}	
	\end{aligned}
\end{equation}
%%%%%%%%%%%%%%%%%%%%%%
%
where $\sum_{\delta}$ sums over the nearest neighbors of site $i$.
For the bond-dependent Kitaev interactions, e.g., on the $x$ bond we obtain
%
%%%%%%%%%%%%%%%%%%%%%%
\begin{equation}
	\begin{aligned}
		\frac{d}{dt}  \mathcal{A}^{\alpha}_{i \beta}\Big|_{\sf K[{\it x}]} 
		=& - i \left[\mathcal{A}^{\alpha}_{i \beta}, \mathcal{H}^{\mathcal{A}}_{\sf K[{\it x}]} \right]   \\
		=& \ i K_x \left[\mathcal{A}^{\alpha}_{i \beta}, (\mathcal{A}^{y}_{i z} - \mathcal{A}^{z}_{i y} ) \right] 
		\left(\mathcal{A}^{y}_{i+x, z} - \mathcal{A}^{z}_{i+x, y} \right) \\
		=& \ i K_x \left( \delta^{y}_{\beta} \mathcal{A}^{\alpha}_{i z} 
		- \delta^{\alpha}_{z} \mathcal{A}^{y}_{i \beta} 
		- \delta^{z}_{\beta} \mathcal{A}^{\alpha}_{i y}  
		+ \delta^{\alpha}_{y} \mathcal{A}^{z}_{i \beta} \right)  \\  
		& \times \left(\mathcal{A}^{y}_{i+x, z} - \mathcal{A}^{z}_{i+x, y} \right)
		 \, ,
	\label{eq:EOM3}
	\end{aligned}
\end{equation}
%%%%%%%%%%%%%%%%%%%%%%
%
which explicitly gives 
%
%%%%%%%%%%%%%%%%%%%%%%
\begin{align}
\begin{rcases}
	\frac{d}{dt}  \mathcal{A}^{\alpha}_{i y}\Big|_{\sf K[{\it x}]}  &=  \mathcal{A}^{\alpha}_{i z} 	\\
	\frac{d}{dt}  \mathcal{A}^{\alpha}_{i z}\Big|_{\sf K[{\it x}]}  &= - \mathcal{A}^{\alpha}_{i y} 	\\
	\frac{d}{dt}  \mathcal{A}^{y}_{i \beta}\Big|_{\sf K[{\it x}]}  &= \mathcal{A}^{z}_{i \beta} 	\\
	\frac{d}{dt}  \mathcal{A}^{z}_{i \beta}\Big|_{\sf K[{\it x}]}  &= - \mathcal{A}^{y}_{i \beta} 	
\end{rcases}  \times    
	i K_x  \left(\mathcal{A}^{y}_{i+x, z} - \mathcal{A}^{z}_{i+x, y} \right) \, .
\end{align}
%%%%%%%%%%%%%%%%%%%%%%
%
Contributions on the $y$ and $z$ bonds can be obtained by cyclic permutations.

The equations of motion for spins respecting the spin space $\mathbb{CP}^{1}$
[spin coherent states of SU(2)] are formulated in terms of classical Heisenberg spins, see 
Eq.~\eqref{eq:sampling.S}.
We obtain from Eq.~\eqref{eq:H.BBQ-K.S} the equations of motion
%
%%%%%%%%%%%%%%%%%%%%%%
\begin{equation}
	\begin{aligned}
		\frac{d}{dt}  S^{\alpha}_i = &- i [S_i^{\alpha}, \mathcal{H}^\mathcal{S}_{\sf BBQ}]		\\
		= 	&- i J_1 \sum_{\langle ij \rangle} \left[S_i^{\alpha}, ({\bf S}_i \cdot {\bf S_j}) \right] 		\\
			&- 2 i J_2 \sum_{\langle ij \rangle} \left[S_i^{\alpha}, \left( {\bf S}_i \cdot {\bf S}_j \right)\right] 
			({\bf S}_i \cdot {\bf S}_j)	\\
			&- i \sum_{\langle ij \rangle_{\gamma}} K_{\gamma} \left[S_i^{\alpha}, S^{\gamma}_i S^{\gamma}_j \right] 
			\, ,
	\end{aligned}
\end{equation}
%%%%%%%%%%%%%%%%%%%%%%
%
which explicitly gives
%
%%%%%%%%%%%%%%%%%%%%%%
\begin{equation}
	\frac{d}{dt} {\bf S}_i = {\bf S}_i  \times \sum_{\delta} \left\{ \left[ J_1
	+ 2 J_2 ({\bf S}_i \cdot {\bf S}_{i+\delta}) \right] {\bf S}_{i+\delta} 
	+ K_{\delta} S^{\delta}_{i+\delta}   \right\}	\, ,
\label{eq:EOM.S}
\end{equation}
%%%%%%%%%%%%%%%%%%%%%%
%
where ``$\times$'' corresponds to the cross product between vector spins on the left
and the local exchange field on the right. 
Throughout this article we choose isotropic Kitaev interactions with 
$K_x = K_y = K_z = K$.

After sufficient thermalization at a specific temperature with MC simulations we evolve $1000$
statistically independent $\mathcal{A}$-matrix (${\bf} S$ vector) configurations in time by numerically 
integrating the equations of motion using a 
fourth-order Runge-Kutta (RK-4) algorithm~\cite{OrdinaryDiffEquations1, NumericalRecipes2007}.
By using a time interval of $\delta t \approx 0.03$, we obtain a time 
series of $\mathcal{A}$ matrices, $\{  \mathcal{A}^{\alpha}_{i \beta} (t) \}$, and
classical vectors, $\{ {\bf S}_i(t) \}$,
which conserve the total energy of the system with a controlled error of 
$\mathcal{O}(\delta t^5)$, as expected from the RK-4 method.
%

%%%%%%%%%%%%%%%%%%%%%%%%%%%%%%%%%%%%%%%%%%
\section{Measurement of observables}	
\label{app:observables.details}	
%%%%%%%%%%%%%%%%%%%%%%%%%%%%%%%%%%%%%%%%%%
%

%%%%%%%%%%%%%%%%%%%%%%%%%%%%%%%%%%%%%
\subsection{Specific heat and entropy}
%%%%%%%%%%%%%%%%%%%%%%%%%%%%%%%%%%%%%

During the MC measurement, we compute the specific heat
%
%%%%%%%%%%%%%%%%%%%%%%
\begin{equation}	
	C = \frac{1}{k_B T^2} \left[ \langle \mathcal{H}^2 \rangle  - \langle \mathcal{H} \rangle^2 \right]	\, ,
\label{eq:spec.heat}
\end{equation}
%%%%%%%%%%%%%%%%%%%%%%
%
where $k_B = 1$ throughout this paper, and $\langle \ldots  \rangle$ represents 
an average over measurements of statistically-independent replica.
In the discretized eight-color model we obtain the thermodynamic entropy, 
by integrating the specific heat over the temperature
%
%%%%%%%%%%%%%%%%%%%%%%
\begin{equation}
	S(T)= S(T_{\infty}) - \int_{T}^{T_{\infty}} \frac{C(T')}{T'} dT'  \, ,
\label{eq:entropy}
\end{equation}
%%%%%%%%%%%%%%%%%%%%%%
%
with $T_{\infty}$ the highest temperature available.

%%%%%%%%%%%%%%%%%%%%%%%%%%%%%%%%%%%%%
\subsection{Correlations in momentum space}
%%%%%%%%%%%%%%%%%%%%%%%%%%%%%%%%%%%%%

Correlations between magnetic moments in momentum space are accessible 
from the equal-time structure factor
%
%%%%%%%%%%%%%%%%%%%%%%
\begin{equation}
	S_{\rm{\lambda}}({\bf q}) = \left\langle \sum_{\alpha\beta}
		 	| {m_{\lambda}}^{\alpha}_{~\beta}({\bf q}) |^2 \ \right\rangle \,  ,
\label{eq:Sq.MC}
\end{equation}
%%%%%%%%%%%%%%%%%%%%%%
%
where the index $\lambda$ denotes the channel for $\mathcal{A}$ matrices, 
dipoles, or quadrupoles as $\lambda = \mathcal{A}$, ${\rm S}$, and ${\rm Q}$, respectively.
Here, ${m_{\lambda}}^{\alpha}_{~\beta}({\bf q})$ is defined by
% 
%%%%%%%%%%%%%%%%%%%%%%
\begin{align}
	{m_{\rm S}}^{\alpha}_{~\alpha}({\bf q})  
		&= -i \sum_{\beta,\gamma} \epsilon^{\alpha \ \gamma}_{\ \beta} {m_{\mathcal{A}}}^{\beta}_{~\gamma}({\bf q})  \, , 
	\label{eq:defn.m.S}	 \\
	{m_{\rm Q}}^{\alpha}_{~\beta}({\bf q})  
		&= -{m_{\mathcal{A}}}^{\alpha}_{~\beta}({\bf q}) - {m_{\mathcal{A}}}^{\beta}_{~\alpha}({\bf q})  
			+\frac{2}{3}  \delta^{\alpha}_{\beta}
			\sum_\gamma {m_{\mathcal{A}}}^{\gamma}_{~\gamma}({\bf q}) \; ,
	\label{eq:defn.m.Q}
\end{align}
%%%%%%%%%%%%%%%%%%%%%%
%
where the Fourier transform of the $\mathcal{A}$ matrices is given as
% 
%%%%%%%%%%%%%%%%%%%%%%
\begin{equation}
	{m_{\mathcal{A}}}^{\alpha}_{~\beta}({\bf q}) 
		= \frac{1}{\sqrt{\ns}} \sum_i^{\ns}  e^{i {\bf q}\cdot{\bf r}_i} \mathcal{\mathcal{A}}^{\alpha}_{i \beta}  \, .
\end{equation}
%%%%%%%%%%%%%%%%%%%%%%
%
For simulations in the $\mathbb{CP}^{1}$ model and the eight-color model, 
we evaluate dipole correlations directly 
% 
%%%%%%%%%%%%%%%%%%%%%%
\begin{equation}
	{m_{\rm S}}^{\alpha}_{~\alpha}({\bf q})  
		= \frac{1}{\sqrt{\ns}} \sum_i^{\ns}  e^{i {\bf q}\cdot{\bf r}_i} S_{i}^{\alpha}  \, ,
\label{eq:defn.m.S2}
\end{equation}
%%%%%%%%%%%%%%%%%%%%%%
%
from spin components sampled via Eq.~\eqref{eq:sampling.S}.

%%%%%%%%%%%%%%%%%%%%%%%%%%%%%%%%%%%%%
\subsection{Scalar spin chirality and semiclassical $\mathbb{Z}_{2}$-flux operator}
\label{app:chirality.flux}
%%%%%%%%%%%%%%%%%%%%%%%%%%%%%%%%%%%%%

%%%%%%%%%%%%%%%%%%%%%%%%%%%%%%%%%%%%%
%  Fig. - Definition chirality & Z2-flux operator
%%%%%%%%%%%%%%%%%%%%%%%%%%%%%%%%%%%%%
%
%%%%%%%%%%%%%%%%%%%%%%%%%%%%%%%%%%%%%
\begin{figure}[t]
	\centering
	\includegraphics[width=0.48\textwidth]{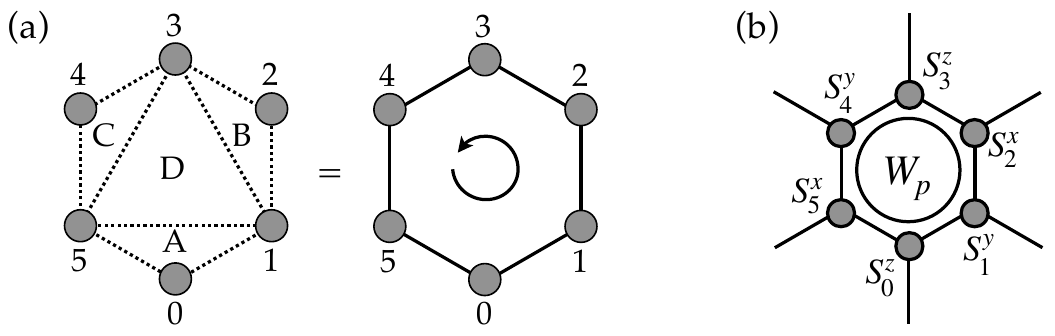}
	\caption{ 
	(a) Definition of the scalar spin chirality on an individual hexagon,
	see Eqs.~\eqref{eq:kappa2}--\eqref{eq:kappa3}.
	(b) The semiclassical analog of the $\mathbb{Z}_{2}$-flux
	operator, as defined in Eq.~\eqref{eq:Wp}.
	}
	\label{fig:def.chirality.Wp}
\end{figure}
%%%%%%%%%%%%%%%%%%%%%%%%%%%%%%%%%%%%%
%

We define the scalar spin chirality on the honeycomb lattice with
%
%%%%%%%%%%%%%%%%%%%%%%%%%%%%%%%%%%%%%
\begin{equation}
	\kappa  = \frac{2}{\ns }\sum_p \kappa_p 	\, ,
\label{eq:kappa1}
\end{equation}
%%%%%%%%%%%%%%%%%%%%%%%%%%%%%%%%%%%%%
%
where the sum is taken over all hexagons $p$.
As shown in Fig.~\ref{fig:def.chirality.Wp}(a) we divide a single hexagon into four triangles which 
contribute to the total scalar chirality of the full hexagon with
%
%%%%%%%%%%%%%%%%%%%%%%
\begin{align}
	\kappa_p = \kappa_p^A + \kappa_p^B + \kappa_p^C + \kappa_p^D  \, ,
\label{eq:kappa2}
\end{align}
%%%%%%%%%%%%%%%%%%%%%%
%
where
%
%%%%%%%%%%%%%%%%%%%%%%
\begin{align}
	\kappa_p^A &= {\bf S}_{p0} \cdot \left({\bf S}_{p1} \times {\bf S}_{p5} \right)	,	\\
	\kappa_p^B &= {\bf S}_{p1} \cdot \left({\bf S}_{p2} \times {\bf S}_{p3} \right) ,		\\
	\kappa_p^C &= {\bf S}_{p3} \cdot \left({\bf S}_{p4} \times {\bf S}_{p5} \right) ,		\\
	\kappa_p^D &= {\bf S}_{p1} \cdot \left({\bf S}_{p3} \times {\bf S}_{p5} \right) .
\label{eq:kappa3}
\end{align}
%%%%%%%%%%%%%%%%%%%%%%
%

We further define the $S=1$ semiclassical analog of the \mbox{``$\mathbb{Z}_{2}$-flux''} 
operator, $W_p$, as 
%
%%%%%%%%%%%%%%%%%%%%%%%%%%%%%%%%%%%%%
\begin{equation}
	W_p =  \prod_{j \in p} \sqrt{3} S_j^{\alpha}   \, ,
\label{eq:Wp}
\end{equation}
%%%%%%%%%%%%%%%%%%%%%%%%%%%%%%%%%%%%%
%
where $\alpha$ denotes the site-dependent label for the Kitaev bond pointing 
outwards from the hexagon $p$ at site $j = 0, \cdots, 5$.
Figure~\ref{fig:def.chirality.Wp}(b) visually represents Eq.~\eqref{eq:Wp}, which, 
in MC simulations, is computed as the spatial average over all plaquettes on the lattice.
We emphasize that $W_p$ differs from the 
conserved flux operator in the quantum \mbox{$S=1$} Kitaev 
model~\cite{Kitaev2006, Baskaran2008}.

%%%%%%%%%%%%%%%%%%%%%%%%%%%%%%%%%%%%%
\subsection{Dynamical structure factor}
%%%%%%%%%%%%%%%%%%%%%%%%%%%%%%%%%%%%%

To obtain the dynamical structure factors for dipole and quadrupole channels, 
we take the Fourier transform of the time-series of $\mathcal{A}$ matrices 
$\{  \mathcal{A}^{\alpha}_{i \beta} (t) \}$ (classical vectors $\{ {\bf S}_i (t) \})$, as 
obtained from from molecular dynamics simulations (see Appendix~\ref{app:u3MD}), 
from real-space into the momentum-space and frequency domain
%
%%%%%%%%%%%%%%%%%%%%%%
\begin{equation}
	S_{\rm{\lambda}}({\bf q},\omega) = \left\langle \sum_{\alpha\beta}
		 	| {m_{\lambda}}^{\alpha}_{~\beta}({\bf q}, \omega) |^2 \ \right\rangle \,  ,
\label{eq:Sqw}
\end{equation}
%%%%%%%%%%%%%%%%%%%%%%
%
with $\lambda = \mathcal{A}$, ${\rm S}$, and ${\rm Q}$.
The Fourier transform 
% 
%%%%%%%%%%%%%%%%%%%%%%
\begin{equation}
	{m_{\mathcal{A}}}^{\alpha}_{~\beta}({\bf q}, \omega) 
		= \frac{1}{\sqrt{N_t}} \sum_{n}^{N_t}  e^{i \omega t_n} {m_{\mathcal{A}}}^{\alpha}_{~ \beta}({\bf q}, t_n)  \, ,
\end{equation}
%%%%%%%%%%%%%%%%%%%%%%
%
for $\mathcal{A}$ matrices is obtained by 
the fast Fourier transform (FFT)~\cite{FFTW05},
after convoluting with a Gaussian envelope to avoid numerical artifacts like the 
Gibbs phenomenon~	\cite{MathematicalPhyics}.
Equivalent equations for dipole moments, ${m_{\rm S}}^{\alpha}_{~\alpha}({\bf q}, \omega)$,
and quadrupole moments, ${m_{\rm Q}}^{\alpha}_{~\beta}({\bf q}, \omega)$,
are defined in analogy to Eqs.~\eqref{eq:defn.m.S} and \eqref{eq:defn.m.Q},
respectively.

In Figs.~\ref{fig:Sqw}, \ref{fig:Sqw.lineCuts}, and 
Appendix~\ref{app:dynamics.CSL.O3},
we set the total number of time steps \mbox{$N_t = 500$} for a maximal frequency of 
\mbox{$\omega_{\rm max} = 10 J$}, 
while in Appendix~\ref{app:dynamics.Kitaev}, we set \mbox{$\omega_{\rm max} = 6 J$}.
Data are sampled over $1000$ statistically independent time series.

The Heisenberg equations of motion [see Eqs.~\eqref{eq:EOM1}--\eqref{eq:EOM3} 
and Eq.~\eqref{eq:EOM.S}] are derived using bosonic commutation relations 
[c.f. Eq.~\eqref{eq:A.com.rel}], resulting in a dispersion relation analogous to 
those found in flavor-wave theory~\cite{Remund2022}. 
However, the molecular dynamics simulations described above are performed for a thermal ensemble 
of classical states, evolving in time while adhering to the bosonic commutation relations. 
This effectively leads to dynamics of a ``mixed ensemble'', encompassing classical 
and bosonic statistics.

To obtain a dynamical structure factor that adheres strictly to quantum 
statistics, it is necessary 
to introduce a prefactor $\omega/T$, which allows to ``correct'' the intensity distribution 
via
%
%%%%%%%%%%%%%%%%%%%%%%
\begin{equation}
	{\tilde S}_{\rm{\lambda}}({\bf q},\omega) = \frac{\omega}{T} S_{\rm{\lambda}}({\bf q},\omega)	\, ,
\label{eq:Dyn.correct}
\end{equation}
%%%%%%%%%%%%%%%%%%%%%%
%
as discussed in great detail in Sec. VII of Ref.~[\onlinecite{Remund2022}], and 
the Supplemental Material of Ref.~[\onlinecite{Zhang2019}].
Throughout the present paper, all dynamical structure factors are presented in 
their ``corrected'' form, ${\tilde S}_{\rm{\lambda}}({\bf q},\omega)$.

%%%%%%%%%%%%%%%%%%%%%%%%%%%%%%%%%%%%%%%%%%
\section{Analytic solutions of phase boundaries for CSL}	
\label{app:PhaseBoundaries}	
%%%%%%%%%%%%%%%%%%%%%%%%%%%%%%%%%%%%%%%%%%
%

In this Appendix we provide an analytical derivation for the phase boundaries of the 
FM CSL in the ground state of the eight-color model, and in the $\mathbb{CP}^{1}$ model,
as discussed in Sec.~\ref{sec:phase_diagrams}. 
%

%%%%%%%%%%%%%%%%%%%%%%%%%%%%%%%%%%%%%%%%%%
\subsection{CSL phase at singular line in the $\mathbb{CP}^{1}$ model}	
%%%%%%%%%%%%%%%%%%%%%%%%%%%%%%%%%%%%%%%%%%
%

In the following we derive the singular line in the phase diagram where the FM CSL becomes 
the ground state in the $\mathbb{CP}^{1}$ model,
$\mathcal{H}^\mathcal{S}_{ {\sf BBQ-K} }$ in Eq.~\eqref{eq:H.BBQ-K.S},
see Fig.~\ref{fig:PD.all}(b).
Without loss of generality, we consider one possible arrangement of spins in the FM CSL 
ground state with nearest-neighbors
(${\bf S}_{\rm g}$, ${\bf S}_{\rm r}$) on the $x$ bond, 
(${\bf S}_{\rm g}$, ${\bf S}_{\rm b}$) on the $y$ bond, and
(${\bf S}_{\rm g}$, ${\bf S}_{\rm r}$) on the $z$ bond, which
all respect the bond-dependent spin constraints of 
Fig.~\ref{fig:8cCSL.bond.constraints}.
We hereby follow the convention of spin states as defined in Table~\ref{tab:8.color.spins} for 
the eight-color states. 

We first consider the energy contribution from bilinear interactions, $J_1$, and biquadratic 
interactions, $J_2$.
These are bond independent, which is why we consider energies only on the $x$ bond for 
simplicity. 
By parametrizing spins on the Bloch sphere [see Eq.~\eqref{eq:sampling.S}], with, 
e.g., $\theta_{\rm g}$ and $\phi_{\rm g}$, the polar and azimuthal angles of a ``green'' spin,
we write:
%
%%%%%%%%%%%%%%%%%%%%%%
\begin{equation}
	\begin{aligned}
		{\bf S}_{\rm g} \cdot {\bf S}_{\rm r} &= 
		\sin{\theta_{\rm g}} \sin{\theta_{\rm r}} \cos{(\phi_{\rm g} - \phi_{\rm r})} 
		+ \cos{\theta_{\rm g}} \cos{\theta_{\rm r}} 
		= -\frac{1}{3}	\, ,
	\end{aligned}
\end{equation}
%%%%%%%%%%%%%%%%%%%%%%
%
and 
%
%%%%%%%%%%%%%%%%%%%%%%
\begin{equation}
	\begin{aligned}
		\left( {\bf S}_{\rm g} \cdot {\bf S}_{\rm r} \right)^2 
		=& \ [\sin{\theta_{\rm g}} \sin{\theta_{\rm r}} \cos{(\phi_{\rm g} - \phi_{\rm r})}]^2   
		+  [\cos{\theta_{\rm g}} \cos{\theta_{\rm r}}]^2  \\
		&+ \sin{\theta_{\rm g}} \sin{\theta_{\rm r}} \cos{\theta_{\rm g}} \cos{\theta_{\rm r}} 
		\cos{(\phi_{\rm g} - \phi_{\rm r})} \\
		=& \ \frac{1}{9}	\, .
	\end{aligned}
\end{equation}
%%%%%%%%%%%%%%%%%%%%%%
%
Considering the spin $\tilde{\bf S}_{\rm g}$, after allowing for a small perturbation 
in $\theta_{\rm g}$ by \mbox{$\delta \ll 1$}, 
gives 
%
%%%%%%%%%%%%%%%%%%%%%%
\begin{equation}
	\begin{aligned}
		\tilde{{\bf S}}_{\rm g} \cdot {\bf S}_{\rm r} 
		&= {\bf S}_{\rm g} \cdot {\bf S}_{\rm r} - 
		\delta \sin{\theta_{\rm g} \cos{\theta_{\rm r}}}	\\
		&= {\bf S}_{\rm g} \cdot {\bf S}_{\rm r} 
		- \frac{\sqrt{2}}{3}\delta	\, .
	\end{aligned}
\label{eq:CSL.GS.O3.J1}
\end{equation}
%%%%%%%%%%%%%%%%%%%%%%
%
Equivalently we obtain:
%
%%%%%%%%%%%%%%%%%%%%%%
\begin{equation}
	\begin{aligned}
		\left( \tilde{{\bf S}}_{\rm g} \cdot {\bf S}_{\rm r} \right)^2 
		&=  	\left( {\bf S}_{\rm g} \cdot {\bf S}_{\rm r} -
			\delta \sin{\theta_{\rm g}} \cos{\theta_{\rm r}} \right)^2 		\\
		&= 	\left( {\bf S}_{\rm g} \cdot {\bf S}_{\rm r} \right)^2  
			-2 \left({\bf S}_{\rm g} \cdot {\bf S}_{\rm r} \right)
			\delta \sin{\theta_{\rm g}} \cos{\theta_{\rm r}} + \mathcal{O}(\delta^2)	\\
		&= 	\left( {\bf S}_{\rm g} \cdot {\bf S}_{\rm r} \right)^2  
			+ \frac{2}{3} \frac{\sqrt{2}}{3}\delta	\, ,
	\end{aligned}
\label{eq:CSL.GS.O3.J2}
\end{equation}
%%%%%%%%%%%%%%%%%%%%%%
%
where we neglect terms of higher order $\mathcal{O}(\delta^2)$.

The energies from bond-dependent Kitaev interactions contribute  
on the $x$ bond:
%
%%%%%%%%%%%%%%%%%%%%%%
\begin{equation}
	\begin{aligned}
		\tilde{{\bf S}}_{\rm g} \cdot {\bf S}_{\rm r} 
		&= {\bf S}_{\rm g} \cdot {\bf S}_{\rm r} - 
		\delta \cos{\theta_{\rm g} \cos{\phi_{\rm g}} \sin{\theta_{\rm r}} \cos{\phi_{\rm r}}}	\\
		&= {\bf S}_{\rm g} \cdot {\bf S}_{\rm r} + \frac{\delta}{2} \frac{\sqrt{2}}{3}\, ,
	\end{aligned}
\end{equation}
%%%%%%%%%%%%%%%%%%%%%%
%
on the $y$ bond:
%
%%%%%%%%%%%%%%%%%%%%%%
\begin{equation}
	\begin{aligned}
		\tilde{{\bf S}}_{\rm g} \cdot {\bf S}_{\rm b} 
		&= {\bf S}_{\rm g} \cdot {\bf S}_{\rm b} + 
		\delta \cos{\theta_{\rm g} \sin{\phi_{\rm g}} \sin{\theta_{\rm b}} \sin{\phi_{\rm b}}} \\
		&= {\bf S}_{\rm g} \cdot {\bf S}_{\rm b} +  \frac{\delta}{2} \frac{\sqrt{2}}{3}\, ,
	\end{aligned}
\end{equation}
%%%%%%%%%%%%%%%%%%%%%%
%
and on the $z$ bond:
%
%%%%%%%%%%%%%%%%%%%%%%
\begin{equation}
	\begin{aligned}
		\tilde{{\bf S}}_{\rm g} \cdot {\bf S}_{\rm r} 
		& = {\bf S}_{\rm g} \cdot {\bf S}_{\rm r} 
		- \delta \sin{\theta_{\rm g} \cos{\theta_{\rm r}}} \\
		&= {\bf S}_{\rm g} \cdot {\bf S}_{\rm r}  - \delta\frac{\sqrt{2}}{3} \, .
	\end{aligned}
\label{eq:CSL.GS.O3.K}
\end{equation}
%%%%%%%%%%%%%%%%%%%%%%
%

The energy difference, $\Delta E$, between the CSL ground state and 
a state with small perturbation, using Eqs.~\eqref{eq:CSL.GS.O3.J1}--\eqref{eq:CSL.GS.O3.K},
is
%
%%%%%%%%%%%%%%%%%%%%%%
\begin{equation}
	\begin{aligned}
		\Delta E = \frac{\sqrt{2}}{3}  \delta \left(-J_1+\frac23 J_2 \right) 
		+ K  \delta \left( \frac{1}{2} \frac{\sqrt{2}}{3} +  \frac{1}{2} \frac{\sqrt{2}}{3} 
		- \frac{\sqrt{2}}{3} \right)	\, .
	\end{aligned}
\label{eq:CSL.GS.O3.DE}
\end{equation}
%%%%%%%%%%%%%%%%%%%%%%
%
Thus, the CSL ground state is realized when $\Delta E = 0$, which occurs for
%
%%%%%%%%%%%%%%%%%%%%%%
\begin{equation}
	J_1 = \frac{2}{3} J_2		\, .
\end{equation}
%%%%%%%%%%%%%%%%%%%%%%
%
This condition corresponds to the model parameters, defined in 
Eq.~\eqref{eq:parametrization.HBBQ-K} 
%
%%%%%%%%%%%%%%%%%%%%%%
\begin{equation}
	\phi/\pi   = \frac{1}{\pi} \arctan{\left( \frac{3}{2} \right)}	\approx 0.312833 \, ,
\end{equation}
%%%%%%%%%%%%%%%%%%%%%%
%
as found from numerics in Fig.~\ref{fig:PD.all}(b), and stated in Eq.~\eqref{eq:O3.CSL.AFM}.
Since the contributions from Kitaev energies perfectly cancel out, 
the CSL phase forms a singular straight line in the phase diagram. 
A similar energy comparison can be applied to the AFM CSL phase, leading to
Eq.~\eqref{eq:O3.CSL.FM}.

%%%%%%%%%%%%%%%%%%%%%%%%%%%%%%%%%%%%%%%%%%
\subsection{Phase boundaries of CSL in the eight-color model}	
%%%%%%%%%%%%%%%%%%%%%%%%%%%%%%%%%%%%%%%%%%
%

The phase boundaries of the FM CSL phase in the eight-color model,
$\mathcal{H}^{ {\sf 8c} }_{{\sf BBQ-K}}$  in Eq.~\eqref{eq:H.BBQ-K.8c}, 
can also be calculated analytically. 
The FM CSL is surrounded by the AFM and AFM 3$Q$ chiral ordered phases
[see Fig.~\ref{fig:PD.all}(c)].
Without loss of generality, we consider the local-bond energy for only one 
bond in the lattice, here the $x$ bond.

In the FM CSL, by respecting the bond-dependent spin constraints in 
Fig.~\ref{fig:8cCSL.bond.constraints}, one obtains for 
allowed nearest-neighbors, e.g., ``bright green'' and ``bright red'', the local bond energy
\mbox{$\boldsymbol \sigma_{\rm g} \cdot \boldsymbol \sigma_{\rm r}= -1/3$}.
On the other side, the AFM phase gives for 
states in the ground state, 
e.g., ``bright green'' and ``dark green'', 
\mbox{$\boldsymbol \sigma_{\rm g} \cdot \bar{\boldsymbol \sigma}_{\rm g} = -1$}.
To obtain the phase boundary between AFM and FM CSL in terms of model parameters 
$\phi$ and $\theta$ [see Eq.~\eqref{eq:parametrization.HBBQ-K}], 
we set their energies equal and obtain the condition
%
%%%%%%%%%%%%%%%%%%%%%%
\begin{equation}
	\begin{aligned}
		- \cos{\phi} + \sin{\phi}  &=	 - \frac{1}{3} \cos{\phi}  + \frac{1}{9} \sin{\phi}		\, ,
	\end{aligned}
\label{eq:boundary.8c.A}
\end{equation}
%%%%%%%%%%%%%%%%%%%%%%
%
which is independent of $\theta$ and therefore results in a straight line as seen 
in the phase diagram of Fig.~\ref{fig:PD.all}(c).
Solving Eq.~\eqref{eq:boundary.8c.A} results in
%
%%%%%%%%%%%%%%%%%%%%%%
\begin{equation}
	\frac{\phi}{\pi} = \frac{1}{\pi}\arctan{\left( \frac{3}{4} \right)} ~ \approx 0.205	\, .
\label{eq:boundary.8c.B}
\end{equation}
%%%%%%%%%%%%%%%%%%%%%%

In the AFM 3$Q$ chiral ordered phase, the bond energy gives, e.g., for 
``bright green'' and ``dark blue'', 
\mbox{$\boldsymbol \sigma_{\rm g} \cdot \bar{\boldsymbol \sigma}_{\rm b} = 1/3$}.
We do the same comparison for the phase boundary between the FM CSL 
and the AFM 3$Q$ chiral phases and obtain the condition
%
%%%%%%%%%%%%%%%%%%%%%%
\begin{equation}
	\begin{aligned}
		\cos{\phi} \sin{\theta}  &=	 - \cos{\phi} \sin{\theta}		\, ,
	\end{aligned}
\end{equation}
%%%%%%%%%%%%%%%%%%%%%%
%
which is only fulfilled for
%
%%%%%%%%%%%%%%%%%%%%%%
\begin{equation}
	\cos{\phi} \sin{\theta} = 0	\, .
\label{eq:boundary.8c.C}
\end{equation}
%%%%%%%%%%%%%%%%%%%%%%
%
Equation~\eqref{eq:boundary.8c.C} is satisfied for $\theta = 0$ or $\pi$, which  
corresponds to the pure Kitaev model, 
and for 
%
%%%%%%%%%%%%%%%%%%%%%%
\begin{equation}
	\frac{\phi}{\pi} = \frac{1}{2} 	\, , 
\label{eq:boundary.8c.D}
\end{equation}
%%%%%%%%%%%%%%%%%%%%%%
%
where the second solution $\phi/\pi = 3/2$ is physically irrelevant. 

The phase boundaries described by Eqs.~\eqref{eq:boundary.8c.B} 
and \eqref{eq:boundary.8c.D}, are well visible in Fig.~\ref{fig:PD.all}(c).
The calculations for phase boundaries in the AFM CSL are straightforward,
and give the symmetric solutions from the self-duality of 
$\mathcal{H}^{ {\sf 8c} }_{{\sf BBQ-K}}$ , as written in Eq.~\eqref{eq:AFM.chiral.PD}.

%%%%%%%%%%%%%%%%%%%%%%%%%%%%%%%%%%%%%%%%%%
\section{Acceptance ratio for single-spin flip and hexagon cluster updates}	
\label{app:acc.SS.hex}	
%%%%%%%%%%%%%%%%%%%%%%%%%%%%%%%%%%%%%%%%%%
%

Single-spin flip MC simulations within the CSL phase suffer from severe slowing down, 
making it impossible to decorrelate the MC samples 
using local spin updates. 
To overcome this issue, we combine single-spin flip updates 
with a hexagon cluster update 
as physically motivated from the nature of the eight-color CSL ground state manifold
(see Fig.~\ref{fig:chiralities}). 
Technical details of the implementation of the cluster update are given in 
Sec.~\ref{sec:CSL.properties} and Appendix~\ref{app:u3MC}

%
%%%%%%%%%%%%%%%%%%%%%%%%%%%%%%%%%%%%%
%. Fig. -- Acceptance ratios single spin / hexagon update
%%%%%%%%%%%%%%%%%%%%%%%%%%%%%%%%%%%%%
\begin{figure}[t]
	\centering
	\includegraphics[width=0.36\textwidth]{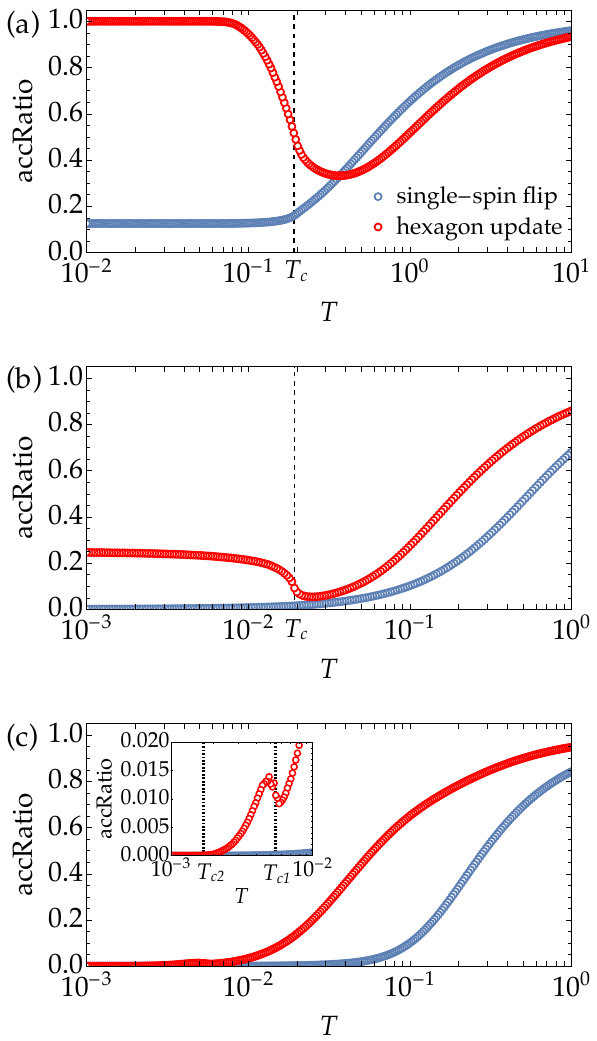}
	\caption{	
	Temperature dependence of the acceptance ratio in the FM CSL as 
	obtained in 
	(a) the eight-color model, 
	(b) the $\mathbb{CP}^{1}$ model,
	and 
	(c) the $\mathbb{CP}^{2}$ model.
	Acceptance ratios for single-spin flip updates are depicted with blue circles, 
	while results for hexagon updates are shown with red circles.
	Model parameters are the same as used in 
	Figs.~\ref{fig:8c.chiral.order.CSL.thermo} (right column),
	\ref{fig:O3.thermo.CSL}, and \ref{fig:U3.thermo.CSL}.
	Simulations were performed for finite-size clusters of $L=48$ 
	($\ns=4608$). 
	}
	\label{fig:acc.ratios}
\end{figure}
%%%%%%%%%%%%%%%%%%%%%%%%%%%%%%%%%%%%%%%
%

In Figs.~\ref{fig:acc.ratios}(a)--\ref{fig:acc.ratios}(c) we show the temperature dependence
of the acceptance ratio for single-spin flip and hexagon cluster updates in the 
eight-color model, the $\mathbb{CP}^{1}$ model, and the $\mathbb{CP}^{2}$ model,  
respectively. 
Model parameters are 
chosen to stabilize the FM CSL at low temperatures, 
as depicted in Figs.~\ref{fig:8c.chiral.order.CSL.thermo} (right column),
\ref{fig:O3.thermo.CSL}, and \ref{fig:U3.thermo.CSL}.
As expected from conventional MC simulations, the efficiency of the single-spin 
flip update monotonically decreases by reducing the temperature, showing a 
basically vanishing acceptance ratio below the transition into the CSL. 
We note that the value of $1/8$ in Fig.~\ref{fig:acc.ratios}(a) comes from the fact that we 
accept a new state even when we sample over the same old state.

The hexagon update follows the same trend in the high-temperature regime.
However, below the phase transition $T_c$ in Figs.~\ref{fig:acc.ratios}(a) and \ref{fig:acc.ratios}(b)
and $T_{c1}$ in Fig.~\ref{fig:acc.ratios}(c), the acceptance ratio for the hexagon
update increases and shows finite values. 
In fact, the hexagon update becomes rejection free in the eight-color model, while 
reaching up to $\approx 25\%$ in the $\mathbb{CP}^{1}$ model.
The efficiency of the cluster update in the $\mathbb{CP}^{2}$ model
is strongly suppressed 
($\lesssim 1.5 \%$), as shown in the inset of Fig.~\ref{fig:acc.ratios}(c),
but is still sufficient to decorrelate the finite-temperature 
FM CSL phase.

%
%%%%%%%%%%%%%%%%%%%%%%%%%%%%%%%%%%%%%
%. Fig. -- Thermodynamics: Kitaev model
%%%%%%%%%%%%%%%%%%%%%%%%%%%%%%%%%%%%%
\begin{figure}[t]
	\centering
	\includegraphics[width=0.45\textwidth]{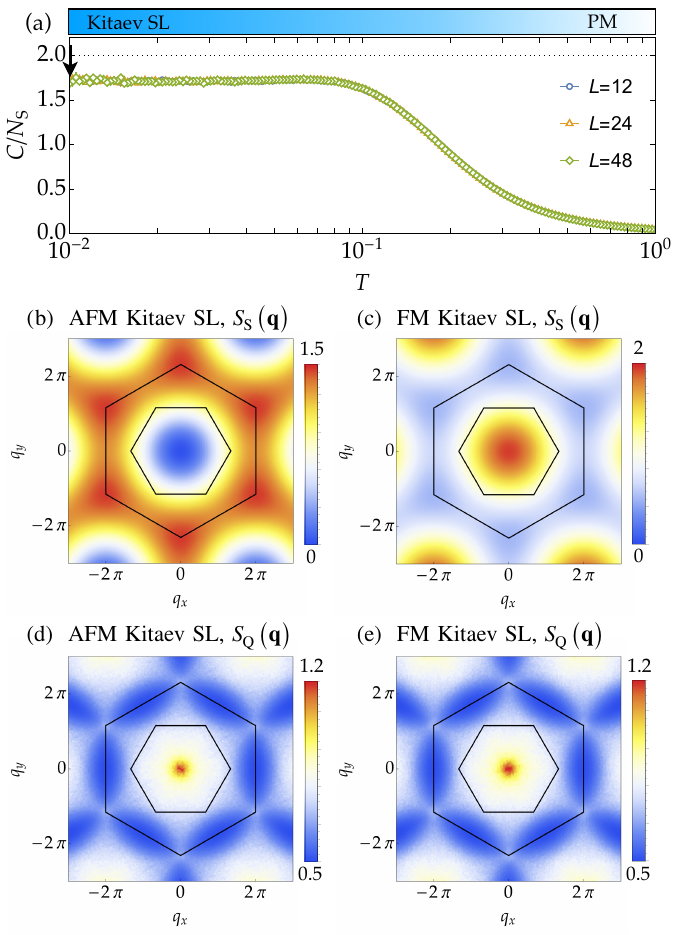}
	\caption{	
	The semiclassical $S=1$ Kitaev SL 
	in the $\mathbb{CP}^{2}$ model, respectively at $\theta/\pi = 0$ and $1$
	of $\mathcal{H}^{\mathcal{A}}_{ {\sf BBQ-K} }$ in Eq.~\eqref{eq:H.BBQ-K.A}.
	The specific heat $C/\ns$ [Eq.~\eqref{eq:spec.heat}] in (a) shows a crossover
	from a high-temperature paramagnet (PM) into the semiclassical analog of the Kitaev 
	spin liquid (SL).  
	The spin structure factors for dipoles, $S_{\rm{S}}({\bf q})$, and quadrupoles, 
	$S_{\rm{Q}}({\bf q})$, [see Eqs.~\eqref{eq:Sq.MC}--\eqref{eq:defn.m.Q}] are 
	shown respectively for the AFM Kitaev SL in (b) and (d), 
	and for the FM Kitaev SL in (c) and (e).
	Simulations were performed at $T = 0.01$ [black arrow in (a)] for finite-size 
	clusters of $L=48$ ($N=4608$).
	}
	\label{fig:KitaevSL.thermo}
\end{figure}
%%%%%%%%%%%%%%%%%%%%%%%%%%%%%%%%%%%%%%%
%

%%%%%%%%%%%%%%%%%%%%%%%%%%%%%%%%%%%%%%%%%%%
%. Fig. --  Dynamics: Kitaev model
%%%%%%%%%%%%%%%%%%%%%%%%%%%%%%%%%%%%%%%%%%%
\begin{figure*}[tbp]
	\centering
	\includegraphics[width=0.98\textwidth]{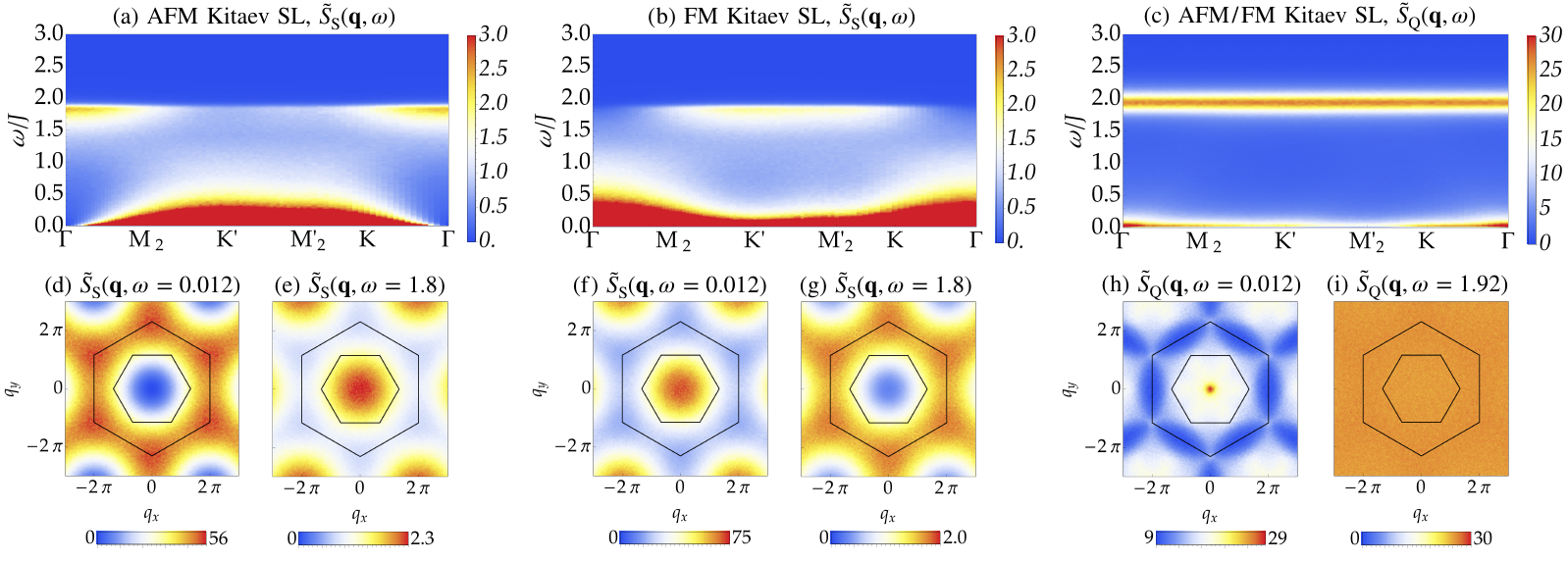}	
	\caption{
	Dynamical structure factors [see Eqs.~\eqref{eq:Sqw}--\eqref{eq:Dyn.correct}] 
	for the semiclassical analog of the $S=1$ AFM and FM 
	Kitaev SL in the $\mathbb{CP}^{2}$ model, respectively at $\theta/\pi = 0$ and $1$ of 
	$\mathcal{H}^{\mathcal{A}}_{ {\sf BBQ-K} }$ in Eq.~\eqref{eq:H.BBQ-K.A}.
	(a) and (b) show dynamical structure factors of dipoles, ${\tilde S}_{\rm{S}}({\bf q}, \omega)$, 
	respectively in the AFM and FM Kitaev SL states, with their energy cross sections in (d)--(g).
	Excitations are very similar to known results for the $S=1/2$ Kitaev model, see 
	Ref.~[\onlinecite{Samarakoon2017}].
	(c) shows dynamical structure factors of quadrupoles, ${\tilde S}_{\rm{Q}}({\bf q}, \omega)$, 
	with energy cross sections in (h)--(i).
	The signals for AFM and FM Kitaev SL are equivalent
	[see Fig.~\ref{fig:KitaevSL.thermo}(d) and \ref{fig:KitaevSL.thermo}(e)], and 
	therefore shown only once. 
	Data were obtained from molecular dynamics simulations (details in 
	Appendix~\ref{app:u3MD}) for finite-size clusters of $L=48$ $(N=4608)$,
	at temperatures $T=0.01$ [black arrow in Fig.~\ref{fig:KitaevSL.thermo}(a)].
	(a)--(c) are plotted along the green path in momentum space, as 
	indicated in the left panel of Fig.~\ref{fig:U3.thermo.CSL}(d).
	}
	\label{fig:KitaevSL.dyn}
\end{figure*}
%%%%%%%%%%%%%%%%%%%%%%%%%%%%%%%%%%%%%%%%%%%
%

%%%%%%%%%%%%%%%%%%%%%%%%%%%%%%%%%%%%%%%%%%
\section{Additional dynamics simulations}	
\label{app:dynamics}	
%%%%%%%%%%%%%%%%%%%%%%%%%%%%%%%%%%%%%%%%%%
%

In this Appendix we provide additional results of nontrivial dynamical structure 
factors for 
the semiclassical $S=1$ AFM and FM Kitaev SL 
and the FM chiral spin liquid (CSL), as discussed in Sec.~\ref{sec:Dynamics},

%%%%%%%%%%%%%%%%%%%%%%%%%%%%%%%%%%%%%%%%%%
\subsection{Semiclassical dynamics of the $S=1$ Kitaev spin liquid}	
\label{app:dynamics.Kitaev}	
%%%%%%%%%%%%%%%%%%%%%%%%%%%%%%%%%%%%%%%%%%
%

In Fig.~\ref{fig:KitaevSL.thermo}, we present thermodynamic properties 
of the semiclassical \mbox{$S=1$} AFM and FM Kitaev models at 
$\theta / \pi = 0$ and $\theta / \pi = 1$ of $\mathcal{H}^{\mathcal{A}}_{ {\sf BBQ-K} }$ in
Eq.~\eqref{eq:H.BBQ-K.A}.
The specific heat $C/\ns$ in Fig.~\ref{fig:KitaevSL.thermo}(a) remains identical for AFM and FM 
Kitaev models, showing a crossover from a high-temperature paramagnet into 
a low-temperature cooperative paramagnet.
Here, and throughout the main text, we refer to this cooperative paramagnetic phase as the 
semiclassical analog of the Kitaev SL.
We numerically estimate a specific heat value in the limit of $T \to 0$, as
%
%%%%%%%%%%%%%%%%%%%%%%
\begin{equation}
	\frac{1}{\ns} C(T\to0) = 1.712(2)		\, ,
\label{eq:specH.KSL}
\end{equation}
%%%%%%%%%%%%%%%%%%%%%%
%
which is smaller than $2$, providing direct evidence of zero-energy modes 
in the semiclassical ground state.
The zero-energy modes are evident in the dynamical structure factors, 
as illustrated in Fig.~\ref{fig:KitaevSL.dyn}.

We further present the equal-time structure factors of dipoles, $\tilde{S}_{\rm{S}}({\bf q})$,
and quadrupoles, $\tilde{S}_{\rm{Q}}({\bf q})$, respectively for the 
AFM Kitaev SL in Figs.~\ref{fig:KitaevSL.thermo}(b) and \ref{fig:KitaevSL.thermo}(d),
and for the FM Kitaev SL in Figs.~\ref{fig:KitaevSL.thermo}(c) and \ref{fig:KitaevSL.thermo}(e),
at $T=0.01$.
The structure factors 
in both liquids are very diffuse and show in the 
dipole channel a strong similarity with the \mbox{$S=1/2$} Kitaev 
spin liquid~\cite{Price2012, Price2013, Samarakoon2017}.
Correlations for quadrupoles show quantitatively the same structure for both
the AFM and the FM Kitaev SL states 
with high intensity at the Brillouin zone 
center and some weak structure reaching out to the Brillouin zone corners.

In Fig.~\ref{fig:KitaevSL.dyn}, we present the dynamical structure factors 
[see Eqs.~\eqref{eq:Sqw}--\eqref{eq:Dyn.correct}] for the AFM and 
FM Kitaev SL phases at $T=0.01$, for the same model parameters as used in 
Fig.~\ref{fig:KitaevSL.thermo}.
The structure factors for dipoles in Figs.~\ref{fig:KitaevSL.dyn}(a) and \ref{fig:KitaevSL.dyn}(b)
and their energy cross sections in Figs.~\ref{fig:KitaevSL.dyn}(d)--\ref{fig:KitaevSL.dyn}(g)
show a strong resemblance to results for the $S=1/2$ case~\cite{Samarakoon2017}, 
indicating that correlations are dominated by spin dipoles. 
Additionally, in Fig.~\ref{fig:KitaevSL.dyn}(c), we show the quadrupole structure factor, 
which is equivalent between the AFM and FM Kitaev SL states, revealing a high-intensity 
zero-energy mode, consistently with the reduced specific heat value in 
Eq.~\eqref{eq:specH.KSL}, and a flat band at $\omega/J \approx 2$.
While the ground state is primarily dipolar, dipole characteristics implicitly influence the 
quadrupole components and lead to dominant zero-energy modes also in the
$S_{\rm{Q}}({\bf q})$.
The flat band, solely associated with quadrupole excitations, is determined by the 
interaction strength $J_2$.
Since $J_2=0$ in the ``pure'' Kitaev model, this quadrupole mode becomes dispersionless. 
The energy cross section in Fig.~\ref{fig:KitaevSL.dyn}(h) captures the dominant quadrupole 
correlations at small $\omega$, as it reproduces the pattern in the static structure factors of
Figs.~\ref{fig:KitaevSL.thermo}(d) and \ref{fig:KitaevSL.thermo}(e).
A cross section through the flat band in Fig.~\ref{fig:KitaevSL.dyn}(i) reveals a perfectly constant 
and featureless intensity, indicating the presence of localized excitations.

%%%%%%%%%%%%%%%%%%%%%%%%%%%%%%%%%%%%%%%%%%
\subsection{Dynamics of the FM CSL -- comparison between $\mathbb{CP}^{2}$
and $\mathbb{CP}^{1}$ models}
\label{app:dynamics.CSL.O3}	
%%%%%%%%%%%%%%%%%%%%%%%%%%%%%%%%%%%%%%%%%%
%

%%%%%%%%%%%%%%%%%%%%%%%%%%%%%%%%%%%%%%%%%%%
%. Fig. -- S(q,w) for AFM CSL in O(3) model 
%%%%%%%%%%%%%%%%%%%%%%%%%%%%%%%%%%%%%%%%%%%
\begin{figure*}[t]
	\centering
	\subfloat[ FM CSL, $\mathbb{CP}^{2}$ model, $T=0.003$ ]{
		\includegraphics[width=0.33\textwidth]{wTSqw_T214.pdf}} 
	\subfloat[ FM CSL, $\mathbb{CP}^{1}$ model, $T=0.004$ ]{
		\includegraphics[width=0.33\textwidth]{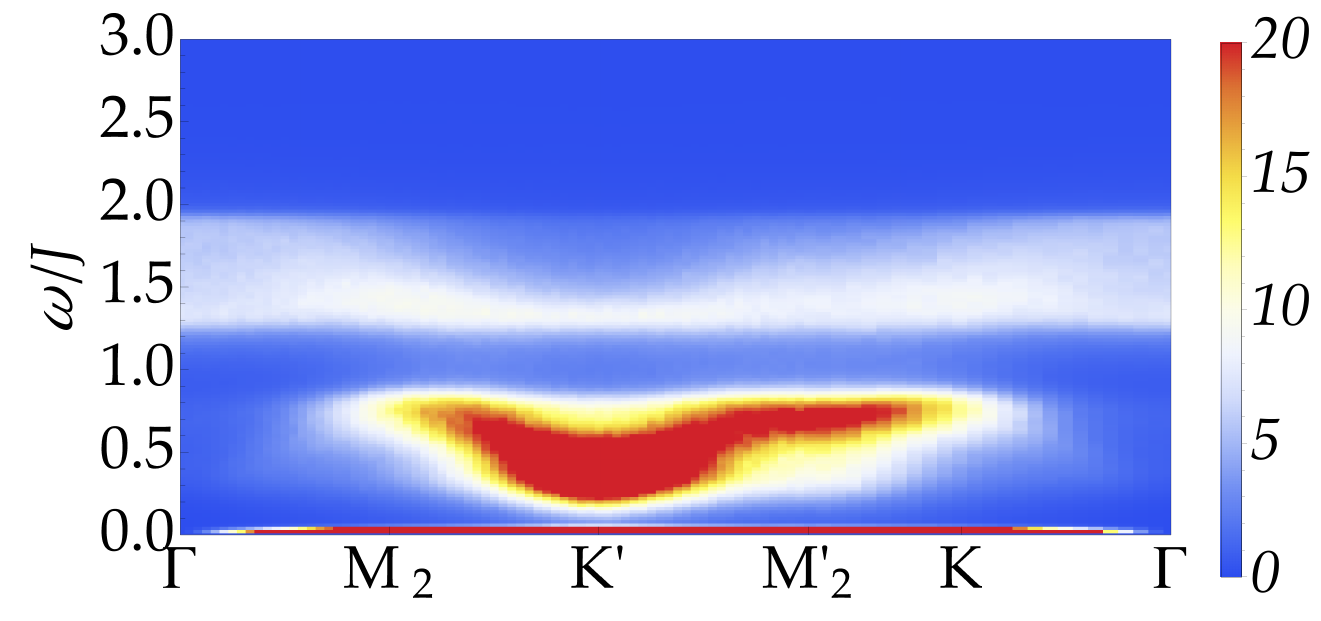}} 
	\subfloat[ FM CSL, $\mathbb{CP}^{1}$ model, $T=0.001$ ]{
		\includegraphics[width=0.33\textwidth]{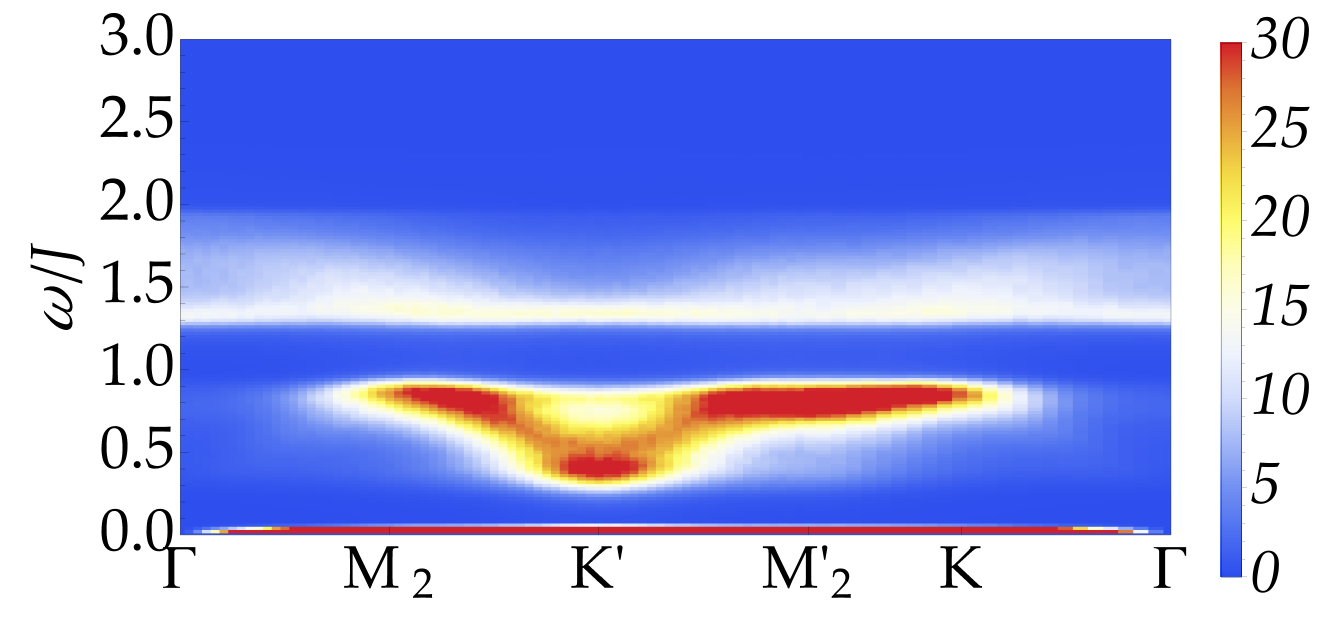}} 	\\
	\subfloat{
		\includegraphics[width=0.33\textwidth]{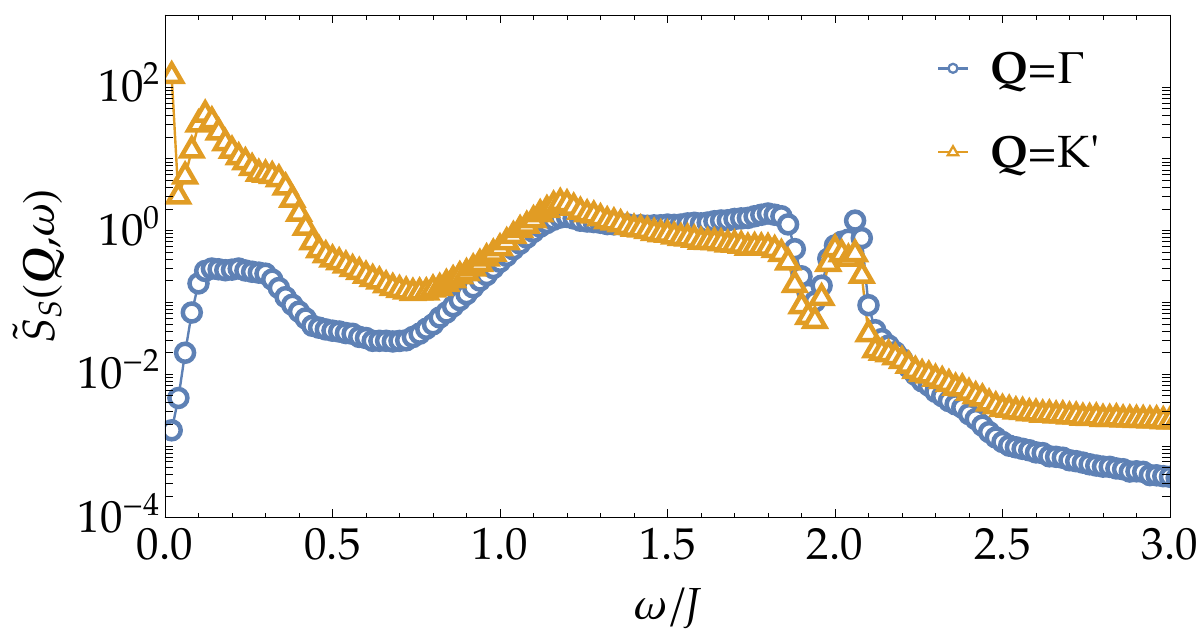}} 
	\subfloat{
		\includegraphics[width=0.33\textwidth]{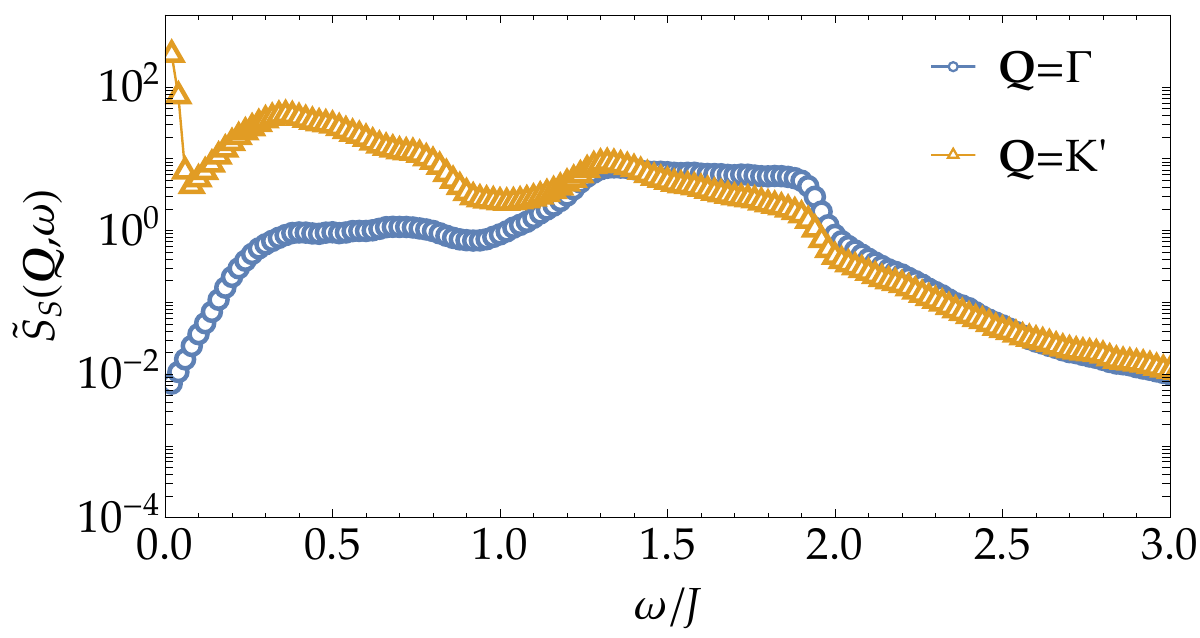}} 
	\subfloat{
		\includegraphics[width=0.33\textwidth]{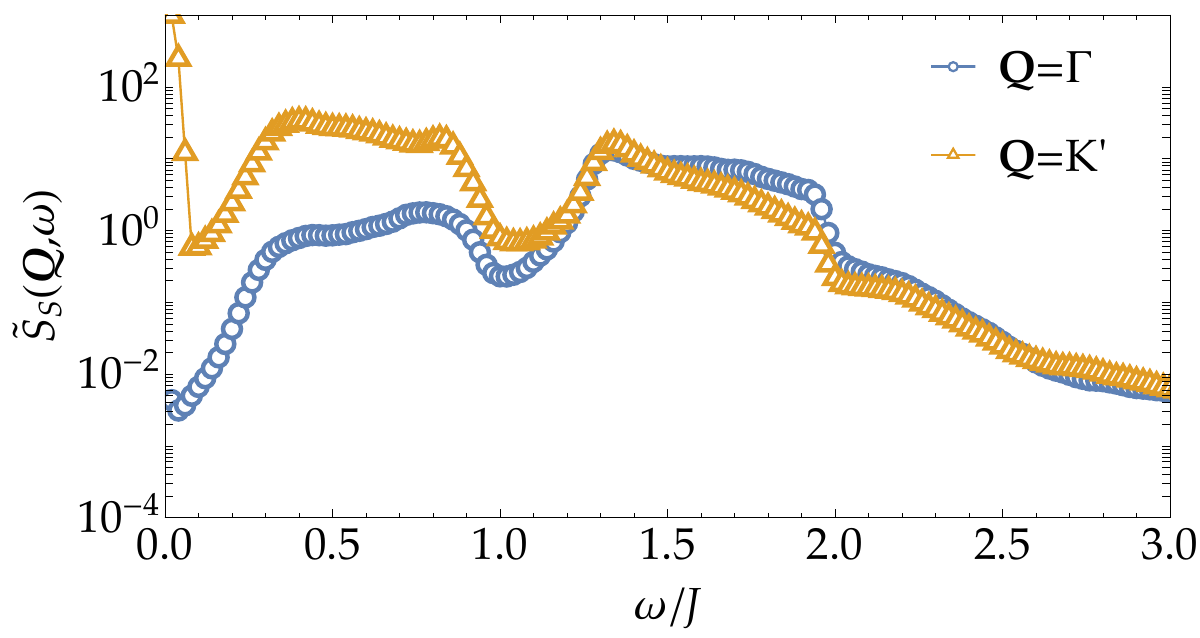}} \\
	\caption{
	Dynamical structure factors for spin-dipole moments, 
	${\tilde S}_{\rm{S}}({\bf q}, \omega)$ [see Eq.~\eqref{eq:Dyn.correct}],
	in the FM CSL phase.	
	Results in the top row are shown along the path indicated by green 
	lines in the left panel of Fig.~\ref{fig:U3.thermo.CSL}(d), while the bottom row
	shows frequency-line cuts at the $\Gamma$ and K$'$ momentum points. 
	All results are presented 
	for the same model parameters:
	$\theta / \pi = 0.06$ and $\phi / \pi = 0.305$ 
	[see Eq.~\eqref{eq:parametrization.HBBQ-K}], and are compared between 
	$\mathbb{CP}^{2}$ model [Eqs.~\eqref{eq:H.BBQ-K.A} and \eqref{eq:EOM1}]
	and  
	$\mathbb{CP}^{1}$ model [Eqs.~\eqref{eq:H.BBQ-K.S} and \eqref{eq:EOM.S}] calculations. 
	(a) Results for the $\mathbb{CP}^{2}$
	model at $T=0.003$ [c.f. Fig.~\ref{fig:Sqw}(b)].
	(b) Results for the $\mathbb{CP}^{1}$
	model at a comparable temperature, $T=0.004$, exhibiting
	essentially the same features as observed for the $\mathbb{CP}^{2}$
	model in (a).
	(c) Results in the $\mathbb{CP}^{1}$ 
	model at very low temperatures, $T=0.001$.
	Data were obtained from molecular dynamics simulations 
	(details in Appendix~\ref{app:u3MD})
	for finite-size clusters of $L=48$ $(N=4608)$.
	}
	\label{fig:Sqw.CSL.O3}
\end{figure*}
%%%%%%%%%%%%%%%%%%%%%%%%%%%%%%%%%%%%%%%%%%%
%

The eight-color model describes a system where the spin degree of freedom is 
discretized, and consequently lacks dynamics. 
Fortunately, as discussed in Secs.~\ref{sec:O3.model} and \ref{sec:SU3.model},  the CSL is 
stabilized in models with continuous spin degree of freedom, which allows us to simulate its 
dynamical properties (see technical details in Appendix~\ref{app:u3MD}). 
Here, we compare the dynamical structure factors of the FM CSL between the 
$\mathbb{CP}^{2}$ model and the $\mathbb{CP}^{1}$ model,  
and demonstrate that their qualitative signatures are essentially the same.

In Fig.~\ref{fig:Sqw.CSL.O3}, we show the dynamical structure factors for spin-dipole 
moments, ${\tilde S}_{\rm{S}}({\bf q}, \omega)$ [see Eqs.~\eqref{eq:Sqw} and 
\eqref{eq:Dyn.correct}], in the FM CSL 
phase for both the $\mathbb{CP}^{2}$ model
and the $\mathbb{CP}^{1}$ model,
at the same model parameters of 
$\theta / \pi = 0.06$ and $\phi / \pi = 0.305$.
The top row shows the spectrum along the path in momentum space as indicated by 
green lines in the left panel of Fig.~\ref{fig:U3.thermo.CSL}(d).
The bottom row depicts frequency cuts at the explicit momentum points 
$\Gamma$ and K$'$.

For better comparison we replot in Fig.~\ref{fig:Sqw.CSL.O3}(a) the result from 
the $\mathbb{CP}^{2}$ model, as previously shown in Fig.~\ref{fig:Sqw}(b).
The spectrum exhibits three energy regimes: a zero-energy flat band, a slightly gapped diffuse band, and a 
largely gapped and diffuse continuum with broad bandwidth at higher energies. 
Additionally, a flat, weak intensity band around \mbox{$\omega/J \approx 2$} is observed, arising 
from the mixing between dipole and quadrupole excitations due to spin anisotropy.
These energy regimes are quantified in the bottom row.

In Fig.~\ref{fig:Sqw.CSL.O3}(b), qualitatively similar features 
are visible in the dispersion obtained from $\mathbb{CP}^{1}$ model
simulations, performed at nearly the same temperature.
Here, the diffuse band at intermediate energies exhibits a larger bandwidth while clearly showing 
an energy gap to the high-intensity zero-energy mode.
The broad continuum is slightly shifted to higher energies compared to the $\mathbb{CP}^{2}$ case 
but demonstrates almost the same intensity distribution. 
The nearly flat band around $\omega/J \approx 2$ is absent in the $\mathbb{CP}^{1}$
case, as quadrupoles 
are strictly excluded in these simulations.

In Fig.~\ref{fig:Sqw.CSL.O3}(c), we additionally present the spectrum for the 
$\mathbb{CP}^{1}$
model at \mbox{$T=0.001$}, a temperature that is not accessible in the 
$\mathbb{CP}^{2}$
model, as the FM CSL transitions into the NC-1 ordered state before reaching such low temperatures 
(see thermodynamics in Sec.~\ref{sec:U3.model.thermo}). 
Since thermal fluctuations are reduced, the spectrum becomes slightly sharper, especially the 
gap of the diffuse intermediate band to the zero-energy mode becomes more pronounced.
However, importantly, qualitative features between $\mathbb{CP}^{1}$ and $\mathbb{CP}^{2}$
model simulations remain the same.

This comparison confirms our intuition that dominant properties of the CSL are 
still captured in the $\mathbb{CP}^{1}$ model,
which includes only dipole spin degree of freedom.
Additionally, simulation results for the dynamics in both the $\mathbb{CP}^{1}$ and 
$\mathbb{CP}^{2}$ models reveal a 
very diffuse and gapped continuum, associated with the excitations of the FM CSL, as discussed 
in the eight-color model in Sec.~\ref{sec:8c.model}.

%%%%%%%%%%%%%%%%%%%%%%%%%%%%%%%%%%%%%
%
%					BIBLIOGRAPHY
%
%%%%%%%%%%%%%%%%%%%%%%%%%%%%%%%%%%%%%
\bibliography{Bibliography}
%%%%%%%%%%%%%%%%%%%%%%%%%%%%%%%%%%%%%

%%%%%%%%%%%%%%%%%%%%%%%%%%%%%%%%%%%%%
\end{document}